\documentclass[12pt,a4paper]{article}
\usepackage{amstext}
\usepackage{graphics}
\newcommand{\dofour}[2]{%
  \begin{figure}[tp]
    \begin{center}
      \includegraphics{yr_#1_161.eps}
      \includegraphics{yr_#1_175.eps}
      \includegraphics{yr_#1_190.eps}
      \includegraphics{yr_#1_205.eps}
    \end{center}
    \caption{\label{fig:#1}#2}
  \end{figure}}
 
\def\ADLO/{\texttt{ADLO/TH}}
\def\ALPHA/{\texttt{ALPHA}}
\def\COMPHEP/{\texttt{CompHEP}}
\def\ERATO/{\texttt{ERATO}}
\def\EXCALIBUR/{\texttt{EXCALIBUR}}
\def\GENTLE/{\texttt{GENTLE}}
\def\GRC4F/{\texttt{grc4f}}
\def\JETSET/{\texttt{JETSET}}
\def\KORALW/{\texttt{KORALW}}
\def\LEPWW/{\texttt{LEPWW}}
\def\LPWW02/{\texttt{LPWW02}}
\def\PYTHIA/{\texttt{PYTHIA}}
\def\WOPPER/{\texttt{WOPPER}}
\def\WPHACT/{\texttt{WPHACT}}
\def\WTO/{\texttt{WTO}}
\def\WWF/{\texttt{WWF}}
\def\WWFTSH/{\texttt{WWFTSH}}
\def\WWGENPV/{\texttt{WWGENPV}}
\def\HIGGSPV/{\texttt{HIGGSPV}}
\newcommand{\GeV}{\mathop{\rm GeV}\nolimits}
\newcommand{\MeV}{\mathop{\rm MeV}\nolimits}
 
\begin{document}
 
%
\def\be{\begin{equation}}
\def\ee{\end{equation}}
\def\ba{\begin{eqnarray}}
\def\ea{\end{eqnarray}}
\def\sw{\mbox{$\sin^2\theta_w$}}
\def\Sw{\sin^2\theta_w}
\def\slash{\hspace{-5pt}/}
\def\el{\mbox{$e_L$}}
\def\elbar{\mbox{$\bar e_L$}}
\def\blrbar{\mbox{$\bar b_{L,R}$}}
\def\blr{\mbox{$b_{L,R}$}}
\def\a2n{${\cal A}(2\nolinebreak \to n)$}
\def\an{${\cal A}(1\to n)$}
\def\cgp{C.G.~Pa\-pa\-do\-pou\-los }
\def\sdv{S.D.P.~Vlas\-so\-pu\-los }
\def\ena{E.N.~Ar\-gy\-res }
\def\rk{R.H.P.~Kleiss}
\def\nga{N.G.~Anto\-ni\-ou}
\def\an{{non-standard couplings}}
\def\ano{{non-standard}}
\def\el{{$e^-\bar{\nu}_e u \bar{d}$}}
\def\mul{{$\mu^-\bar{\nu}_{\mu} u \bar{d}$}}
\def\ert{{\tt ERATO}}
\def\exc{{\tt EXCALIBUR}}
\def\exca{{\tt EXCALIBUR1}}
\def\pb{\parbox}
\def\va{\vspace*{-4pt}}
\def\Ord{\buildrel{\scriptscriptstyle <}\over{\scriptscriptstyle\sim}}
\def\OOrd{\buildrel{\scriptscriptstyle >}\over{\scriptscriptstyle\sim}}
\def\zp #1 #2 #3 {{\it Z.~Phys.} {\bf#1} (#2) #3}
\def\rmp #1 #2 #3 {{\it Rev. Mod. Phys.} {\bf#1} (#2) #3}
\def\xx #1 #2 #3 {{\bf#1}, (#2) #3}
\def\xx #1 #2 #3 {{\bf#1}, (#2) #3}
\def\z0{Z}
\def\gf{G_{\mu}}
\def\zm{M_{_Z}}
\def\bm{m_b}
\def\cm{m_c}
\def\gev{{\hbox{GeV}}}
\def\tev{{\hbox{TeV}}}
\def\nb{{\hbox{nb}}}
\def\msb{{\overline{MS}}}
\def\als{\alpha_{_{S}}}
\def\tm{m_{t}}
\def\hm{M_{_H}}
\def\wm{M_{_W}}
\def\gn{\Gamma_{\nu}}
\def\ge{\Gamma_{e}}
\def\gmu{\Gamma_{\mu}}
\def\gt{\Gamma_{\tau}}
\def\gl{\Gamma_{l}}
\def\gu{\Gamma_{u}}
\def\gd{\Gamma_{d}}
\def\gc{\Gamma_{c}}
\def\gs{\Gamma_{s}}
\def\gb{\Gamma_{b}}
\def\gz{\Gamma_{_Z}}
\def\gh{\Gamma_{h}}
\def\gi{\Gamma_{\rm {inv}}}
\def\afb{A_{_{\rm {FB}}}}
\def\alr{A_{_{\rm {LR}}}}
\def\gv{g_{_V}}
\def\ga{g_{_A}}
\def\barf{\overline f}
\def\barq{\overline q}
\def\barb{\overline b}
\def\bart{\overline t}
\def\barc{\overline c}
\def\gvf{g^f_{_{V}}}
\def\gaf{g^f_{_{A}}}
\def\gvl{g^l_{_{V}}}
\def\gal{g^l_{_{A}}}
\def\gvb{g^b_{_{V}}}
\def\gab{g^b_{_{A}}}
\def\gsvb{g^b_v}
\def\gsab{g^b_a}
\def\ord {\cal O}
\def\ical{\cal I}
\def\shat{\hat s}
\def\chat{\hat c}
\def\vhat{\hat v}
\def\thetahat{\hat \theta}
\def\alphahat{\hat \alpha}
\def\fvf{F_{_V}^f}
\def\faf{F_{_A}^f}
\def\fvl{F_{_V}^l}
\def\fal{F_{_A}^l}
\def\acal{\cal A}
\def\ste{\sin\theta}
\def\stes{\sin^2\theta_{\rm{eff}}}
\def\xhat{\hat x}
\def\piv{\Pi_{_V}}
\def\pia{\Pi_{_A}}
\def\dr{\Delta r}
\def\drl{\Delta r_{_L}}
\def\dgvf{\delta g_{_V}^f}
\def\dgaf{\delta g_{_A}^f}
\def\dgvl{\delta g_{_V}^l}
\def\dgal{\delta g_{_A}^l}
\def\i3f{I^{(3)}_f}
\def\pih{{\hat\Pi}}
\def\sgh{{\hat\Sigma}}
\def\osp2{16\,\pi^2}
\def\ap2{\left(p^2\right)}
\def\stw{s_{\theta}}
\def\ctw{c_{\theta}}
\def\stws{s_{\theta}^2}
\def\stwf{s_{\theta}^4}
\def\ctws{c_{\theta}^2}
\def\Szg{\Sigma_{_{Z\gamma}}}
\def\Szz{\Sigma_{_{ZZ}}}
\def\Sww{\Sigma_{_{WW}}}
\def\Swwg{\Sigma_{_{WW}}^{^G}}
\def\Stg{\Sigma_{_{3Q}}}
\def\Stt{\Sigma_{_{33}}}
\def\Pgg{\Pi_{\gamma\gamma}}
\def\Pf{\Pi_{_F}}
\def\rhou{\rho_{_U}}
\def\ku{\kappa_{_U}}
\def\rhoz{\rho_{_Z}}
\def\rhozr{\rho^{\scriptscriptstyle R}_{_Z}}
\def\gfd{\gamma_5}
\def\gau{\gamma^{\alpha}}
\def\gad{\gamma_{\alpha}}
\def\mev{{\hbox{MeV}}}
\def\gev{{\hbox{GeV}}}
\def\tmo{\times 10^{-1}}
\def\tmt{\times 10^{-2}}
\def\tmth{\times 10^{-3}}
\def\tmf{\times 10^{-4}}
\def\tmfv{\times 10^{-5}}
\def\srt{\sqrt{2}}
\def\xsf{\sigma_{_F}}
\def\xsb{\sigma_{_B}}
\def\chig{\chi_{\gamma}}
\def\chiz{\chi_{_Z}}
\def\s0h{\sigma^h_0}
\def\ea{\end{eqnarray}}
\def\nl{\nonumber \\}
\def\baral{\bar{\alpha}}
\mark{{}{}}
\newcommand{\nn}{\nonumber}
\newcommand{\lb}{\linebreak}
\newcommand{\fig}[1]{Fig.\ref{#1}}
\newcommand{\bq}{\begin{equation}}
\newcommand{\bc}{\begin{center}}
\newcommand{\ec}{\end{center}}
\renewcommand{\theenumiii}{\arabic{enumiii}}
\newcommand{\eq}{\end{equation}}
\newcommand{\bqa}{\begin{eqnarray}}
\newcommand{\eqa}{\end{eqnarray}}
\newcommand{\ben}{\begin{enumerate}}
\newcommand{\een}{\end{enumerate}}
\renewcommand{\theenumii}{ \arabic{enumii} }
\newcommand{\eqn}[1]{Eq.(\ref{#1})}
\newcommand{\itema}{\item\addtocounter{cit}{1}}
\newcommand{\itemb}{\item\addtocounter{cit}{1}\addtocounter{cita}{1}}
\newcommand {\alphaqed} {\alpha_{\mathrm Q\mathrm E\mathrm D}}
\newcommand {\alphas} {\alpha_{\mathrm s}}
\newcommand {\ptee} {P_T}
\newcommand {\mw} {M_{\mathrm{W}}}
\newcommand {\mz} {M_{\mathrm{Z}}}
\newcommand {\gw} {\Gamma_{\mathrm{W}}}
\newcommand {\munudu} {\mu^-\bar{\nu}_\mu\bar{\mathrm{d}}\mathrm{u}}
\newcommand {\mumuuu} {\mu^-\mu^+\bar{\mathrm{u}}\mathrm{u}}
\newcommand {\uuuu}
{\mathrm{u}\bar{\mathrm{u}}\bar{\mathrm{u}}\mathrm{u}}
\newcommand {\uddu}
{\mathrm{d}\bar{\mathrm{u}}\bar{\mathrm{d}}\mathrm{u}}
\newcommand {\enuud}
{\mathrm{e}^-\bar{\nu}_{\mathrm{e}}\bar{\mathrm{d}}\mathrm{u}}
\newcommand {\ie} {{\it i.e.}}
\newcommand {\eg} {{\it e.g.}}
\newcommand {\etal} {{\it et al }}
\newcommand {\goto} {\rightarrow}
\newcommand {\lepww} {{\tt LPWW}}
\newcommand {\gentle} {{\tt GENTLE}}
\newcommand {\excalibur} {{\tt EXCALIBUR}}
\newcommand {\pythia} {{\tt PYTHIA}}
\newcommand{\br}{\begin{eqnarray}}
\newcommand{\er}{\end{eqnarray}}
\newcommand{\barr}{\begin{array}}
\newcommand{\earr}{\end{array}}
\newcommand{\bi}{\begin{itemize}}
\newcommand{\ei}{\end{itemize}}
\newcommand{\bn}{\begin{enumerate}}
\newcommand{\en}{\end{enumerate}}
\newcommand{\ul}{\underline}
\newcommand{\ol}{\overline}
\newcommand{\eebbww}{$e^+e^-\rightarrow b\bar b W^+W^-$}
\newcommand{\bb}{$ b\bar b \ $}
\newcommand{\ttb}{$ t\bar t \ $}
\newcommand{\ar}{\rightarrow}
\newcommand{\sm}{${\cal {SM}}\ $}
\newcommand{\Dir}{\kern -7.4pt\Big{/}\kern 1.pt}
\newcommand{\Dirin}{\kern -13.4pt\Big{/}\kern 7.4pt}
\newcommand{\DDir}{\kern -7.6pt\Big{/}}
\newcommand{\DGir}{\kern -6.0pt\Big{/}}
\newcommand{\sla}{\kern -5.4pt /}
\newcommand{\dotp}{\!\cdot\!}
\newcommand{\bea}{\begin{eqnarray}}
\newcommand{\eea}{\end{eqnarray}}
\newcommand{\hed}[1]{\noindent{\bf #1 \\}}
\def\Was{W\c as}
\def\Order#1{${\cal O}(#1$)}
\def\lint{\int\limits}
\def\bbeta{\bar{\beta}}
\def\tbeta{\tilde{\beta}}
\def\talpha{\tilde{\alpha}}
\def\tomega{\tilde{\omega}}
%
\newcommand{\bfig}{\begin{center}\begin{picture}}
\newcommand{\efig}[1]{\end{picture}\\{\small #1}\end{center}}
\newcommand{\flin}[2]{\ArrowLine(#1)(#2)}
\newcommand{\wlin}[2]{\DashLine(#1)(#2){2}}
\newcommand{\zlin}[2]{\DashLine(#1)(#2){5}}
\newcommand{\glin}[3]{\Photon(#1)(#2){2}{#3}}
\newcommand{\lin}[2]{\Line(#1)(#2)}
\newcommand{\sof}{\SetOffset}
\newcommand{\bmip}[2]{\begin{minipage}[t]{#1pt}\bfig(#1,#2)}
\newcommand{\emip}[1]{\efig{#1}\end{minipage}}
\newcommand{\putk}[2]{\Text(#1)[r]{$p_{#2}$}}
\newcommand{\putp}[2]{\Text(#1)[l]{$p_{#2}$}}
\newcommand{\ibidem}{{\it ibidem\/},}
\newcommand{\into}{\;\;\to\;\;}
\newcommand{\epl}{e^+}
\newcommand{\emn}{e^-}
\newcommand{\nue}{\nu_e}
\newcommand{\nueb}{\bar{\nu}_e}
\newcommand{\mpl}{\mu^+}
\newcommand{\mmn}{\mu^-}
\newcommand{\num}{\nu_{\mu}}
\newcommand{\numb}{\bar{\nu}_{\mu}}
\newcommand{\tpl}{\tau^+}
\newcommand{\tmn}{\tau^-}
\newcommand{\nut}{\nu_{\tau}}
\newcommand{\nutb}{\bar{\nu}_{\tau}}
\newcommand{\ubar}{\bar{u}}
\newcommand{\dbar}{\bar{d}}
\newcommand{\cbar}{\bar{c}}
\newcommand{\sbar}{\bar{s}}
\newcommand{\bbar}{\bar{b}}
\newcommand{\ww}[2]{\langle #1 #2\rangle}
\newcommand{\wws}[2]{\langle #1 #2\rangle^{\star}}
\newcommand{\smod}{\tilde{\sigma}}
\newcommand{\dilog}[1]{\mbox{Li}_2\left(#1\right)}
\newcommand{\umu}{^{\mu}}
\newcommand{\cjg}{^{\star}}
\newcommand{\lgn}[1]{\log\left(#1\right)}
\newcommand{\si}{\sigma}
\newcommand{\sit}{\sigma_{tot}}
\newcommand{\sqs}{\sqrt{s}}
\newcommand{\sih}{\hat{\sigma}}
\newcommand{\sith}{\hat{\sigma}_{tot}}
\newcommand{\p}[1]{{\scriptstyle{\,(#1)}}}
\newcommand{\res}[3]{$#1 \pm #2~~\,10^{-#3}$}
\newcommand{\rrs}[2]{\multicolumn{1}{l|}{$~~~.#1~~10^{#2}$}}
\newcommand{\err}[1]{\multicolumn{1}{l|}{$~~~.#1$}}
\newcommand{\ru}[1]{\raisebox{-.2ex}{#1}}
%
%
\def\diagram#1{%
  \bgroup
    \def\fmfL(##1,##2,##3)##4{\put(##1,##2){\makebox(0,0)[##3]{##4}}}
    \unitlength=1mm
    \begin{picture}(35,20)%
      \put(0,0){\includegraphics{#1.eps}}%
      \input{#1.tex}%
    \end{picture}%
  \egroup}%
\begin{center}
{\large {\bf EVENT GENERATORS FOR WW PHYSICS}}
\end{center}
\begin{center}
{\it Conveners}: D.~Bardin and R.~Kleiss
\end{center}
\begin{center}
{\it Working group}:
E.~Accomando,
H.~Anlauf,
A.~Ballestrero,
F.A.~Berends,
E.~Boos,
F.~Caravaglios,
D.~van Dierendonck,
M.~Dubinin,
V.~Edneral,
F.C.~Ern\'e,
J.~Fujimoto,
V.~Ilyin,
T.~Ishikawa,
S.~Jadach,
T.~Kaneko,
K.~Kato,
S.~Kawabata,
Y.~Kurihara,
D.~Lehner,
A.~Leike,
R.~Miquel,
G.~Montagna,
M.~Moretti,
T.~Munehisa,
O.~Nicrosini,
T.~Ohl,
A.~Olchevski,
G.J.~van Oldenborgh,
C.G.~Papadopoulos,
G.~Passarino,
D.~Perret-Gallix,
F.~Piccinini,
R.~Pittau,
W.~P\l{}aczek,
A.~Pukhov,
V.~Savrin,
M.~Schmitt,
S.~Shichanin,
Y.~Shimizu,
T.~Sj\"ostrand,
M.~Skrzypek,
H.~Tanaka,
Z.~W\c{a}s
\end{center}
 
\vspace*{1.0cm}
 
\tableofcontents
 
\newpage
 
\section{Introduction: the need for Monte Carlo}
 
\noindent
In this report we shall deal with the practical
implementation of the theoretical results described in the
WW study group report. There, many important
results and formulae have been given which have to find
their way into the analysis of the LEP2 data, in particular
those dealing with the measurement of the W mass and couplings.
It is our aim to describe the current state of the art
of this implementation.
 
The simplest detectable final states
of relevance are those consisting of four fermions (when we
disregard the complications arising from photon bremsstrahlung,
gluon bremsstrahlung and hadronization effects), and consequently
the phase space has seven dimensions (eight, if we also
include the overall azimuthal distribution of events around
the beam axis -- this distribution, however, is trivial
as long as no transversely polarized beams are considered).
Obviously, the sets of diagrams that contribute to a given
final state is also quite complicated. Below, we shall
present a classification of the various sets of diagrams
that we have found useful in discussing and comparing results.
When we also take into account the complicated peaking structures
resulting from the many different Feynman diagrams, it becomes clear
that the only way in which we can arrive at experimentally
meaningful results in which all cuts can be accommodated is that
of Monte Carlo simulation of the full event. This feature is
even more pronounced than at LEP1, where the important events
have a two-fermion final state, with only one relevant
angular variable, and little peaking structure at given energy.
There are, of course, processes such as
$e^+ e^- \to W^+ W^- \to q \bar{q} \mu \nu_{\mu}$
where experimental cuts tend to be not very drastic, but even
is such cases the estimate of a given experiment's acceptance
and efficiency will probably have to rely on Monte Carlo
simulation, even if the final fits are performed in some
semi-analytic fashion. This is even more the case if in the
above process we replace the muon by the electron.
 
\subsection{Semianalytics versus event generators}
 
\noindent
Notwithstanding all this, it is very desirable to have
at our disposal also calculations that do not rely on
explicit event generation. As is the case in LEP1 physics,
a number of semi-analytical results have been obtained,
mainly in the form of the {\tt GENTLE} code, which extends
the formalism of~\cite{muta} to integrate analytically
over a number of variables, and performs the few remaining
integrations using standard numerical packages
(see~\cite{gentle_unicc11} and references therein).
Although in this
way neither all diagrams nor all possible experimental cuts
can be incorporated, we feel that the existence of such
results, with an inherently much smaller numerical error
as well as excellent control over the theoretical input,
establishes an important benchmark for the Monte Carlo
programs. As will be clear from our comparisons of the
results of the various programs, {\tt GENTLE} indeed
serves, in many cases, as such a benchmark,
especially in the `tuned comparisons' we describe below.
 
Essentially all Monte Carlo codes presented here consist of
two main ingredients, incorporated in (usually) three steps
to produce numerical output.
The ingredients are:
\begin{itemize}
\item a set of routines that, for given values of
  the fermions' four-momenta, produce the value of the
  matrix element, squared, and summed/averaged over the
  appropriate spins and colors. A wide number of techniques
  are used to obtain the matrix elements. For example,
  the {\tt ALPHA} code takes as input the effective action
  of the theory, and numerically computes the saddle point
  of the path integral for given external momenta, without
  explicit reference to Feynman diagrams.
  The {\tt ERATO, EXCALIBUR, WTO, WPHACT}, and {\tt WWGENPV} codes
  (among many) use different kinds of helicity techniques,
  where the relevant diagrams are either put in `by hand'
  or generated by some semi-automatic procedure. Yet other
  codes such as the {\tt CompHEP} and {\tt grc4f}
  programs employ a fully automated diagram-generating-and-evaluating
  code. The fact that such disparate treatments manage to
  come up with agreeing numbers can be viewed as important
  checks on the correctness of the various individual procedures.
  Some programs (in particular {\tt ALPHA} and {\tt WWFT})
  also incorporate explicit photons
  into the computation of the matrix element, while the {\tt grc4f},
  {\tt PYTHIA} and {\tt WOPPER} programs use `parton shower'
  techniques to generate photons, the {\tt KORALW} code
  employs the so-called YFS approach,
  and WWGENPV uses a $p_T$-dependent
  structure-functions-inspired formulation.
  It should also be stressed that not all programs can
  compute all contributing Feynman diagrams: this important
  fact should be kept  in mind when we discuss the results.
\item a set of routines that transform uniformly distributed
  pseudo-random numbers into phase space variables, taking
  as much of the peaking structure as possible into account
  by a number of mappings and branch choices. Again, different
  programs employ widely different techniques to this end.
  In particular for processes with electrons or positrons
  in the final state the occurrence of $t$-channel photon
  exchange calls for a very careful treatment.
\end{itemize}
Obviously, the distinction between these two ingredients
is not always completely straightforward, especially in
codes that employ `showering', where the phase space generation
should itself induce the correct matrix elements. Also,
not all programs use pseudo-random numbers as a basis for
the phase space generation: some codes employ `black box'
integrators such as provided by the NAG library, while
the {\tt WTO} uses quasi-random, deterministic number sets
(technically known as shifted Korobov sets).
 
The running of a typical Monte Carlo consists of three steps:
\begin{itemize}
\item initialization: here the input parameters are read in,
and various preparatory steps are undertaken. For instance,
{\tt EXCALIBUR} will, at this stage, determine the contributing
Feynman diagrams and print them, and work out which peaking
structures contribute.
\item generation: here a event-generating routine is called
 the desired number of times to arrive at a phase space point
 together with its matrix element. Also the necessary filling
 of histograms and other bookkeeping is performed in this step.
\item evaluation: when the desired number of events has been
 produced, the total cross section is computed as the
 average event weight, where the event weight is defined as the
 ratio of the matrix element squared over the phase space Jacobian.
\end{itemize}
For details about the workings of the various different programs
we refer to the next subsection, where more information is given
for each individual program, together with the necessary references.
 
\subsection{The Ultimate Monte Carlo}
The above rough description does, of course, no justice to the
effort that has already gone into all the existing codes: but it
is only fair to say that, at  this moment, none of them
can be considered as the definitive program. This `Ultimate
Monte Carlo' (which may remain out of reach) is approached, by
different authors, in different ways, and some programs have
desirable features (for instance, explicit, finite-$p_T$
photons), that are not shared by other programs, which however
have their own attractions (for instance, inclusion of
all Feynman diagrams). As we have already indicated, it must be
always kept in mind, when comparing programs, that such differences
in approach will unavoidably result in differences in results;
{\em but such differences should {\bf not} be regarded as
any kind of theoretical uncertainty, but rather as an indication
of the importance of the different ingredients}.
In fact, the real theoretical uncertainty (due, for example,
to unknown higher-order corrections) is quite distinct from the
differences between programs.
It may be
instructive to give a list of the features of the Ultimate
Monte Carlo, in order for the user to appreciate to what extent
a given program satisfies her/his needs in a particular analysis.
The Ultimate Monte Carlo should:
\begin{itemize}
\item treat all possible four-fermion final states, with
  all relevant Feynman diagrams (possibly with the option
  to restrict the set of diagrams).
\item produce gauge-invariant results. If one
  describes off-shell, unstable W pair production using only
  the three Feynman diagrams in the \emph{CC03} sector, then gauge
  dependence will result. Fortunately, at LEP2 energies
  these effects are very small provided a suitable gauge such
  as the unitary or 't Hooft-Feynman gauge is chosen: but,
  especially when t-channel photon exchange takes place,
  the gauge cancellations can be very delicate.
  Related to this is the requirement that the various
  coupling constants are chosen in a consistent manner.
\item have a correct treatment of the bosonic widths. This
  is closely related to the previous point: if one just
  inserts a running width, gauge invariance is lost,
  with dramatic results for final states with electrons or positrons.
  This problem, and its various possible resolutions,
  are described in detail in~\cite{bhfpaper}.
\item have the fermion masses taken into account. For instance,
 {\tt EXCALIBUR} treats the fermions as strictly massless, which
 accelerates the computation of the matrix elements considerably,
 but imposes the need to avoid phase space singularities by
 explicit cuts, and makes it impossible to incorporate Higgs
 production and decay consistently.
\item have explicit, $p_T$-carrying photons. This is
 of particular importance for a distinction of ``initial'' and ``final''
 state radiation in an $M_W$ measurement,
 as well as the search for anomalous couplings.
\item have the higher-order photonic radiative corrections taken
 into account properly. This probably does not mean, given the
 experimental accuracy to be expected at LEP2, that very high
 orders or very high precision are required, but it would
 be very useful to be able to prove that radiative effects
 are small for a particular quantity.
 For instance, the Coulomb singularity which modifies the WW
 intermediate state is an important effect.
\item should have good control over the non-QED radiative correction,
 preferably in the form of the complete ${\cal O}(\alpha)$
 corrections, and resummed higher-order effects where necessary.
\item incorporate QCD effects, both in the W self-energy and in
 the gluonic corrections to quark final states. Also relevant
 is the interference between electroweak and QCD channels in
 the production of four-quark final states.
 In this place it should be remarked that it is of course
 trivial to add the `naive' QCD correction $1+\alpha_s/\pi$
 to the total cross section, but in the presence of cuts
 this may be less appropriate: the particular strategy adopted
 must depend on the interface with a hadronization routine.
\item have a good interface to hadronization packages. This
 is especially relevant to the W mass measurement, together with
 the next point:
\item give information, for each generated event, on how
 much of the matrix element is contributed by each subset
 of Feynman diagrams, and/or each
 color configuration. This is important for problems of
 color reconnection and Bose-Einstein effects.
\item have Higgs production and decay implemented.
\item have the possibility of anomalous couplings. This
 allows for the study of the effects of such couplings to
 good precision using control-variate techniques (that is,
 switching the anomalous couplings on and off for a given
 event sample, thereby avoiding statistical fluctuations
 that might wash out the small anomalous effects).
\end{itemize}
 
\subsection{Comparison generalities}
The rest of this contribution deals with the description
and the comparison of the different codes and their results.
It must again be stressed, that the field is still in a
state of flux, and probably not one of the programs has taken on
its final form. We can, therefore, only present results as they
are at this particular moment (December 1995), with the remark that
most of the discrepancies are well-understood and are expected
to decrease significantly in the near future.
There are several ways in which we have compared the
various codes:
\begin{itemize}
\item {\bf by ingredients}\\
 To this end, we just compare which of the features of the
 Ultimate Monte Carlo are part of the different codes. Again,
 we stress that the choice of code depends to a large extent on
 the user's particular problem. For instance, background studies
 will require a code that contains all Feynman diagrams, while
 high-precision studies of inclusive quantities may be better
 of with a semi-analytical program such as {\tt GENTLE}.
 In the next section we present what we feel to be the most
 relevant information on each program.
\item {\bf by `tuned' comparison}\\
 This means that we have chosen a minimal process described
 by a minimal set of diagrams (\emph{CC03} and \emph{CC10}), for which we
 have computed several quantities. The idea of this exercise
 is that {\em all programs should agree on these numbers}.
 Of course, one must make sure that the physical parameters
 of the theory such as masses and widths in propagators, and the
 coupling constants in the Lagrangian, are constructed to be
 identical in all codes. The aim is twofold.
 In the first place it allows to establish the {\em technical
 precision\/} of the various codes, and we have come (as will
 be shown) to a satisfactory number of one per mille or better,
 at least for a large cluster of dedicated codes.
 In the second place, such a tuned comparison is a good bug hunting
 ground, as we have found. Many small differences usually can be
 traced back either to small bugs or small differences in
 input parameters or cuts.
\item {\bf by `best you can do' comparison}\\
 The tuned comparison, useful as it is, is not of direct
 experimental relevance since it relies on switching off all
 features in which one program is better than another. The
 real physics results must of course incorporate more than
 this bare minimum, and therefore we have computed a number of
 quantities, for one class of processes,
 in which (apart from agreed-upon input parameters)
 each code provides us with its own `best answer'. Again, we
 want to stress that these results do not agree, nor should they
 be expected to: differences in these results reflect differences
 in the physics approach. Comparisons apart, in the end
 the programs will have to provide the community with explicit
 predictions, and this `best you can' should give an idea
 of the extent to which these predictions depend on the
 various pieces of physics input.
 Whereas the results of the tuned comparison are not expected
to change appreciably in the near future, the `best you can'
results must, and probably will, converge over time as more
physics input is incorporated into more programs.
\item {\bf by `all you can do' comparison}\\
finally, we have let the programs pass an `all you can do'
 comparison phase, where each program has computed essentially all
 the processes it is able to treat. Of course, only some out of all
 the codes can do {\em all\/} four-fermion processes: but from such
 a game should arise a coherent picture of what the
 current state-of-the-art is. Another goal of the `all you can do'
 comparison, which is also `tuned', is to provide
 precision benchmarks for {\it all four-fermion processes}.
\end{itemize}
 
\subsection{A classification of 4-fermion processes}
For the various four-fermion final states produced in $e^+e^-$ annihilation,
the numbers of contributing Feynman diagrams are quite different.
On top of double-pole (WW or ZZ) diagrams there are, in general,
a lot of so-called background diagrams with different
intermediate states, which are single-resonant or non-resonant.
In this section we present a classification of all
four-fermion final states in the Standard Model
\footnote{The classification is done with the help of
{\tt CompHEP}~\cite{comphep}.}.
This classification was originally proposed in \cite{wwteup}.
The tables presented below are borrowed from
papers \cite{gentle_unicc11} and \cite{gentle_nc24h}, while their
description is updated.
 
In general all possible final states can be subdivided into
two classes.
The first class comprises
production of (up, anti-down) and
(down, anti-up) fermion pairs,
\begin{eqnarray}
  (U_i~{\bar D_i })~ +~(D_j~{\bar U_j })~,
  \nonumber
\end{eqnarray}
where $i,j$ are generation indices.
The final states produced via virtual W-pairs
belong to this class.
Therefore, we will call these \emph{`CC'}-type final states.
The second class is the production of two fermion-antifermion pairs,
\begin{eqnarray}
  (f_i~~{\bar f_i })~ +~(f_j~~{\bar f_j })~,\;\;f=U,~D.
  \nonumber
\end{eqnarray}
As it is produced via a pair of two virtual neutral vector
bosons we will call this a final state
of `NC'-type. Obviously these two classes overlap for certain
final states.
\par
The number of Feynman diagrams in the \emph{CC} classes are
shown in table~1.
 
\begin{table}[ht]
\begin{center}
\begin{tabular}{|c|c|c|c|c|c|}
\hline
             &
\raisebox{0.pt}[2.5ex][0.0ex]{${\bar d} u$}
& ${\bar s} c$ & ${\bar e} \nu_{e}$ &
              ${\bar \mu} \nu_{\mu}$ & ${\bar \tau} \nu_{\tau}$   \\
\hline
$d {\bar u}$            &{\it  43}& {\bf 11} &  20 & {\bf 10} & {\bf 10} \\
\hline
$e {\bar \nu}_{e}$      &  20 &  20 &{\it 56}&  18 &  18 \\
\hline
 $\mu {\bar \nu}_{\mu}$ & {\bf 10} & {\bf 10} &  18 & {\it 19} & {\bf 9}  \\
\hline
\end{tabular}
\caption
{\it
Number of Feynman diagrams for \emph{`CC'} type final states.
\label{tab1}
}
\end{center}
\end{table}
 
\vspace*{-0.5cm}
 
Three different cases occur in the table \ref{tab1}
\footnote{
In~\cite{excalit},
a slightly different classification has been introduced;
the relation of both schemes is discussed in~\cite{wwteup}.
}:
\begin{itemize}
\item[(i)]
    The \emph{CC11} family.
\\
    The two fermion pairs are different, the final state does not
    contain
    identical particles nor electrons or electron neutrinos
    (numbers in table~1 in {\bf boldface}). The corresponding eleven
    diagrams
    are shown in figures~1 and~2. There are less diagrams if neutrinos
    are produced    ({\tt   CC9, CC10} processes).
 
\item[(ii)]
    The \emph{CC20} family.
\\
    The final state contains one $e^{\pm}$ together
    with its neutrino (Roman numbers in table~1);
    compared to case~(i), the additional diagrams have a $t$ channel
    gauge boson exchange. For a purely leptonic final state, a \emph{CC18}
    process results.
 
\item[(iii)]
    The \emph{CC43}/\emph{mix43} family and \emph{CC56}/\emph{mix56} process.
\\
    Two mutually charge conjugated fermion pairs are produced ({\it
      italic} numbers in table~1).
    Differing from cases~(i) and~(ii), the diagrams may
                   proceed via both, $WW$- and $ZZ$-exchanges.
    For this reason, we will also call them {\it mix}-ed class.
    There are less diagrams in the \emph{mix43} process if neutrinos
    are produced (\emph{mix19} process). With the two charge conjugated
   ($\bar{e}\nu_e$) doublets, one has \emph{mix56} process.
\end{itemize}
 
\begin{figure}[thb]
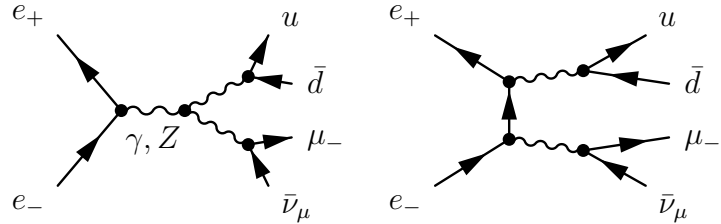

  \begin{center}
    \vspace*{\baselineskip}
    \input{diag_cc03_a.tex}
    \qquad\quad
    \input{diag_cc03_b.tex}
    \vspace*{\baselineskip}
  \end{center}
  \caption{\label{fig:CC03}The \emph{CC03} set of Feynman diagrams}
\end{figure}

\begin{figure}[thb]
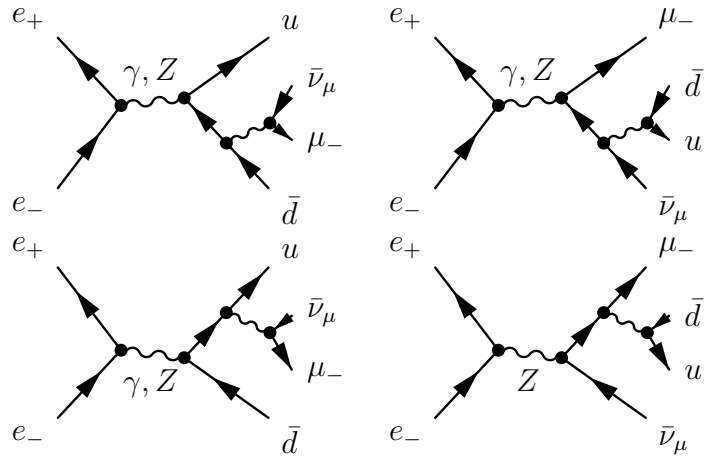

  \begin{center}
    \vspace*{\baselineskip}
    \input{diag_cc11_a.tex}
    \qquad\quad
    \input{diag_cc11_b.tex}\\
    \vspace*{2\baselineskip}
    \input{diag_cc11_c.tex}
    \qquad\quad
    \input{diag_cc11_d.tex}\\
    \vspace*{\baselineskip}
  \end{center}
  \caption{\label{fig:CC11}The \emph{CC11} set of Feynman diagrams}
\end{figure}

 
Each of these classes
contains the \emph{CC03} process, which is described by the usual three
`double W-pole' Feynman diagrams, figure \ref{fig:CC03}.
{}From the \emph{CC11} set of diagrams only 10 contribute to the process
$e^+e^- \to \mu^-\bar\nu_\mu u\bar d$, because the photon doesn't
couple to the neutrino (cf.~fig.~\ref{fig:CC11}).
 
For the final states corresponding to the NC class
the number of Feynman diagrams is presented in table \ref{tab2}.
%
\begin{table}[ht]
\begin{center}
 \begin{tabular}{|c|c|c|c|c|c|c|}
\hline
&
\raisebox{0.pt}[2.5ex][0.0ex]{${\bar d} d$}
&${\bar u} u$
&${\bar e} e$
&${\bar \mu} \mu$
&${\bar \nu}_{e} \nu_{e}$
&${\bar \nu}_{\mu} \nu_{\mu}$
\\
\hline
\raisebox{0.pt}[2.5ex][0.0ex]{${\bar d} d$}
 & {\tt 4$\cdot $16} & {\it 43} & {48}
             & {\bf 24} & 21 & {\bf 10} \\
\hline
\raisebox{0.pt}[2.5ex][0.0ex]
{${\bar s} s, {\bar b} b$} & {\bf 32} & {\it 43} & {48}
             & {\bf 24} & {21} & {\bf 10} \\
\hline
${\bar u} u$ & {\it 43} & {\tt 4$\cdot$16} & {48}
             & {\bf 24} & {21} & {\bf 10} \\
\hline
${\bar e} e$ &{48} &{48} & \textsf{4$\cdot$36} &{48}
& {\it 56} & {20}
\\
\hline
${\bar \mu} \mu$  & {\bf 24} & {\bf 24} & {48} & {\tt 4$\cdot$12}
                  & {19} & {\it 19}         \\
\hline
${\bar \tau} \tau$& {\bf 24} & {\bf 24} & {48} & {\bf 24}
                  & {19} & {\bf 10}         \\
\hline
${\bar \nu}_e \nu_{e}$  & {21} & {21} & {\it 56} & {19}
                  & \textsf{4$\cdot$9} & {12}                   \\
\hline
${\bar \nu}_{\mu} \nu_{\mu}$ & {\bf 10} & {\bf 10} & {20}
             & {\it 19} & {12} & {\tt 4$\cdot$3}  \\
\hline
${\bar \nu}_{\tau} \nu_{\tau}$ & {\bf 10} & {\bf 10} & {20}
             & {\bf 10} & {12} & {\bf 6}  \\
\hline
\end{tabular}
\caption[]
{\it
Number of Feynman diagrams for
`NC' type final states.
\label{tab2}
}
\end{center}
\end{table}
 
\begin{itemize}
\item[(i)]
    The \emph{NC32} family.   \\
The simplest case (numbers in
{\bf boldface}) does not contain electrons or identical
fermions\footnote{
We exclude the Higgs boson exchange diagrams from the classification
in the tables.}.
 
\item[(ii)]
    The \emph{NC48} and \emph{NC21} families.\\
The numbers in roman
correspond to the final states
which include $f=e,\nu_e$ except for cases covered by item {(iv)}.
The large number of diagrams here is due
to additional $t$-channel exchange.
 
\item[(iii)]
    The \emph{NC4}$\cdot$\emph{16} family. \\
With identical fermions $f$ ($f \neq e,\nu_e$),
the number of diagrams
grows drastically due to the necessity to satisfy the Pauli principle,
i.e. to anti-symmetrize the amplitude. For purely leptonic processes this
number of diagrams reduces to \emph{4}$\cdot$\emph{12} since the gluon exchange
doesn't contribute.
 
\item[(iv)]
    The \emph{NC4}$\cdot$\emph{36} and \emph{NC4}$\cdot$\emph{9} processes,
with the two $e^+e^-$ or $\bar{\nu_e}\nu_e$ pairs in the final state.
    The corresponding numbers are shown \textsf{sans serif}.
 
\item[(v)]
  The \emph{mix43} and \emph{mix56} processes.\\
The numbers in {\it italic} correspond to final states which are also
present in table~1, case~(iii).
 
\end{itemize}
%
%
\section{Descriptions of 4-fermion codes}
\subsection{ALPHA}
\leftline{\bf Authors:}
\begin{tabular}{ll}
Francesco Caravaglios & 
caravagl@thphys.ox.ac.uk \\
Mauro Moretti & 
moretti@hep1.phys.soton.ac.uk
\end{tabular}
 
 
\noindent{\bf Description}
 
\noindent In ref.\cite{alpha},
we suggested an {\it iterative} algorithm
 to compute automatically  the scattering matrix elements
of  any given effective
Lagrangian, $\Gamma$. By exploiting the relation between
$\Gamma$
 and the connected Green's function generator, $Z$, we obtained a formula
which   does not require the use of
Feynman graphs, and is suitable to implementation in a numerical routine.
The problem of computing the scattering matrix element
can be reformulated
as the problem of finding the minimum of $Z$ with
respect to a {\it finite} set of variables. Once the stationary conditions
for $Z$ are written down, they can be solved iteratively and,
truncating  the series after a proper number of steps, one
obtains the solution.
Using this algorithm we have been able to build a Fortran code, {\tt ALPHA},
for the automatic computation of matrix elements.
When the initial and final states of the process are specified
(type,  momenta and spin of the external particles)
the program  prepares an array $b_j$ for all the possible
degrees of freedom ( the label $j$ refers to internal
and external momenta  and
to the particles type, color  and spin).
As shown in \cite{alpha}, the scattering matrix
element ${\cal A}$ is obtained
as
\be
{\cal A} =  a_i b_i +{1\over 2} K_{lm} b_l
b_m +{1\over 6} O_{ijk} b_i b_j b_k.
\ee
where the $b_j$ are obtained
 from the equation of motion in presence of a source term $a_i$.
\be
 a_i=K_{im} b_m +{1\over 2} O_{ijk} b_j b_k,
\ee
which can be solved iteratively.
 
The  matrix $O_{ijk}$ contains  the physical couplings between
the degrees of freedom $b_{j}$ of the fields
entering the scattering process and the matrix $K_{lm}$ accounts
for the kinetic terms in the Lagrangian.
In the Fortran code the matrix elements
$O_{ijk}$ and $K_{lm}$ are returned by some
subroutines
as
 a function of  the finite set of possible momenta $P_m$.
 
The {\tt ALPHA} code includes all the electroweak interactions and
the whole flavor content of the Standard Model (SM)
(presently it does not account for the strong interactions)
and it can perform
all possible electroweak matrix elements in the SM regardless of the initial
or final state type.
In addition, due to its simple logic, it
allows for modification of  the Lagrangian
with no excessive effort (by adding the proper subroutines
to compute the new $O_{ijk}$
interactions  and/or adding the relevant variables
for the new particles).
Since the algorithm is purely numerical,  the output can be
immediately used for an integration procedure.
 
\noindent{\bf Features of the program}
 
\noindent
The numerical integration is performed by mean of the package {\tt VEGAS}
\cite{vegas}.
The variables  have been chosen in such a way that  each
singularity corresponds to an integration  variable
allowing  {\tt VEGAS} to
cope effectively with the pole structure of the
physical process.
The phase space is factorized as a multiple decay process using the
formula
\be
d \Phi(P;q_1,q_2,q_3,...,q_n)=d \Phi(Q=q_1+q_2;q_1,q_2)
d \Phi(P;Q,q_3,...,q_n) (2\pi)^3 d^2 Q
\ee
where the squared momenta $Q^2$ corresponds to the physical
 singularities.
For some final states there are multiple channels exhibiting a
pole structure.
In these cases it is difficult  to obtain a good convergence
of the integral with a single choice
of phase space variables. Therefore we split
the integration domain in different regions, and in  each of them
we make a different choice   of physically motivated  variables.
One additional real variable is used to map the discrete set of spin
configurations.
At least for the processes we have considered, the {\tt VEGAS} algorithm
has adequately performed a selection of  the relevant spin configurations.
 
In principle, all possible final states can be treated. For
most of them the corresponding phase space routines are also
implemented: an exception being processes with electrons in the
final state.
All possible  choices of spin configurations can be selected,
for instance polarized initial states are immediately available.
 
The Monte Carlo  does not include initial/final-state radiation (ISR/FSR).
We have instead used {\tt ALPHA} to compute the rates for the process
$e^+ e^- \rightarrow 4 \ fermions \ + \gamma$;
 all the Standard Model diagrams are evaluated with a finite (constant)
 width of the  electroweak gauge bosons and the  physical   fermion masses.
 
Anomalous couplings can be  easily  added, even with momentum
dependent form factors, running widths etc.
 
Since the method of calculation does not rely on Feynman graphs
technique it is not possible, in general, to isolate the contribution
of a single graph.
Turning on/off each single interactions, the  contribution of
many  subsets of diagrams can be extracted but this might be not
practical enough.
 
\noindent{\bf Program layout}
 
\noindent
The program requires as input the center of mass energy and
the number of external particles:
for each type ({\it i.e.} top, strange,...Z)
 we have to enter  a number which can be 0 if no particle
of that type exists, or 1,2,...  as required.
A subroutine generates the momenta and the
spin configurations according
to a phase space  preselected
among a list of prepared ones.
All the couplings of the theory are collected in a single subroutine
which is adequately commented and is called only once at the
beginning of the run. A subroutine is provided which has as input
the external momenta and as output a flag which when set to zero
forces the program to ignore the given phase space point,
thereby allowing for any kind of cut.
Another subroutine is provided to make it possible to produce plots.
Each variable to be plotted must be normalized between 0 and 1
and as output a file is produced which registers for each variable
N (input number)
equispatiated bins containing the (unnormalized) integral and
variance.
As output the cross section (in picobarn) is also given with its
statistical  error.
 
With few modifications, we can therefore provide a code for the computation
of {\it all} processes listed in tables \ref{tab1} and \ref{tab2}
allowing the user to implement any cuts
to change the numerical values of the electroweak couplings
and to record all the data required to produce a plot.
 
Other operations, like allowing the user to compute an arbitrary process
or to change the Lagrangian of the model are not completely user-friendly
at the moment.
 
\clearpage    
 
\noindent{\bf Input parameters and the Lagrangian}
 
\noindent
We used the common set of Standard Model parameters (as discussed in section 3).
All the fermions are massive.
%
The gauge boson propagators include  the  width, which
 is constant in order to obtain  gauge invariant matrix elements.
The inclusion of the proper, physical, running width for the gauge
bosons in  a gauge invariant way, namely including the relevant
corrections to the three and four point Green Functions, is
straightforward in our approach and it will be done in a near future.
The cuts
applied to
the four final fermions
are the common one used for
the comparison tests.
 
\noindent{\bf Availability:}                        \\
The program is available upon e-mail request from the authors.
 
 
\subsection{CompHEP 3.0}
 
\leftline{\bf Authors:}
\begin{tabular}{ll}
E.Boos        &        boos@theory.npi.msu.su \\
M.Dubinin     &     dubinin@theory.npi.msu.su \\
V.Edneral     &     edneral@theory.npi.msu.su \\
V.Ilyin       &       ilyin@theory.npi.msu.su \\
A.Pukhov      &      pukhov@theory.npi.msu.su \\
V.Savrin      &      savrin@theory.npi.msu.su \\
S.Shichanin   &   shichanin@m9.ihep.su
\end{tabular}
 
 
\noindent{\bf Description}
 
\noindent
The main idea in
{\tt CompHEP} \cite{r1} was to enable on to go directly
from the Lagrangian to cross sections and distributions effectively,
with a high level of automation.
 
Version 3.0 has 4 built-in physical models. Two of them are
versions of the Standard Model (SU(3)xSU(2)xU(1)) in the unitary and
't Hooft-Feynman gauges with the parameters corresponding to the standard
LEP2 input.
 
   The general structure of the {\tt CompHEP} package is represented
in Figures \ref{comfig2}, \ref{comfig4}.
It consists of symbolical and numerical modules. The main tasks solved by
the symbolical module (written in C) are :
 
          1.  to select a process  by specifying  {\it in-}  and {\it out-}
              particles. Any type of five particle final state for decays and
              five particle final state for collisions can be defined;
 
          2.  to generate and display Feynman diagrams. It is
              possible to delete some diagrams from the further
              consideration, leaving only limited subsets;
 
          4.  to generate and display squared Feynman diagrams
             (corresponding to squared S-matrix elements);
 
          5. to calculate analytical expressions corresponding to squared
             diagrams with the help of a  fast built-in symbolic calculator.
             Traces of gamma matrices products are calculated, summing
             over the final state polarizations. Masses of initial
             and final particles can be kept nonzero in the squared
             amplitude calculation and phase space integration;
 
          6.  to save symbolic results corresponding to the squared diagrams
          calculated in the {\tt REDUCE} and {\tt MATHEMATICA} codes for further
              symbolical manipulations;
 
          7. to generate the optimized {\tt FORTRAN}
 code for the squared matrix
             elements for further numerical calculations.\\
 
 
 
\begin{figure}[tbhp]
\small
\begin{picture}(400,500)(0,0)
 
\put(125,465){
\vbox{
\begin {tabular}{|l|}
 \multicolumn{1}{c}{\rm menu 1}\\ \hline
                 QED  \\
                 Fermi model\\
                 St. model (unit. gauge)\\
                 St. model (Feyn. gauge) \\
                 NEW MODEL \\
                \hline
\end{tabular}
 
    }
}
 
\put(200,425){\vector(0,-1){25}}
 
\put(140,375){
\vbox{
\begin {tabular}{|l|}
\multicolumn{1}{c}{\rm menu2}\\ \hline
                 Enter process   \\
                 Edit model  \\
                 Delete changes \\
                \hline
\end{tabular}
 
}
 
}
 
\put(150,372){\vector(-1,0){80}}
\put(243,382){\vector(1,0){65}}
 
\put(0,345){
\begin {tabular}{|l|}
 \multicolumn{1}{c}{\rm menu3}\\ \hline
                 Variables \\
                 Constraints\\
                 Particles\\
                 Lagrangian \\
                \hline
\end{tabular}
}
 
\put(300,367){
\begin {tabular}{|l|}
 \multicolumn{1}{c}{\rm menu 4}\\ \hline
                 Squaring  \\
                 View diagrams\\
                \hline
\end{tabular}
}
 
\put(297,370){\line(-1,0){10}}
\put(287,370){\vector(0,-1){50}}
\put(235,265){
\begin {tabular}{|l|}
 \multicolumn{1}{c}{\rm menu 5}\\ \hline
                 View squared diagrams  \\
                 Symbolic calculation\\
                 Write results\\
                 REDUCE program \\
                 Numerical calculator\\
                 Enter new process\\
                 Interface \\
                \hline
\end{tabular}
}
 
\put(235,278){\vector(-1,0){150}}
\put(235,250){\line(-1,0){15}}
\put(220,250){\vector(0,-1){45}}
 
\put(0,250){
\begin {tabular}{|l|}
 \multicolumn{1}{c}{\rm menu 6}\\ \hline
                 FORTRAN code  \\
                 REDUCE code\\
                 MATHEMATICA code\\
                \hline
\end{tabular}
}
 
\put(150,165){
\begin {tabular}{|l|}
 \multicolumn{1}{c}{\rm menu 7}\\ \hline
                 View/change data \\
                 (Set angular range) \\
                 (Set precision) \\
                 (Angular dependence) \\
                 Parameter dependence\\
                \hline
\end{tabular}
}
 
\put(150,136){\vector(-1,0){55}}
\put(275,149){\vector(1,0){60}}
 
\put(315,129){
\begin {tabular}{|l|}
 \multicolumn{1}{c}{\rm menu 8}\\ \hline
                 Show plot \\
                 Save results in a file\\
                 Recalculate \\
                \hline
\end{tabular}
}
 
\put(0,120){
\begin {tabular}{|l|}
 \multicolumn{1}{c}{\rm menu 9}\\ \hline
                 (Total cross section) \\
                 (Asymmetry) \\
                \hline
\end{tabular}
}
 
\put(50,100){\vector(0,-1){30}}
\put(0,50){
\begin {tabular}{|l|}
 \multicolumn{1}{c}{\rm menu 10}\\ \hline
                 Show plot   \\
                 Save results in a file\\
                \hline
\end{tabular}
}
 
 
\end{picture}
\caption[.]{The menu system for the {\tt CompHEP} symbolic part}
\label{comfig2}
\end{figure}
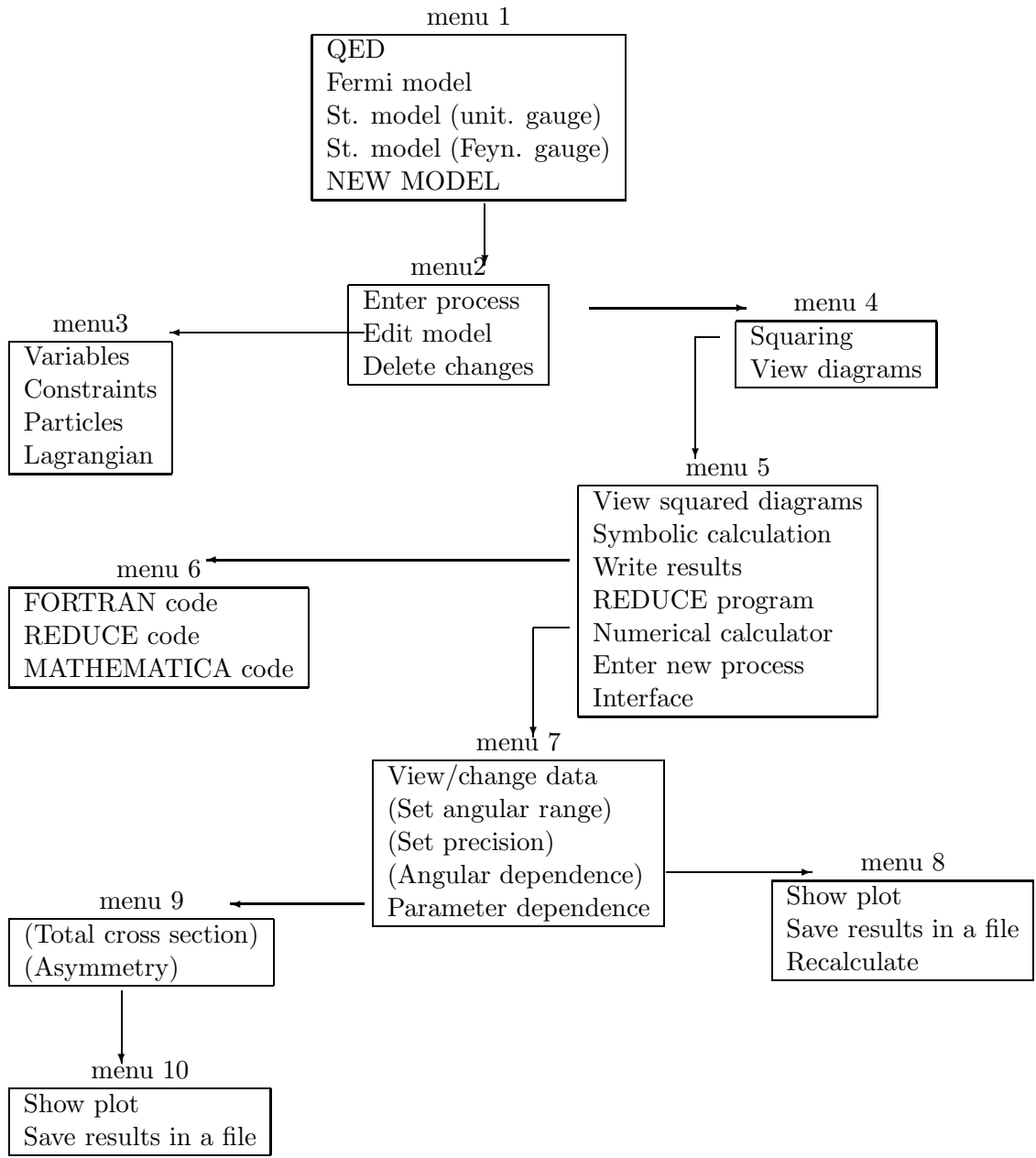
 
 
 
 
\begin{figure}[tbhp]
\vbox{
\centerline{
\begin {tabular}{|ll|}
 \multicolumn{2}{c}{\rm Main menu }\\ \hline
1. Calculation      &  2. IN state \\
3. Model parameters &  4. Invariant cuts \\
5. Kinematics       &  6. MC parameters \\
7. Regularization   &   8. Task formation \\
9. View results     & 10. User's menu \\
 \hline
\end{tabular}
}
  }
 
\centerline{
\begin{picture}(160,100)
\put(0,20){
\vbox{
\begin {tabular}{|l|}
 \multicolumn{1}{c}{\rm In state }\\ \hline
1. StructF(1) = {\it OFF } \\    2. SQRTS = {\it 1000 }   \\
3. StructF(2) = {\it OFF }                 \\
 \hline
\end{tabular}
   }
   }
\end{picture}
\begin{picture}(160,100)
\put(0,20){
\vbox{
\begin {tabular}{|l|}
\multicolumn{1}{c}{\rm Invariant cuts}\\ \hline
1. Insert new cut \\ 2. Delete cut \\
3. Change cut        \\
 \hline
\end{tabular}
  }
  }
\end{picture}
}
 
\vskip 1.0 cm
\vbox{
\centerline{
\begin {tabular}{|ll|}
 \multicolumn{2}{c}{\rm MC parameters }\\ \hline
1. Ncall = {\it 10000} &  2. Acc1= {\it 0.1 }\\
3. Itmx1= {\it 5 }   & 4. Acc2= {\it 0.1 } \\
5. Itmx2={\it 0}   & 6. Event generator {\rm OFF} \\
7. Number of events = {\it 1000 } &  \\
 \hline
\end{tabular}
  }
}
 
\centerline{
\begin{picture}(180,100)
\put(0,20){
\vbox{
\begin {tabular}{|l|}
 \multicolumn{1}{c}{\rm Regularization }\\ \hline
1. Insert new regularization \\  2. Delete regularization \\
3. Change regularization       \\
 \hline
\end{tabular}
}
}
\end{picture}
\begin{picture}(160,100)
\put(0,20){
\vbox{
\begin {tabular}{|l|}
 \multicolumn{1}{c}{\rm Task formation }\\ \hline
1. Table parameters  \\  2. Set default session \\
3. Add session to batch  \\
 \hline
\end{tabular}
}
}
\end{picture}
}
\vskip 1.0 cm
\vbox{
\centerline{
\begin {tabular}{|ll|}
 \multicolumn{2}{c}{\rm View results}\\ \hline
1.  session \# to view - {\it 3 } &  2. View result file \\
3. View protocol file      &  4. View histogram file \\
 \hline
\end{tabular}
  }
}
\vskip 1.0 cm
\caption[.]{The menu system for the {\tt CompHEP} numerical part}
\label{comfig4}
\end{figure}
 
\noindent{\bf Program layout}
 
\noindent
     The numerical part of the
{\tt CompHEP} package is written in {\tt FORTRAN}. It
uses the {\tt CompHEP
FORTRAN} output, the {\tt BASES}\&{\tt SPRING} package \cite{r2} for
adaptive Monte-Carlo integration and unweighted event generation. The main
tasks solved by the numerical module are :
 
    1. to choose phase-space kinematical variables. Exact
       parameterizations of three, four and five particle phase space in the
       case of massive particles are used \cite{r3};
 
    2. to introduce kinematical cuts over any squared momenta transferred
       and squared masses for any groups of outgoing particles. Any
       kinematical cuts for noninvariant variables can be introduced
       using explicit restrictions on the four-momenta;
 
    3. to perform a kinematical regularization (mapping) to remove sharp
       peaks in the squared matrix elements. The package has a rich
       choice of optimizing possibilities (various combinations of
       phase space parameterizations and mappings);
 
    4. to change the {\tt BASES} parameters for Monte-Carlo integration;
 
    5. to change numerical values of model parameters;
 
    6.  to calculate distributions, cross sections or
        particle widths by the Monte-Carlo method. The output
        for a cross section value (sequence of MC iterations)
        and distributions (set of histograms) has the standard {\tt BASES}
form;
 
    7. to perform the same integration taking into account structure function
       for incoming particles. Initial state radiation (ISR) is
       implemented in the structure function approach \cite{r4}. An interface
       to the standard PDF library is available. Final state radiation
       and the Coulomb term are not implemented.
       Photon radiation from the initial and final states can be introduced
       by calculation of exact amplitude for $2 \rightarrow 5$ process
       (4 fermions + photon).
 
    8. to generate events and to get histograms simulating the signal
       and background. {\tt SPRING} \cite{r2} is used for unweighted
       event generation.
 
{\tt CompHEP} is a menu-driven program with a context HELP facility.
Each of two variants of the Standard Model (unitary or         and
`t Hooft-Feynman gauges) is defined by four tables:   
 
\begin{tabular}{ll}
Variables & list of parameters (masses, widths, couplings,
mixings) \\
Constraints & list of functionally dependent parameters \\
Particles  & list of particles and quantum numbers \\
Lagrangian & list of Feynman rules for vertices
\end{tabular}
 
At present, versions for different platforms exist: HP Apollo
9000, IBM RS 6000, DECstation 3000, SPARC station, Silicon Graphics
and VAX.
 
\noindent{\bf Availability}                         \\
The package is available from \\
internet host: theory.npi.msu.su \\
 directory: pub/comphep-3.0 \\
 files: 30.tar.Z, install.doc, manual.ps.Z
 
 
\subsection{ERATO}
 
\leftline{\bf Author:\footnote{
 In several aspects of the program
 the following people have contributed: \\
\begin{tabular}{ll}
Mark Gibbs, Liverpool & {\tt gibbs@afsmail.cern.ch}   \\
Robert Sekulin, DRAL & {\tt robert@vax2.rutherford.ac.uk}  \\
Spyros Tzamarias, Liverpool & {\tt tzamaria@cernvm.cern.ch}\\
\end{tabular}
}}
 
\begin{tabular}{ll}
Costas G Papadopoulos & papadopo@cernvm.cern.ch \\
 &   C.G.Papadopoulos@durham.ac.uk  and\\
 &   papadopo@alice.nrcps.ariadne-t.gr
\end{tabular}
 
 
\noindent{\bf Description}
 
\noindent
{\ert}\cite{ert1}-\cite{ert2}
is a four-fermion Monte Carlo\footnote{
In ancient Greek mythology $\mbox{EPAT}\Omega$ was the muse of Music.
By accident the name of the program is also part of the gen{\tt ERATO}r
group.}
. This program is an
 evolution of an older code
where single-$W$ production, $e^- e^+ \to e^- \bar{\nu}_{e} W$ was
 calculated
including all possible {\an} of the three-boson interactions\cite{ert1},
$WW\gamma$ and $WWZ$.
This code has now been updated in order to include all
background graphs for the processes $e^- e^+ \to \ell \bar{\nu}_{\ell} u
 \bar{d}$
with $\ell=e,\mu,\tau$.
The actual version of the program can now produce results for any four-fermion
final state.
As far as the matrix element calculation is concerned, the program uses a
representation of the basic fermion current
$\bar{u}_\lambda (p_1)\gamma^\mu u_\lambda(p_2)$, the `E-vector',which
is given as follows:
\bq E_\lambda^\mu(p_1,p_2)\equiv\bar{u}_\lambda (p_1)\gamma^\mu u_\lambda(p_2)
\eq
where
\bqa
E_-^0&=& \sqrt{p_1^+ p_2^+}+
\frac{(p_{1x}+i p_{1y})(p_{2x}-i p_{2y})}{\sqrt{p_1^+ p_2^+}}\nn \\
E_-^x&=& \sqrt{\frac{p_2^+}{p_1^+}}(p_{1x}+i p_{1y})
+\sqrt{\frac{p_1^+}{p_2^+}}(p_{2x}-i p_{2y})\nn \\
E_-^y&=& -i\biggl( \sqrt{\frac{p_2^+}{p_1^+}}(p_{1x}+i p_{1y})
-\sqrt{\frac{p_1^+}{p_2^+}}(p_{2x}-i p_{2y})\biggr) \nn \\
E_-^z&=& \sqrt{p_1^+ p_2^+}-
\frac{(p_{1x}+i p_{1y})(p_{2x}-i p_{2y})}{\sqrt{p_1^+ p_2^+}}
\eqa
with $p^\pm=p^0\pm p^3$.
The above representation is valid only for massless fermions.
All matrix elements have been tested against {\tt MadGraph}\cite{madg}
 calculations
under the same conditions, and the agreement was at least 13 digits using
a {\tt REAL*8} declaration.
 
In addition to the
amplitude calculation, we have implemented a
Monte Carlo integration algorithm which is essentially
identical to the multichannel approach of references
\cite{excalit,optimi}. The problem is that
the amplitude we have to integrate over is a very complicated
function of the kinematical variables, peaking at different regions
of phase space. The idea is to define different kinematical
mappings, corresponding to different peaking structures
of the amplitude and
then use an optimization procedure to adjust the percentage
of the generated phase-space points, according to any specific
mapping, in such a way that the total error is minimized.
 
\par
Special care has also been taken in order to include in a
gauge-invariant way the width effects. As is well known the introduction
of an $s$-dependent width leads to gauge-violation in the $s-$ and $t-$ channel.
This is because the $s$-dependent width violates the Ward identities at the
one loop. The solution is to include consistently all one loop corrections.
More precisely, if one restricts oneself to fermionic corrections, one has
to include the one-loop fermion `triangle' to the three-boson vertex function.
This way, the gauge-invariance is restored.
Bosonic corrections are much more subtle due to the gauge-parameter
dependence, but in the case of $W$ and $Z$ line-shape parameters
their contribution is suppressed compared to the fermionic one, due to simple
kinematical reasons. In {\ert} the imaginary part at the one-loop level of both
two-point and three-point functions of vector bosons is implemented in a very
compact analytic form\cite{bhfpaper}.
 
\par
Leading higher order corrections are also included in {\ert}, in the form
of initial-state radiation (ISR), using the structure function approach
with all possible ISR-radiator functions available ($\beta$ or $\eta$ option).
 
\par
An other important feature of {\ert} is the incorporation of all CP
conserving {\an}. In fact the way the program is written
enables us to include any {\an}, for instance $ZZ\gamma$ or
CP-violating $WW\gamma$ and $WWZ$ parameters.
 
\noindent{\bf Features of the program} \\
%
The main features of the program are the following:
it can be used both as an event generator and as an integrator:
all final states, and all possible cuts, are in principle allowed.
Initial-state radiation is implemented using structure functions;
final-state radiation and the Coulomb correction are not
implemented. All possible anomalous couplings are implemented,
the fermions are assumed to be massless, with a leading-log
approximation for the structure functions.
 
\noindent{\bf Interface}                       \\
The output from the {\ert} generator for the semi-leptonic and four-jet channels
contains colored partons, and consequently it is desirable to include
models of QCD effects such as hadronization in the simulation procedure.
One way to include these phenomena is to pass the four-momenta generated
by ERATO to an existing simulation package. This approach is attractive
as there are a number of such packages in existence.
\par
The {\ert} generator has been interfaced successfully to the
{\tt JETSET} \cite{jetset} and
{\tt HERWIG} \cite{herwig} packages. T
he procedure is the same in both cases and can be
easily extended to other simulation packages.
\par
Firstly, the event configurations produced by {\ert} are not of equal
probability and have to be selectively used in such a way so as to
respect the correct distributions of kinematic variables. This is
achieved by unweighting the events; events are used at random with
a probability given by the weight of the event divided by the maximum
weight. The efficiency of this procedure is typically of order
0.1\%.
\par
Secondly, the particle content of the {\ert} final state has to be
selected. At present, this is determined at the start of a simulation
run but in principle can be performed on an event- by event- basis.
\par
Thirdly, the {\ert} program assumes that all the fermions are massless. As
a result, the four-momenta of a final state configuration have to be
shifted in order to place massive fermions on shell. This is achieved
by shifting the three-momenta slightly. As the energies in a typical
LEPII event are high compared to the particle masses the change
in momenta is a negligible effect.
Following these steps, the simulation package is then used for
the parton showering and hadronization stages of event generation.
 
\noindent{\bf Program layout}                       \\
The structure of the program will be described in detail in a future
publication in CPC.
 
\noindent{\bf Input parameters}                     \\
\noindent
Any set of input parameters can be implemented. In the most usual version
the LEP2 standard input is used. Preferred and comparison values are
identical.
 
\noindent{\bf Output}                               \\
\noindent
In the present form of the program any histogram can be obtained very easily.
Cross sections for left and right incoming electrons are given separately.
Error estimates are the standard ones.
 
\noindent{\bf Availability}                         \\
\noindent
{}From ftp://alice.nrcps.ariadne-t.gr/pub/papadopo/erato/
 
\clearpage    
 
\subsection{EXCALIBUR}
 
\leftline{\bf Authors:}
\begin{tabular}{ll}
F.A.~Berends & berends@rulgm0.leidenuniv.nl \\
R.~Kleiss    & t30@nikhefh.nikhef.nl        \\
R.~Pittau    & pittau@psw218.psi.ch
\end{tabular}
 
\noindent{\bf Short description:}                   \\
The program {\tt EXCALIBUR}~\cite{excalit,optimi}
evaluates cross sections for
electron-positron scattering into four final-state fermions.
This is done by Monte Carlo simulation, in which events are
generated over a phase space determined by a number of
a-priori cuts (in many cases, the whole phase space is
accessible). Each event carries a weight such that the
average event weight gives the total cross section. The
distribution of events over the phase space is generated
by employing a large number of mappings of random numbers.
Given an event, additional cuts can be imposed by hand by
setting the weight of unwanted events to zero; and, of course,
any number of differential distributions can also be constructed.
Since the matrix elements are computed on the level of
helicity amplitudes, as sums of distinct diagrams, the
contributions of subsets of diagrams and of particular helicity
configurations can also be studied.
 
\noindent{\bf Program features:}
\begin{enumerate}
 
\item\hed{method of integration:}
the program is a strict Monte-Carlo one, in the sense that
no phase space variables are integrated over analytically.
This means that {\em all\/} phase space variables are amenable
to any kind of cut. The generated events come with a non-constant
weight: a sample of unweighted events can be selected from the
generated sample by the usual rejection techniques. The efficiency
of this procedure is in many cases of the order of a few per cent,
depending on the final state of choice and the phase space cuts.
 
\item\hed{possible final states:}
all possible four-fermion final states are included: the user
supplies the choice in the input file. An important restriction
is that the fermions are considered to be strictly massless,
and therefore Higgs exchange is not included.
 
\item\hed{possible cuts:}
since every event is completely specified, in principle any
conceivable phase space cut can be implemented. It must be
noted that, since all fermion masses are taken to be zero,
singularities can occur in photon exchange channels, and these
have to be excised by user-supplied a-priori cuts. Therefore,
when a final state $e^+$ or $e^-$ occurs, a cut on its
scattering angle and energy is necessary, and when a charged
particle-antiparticle pair is produced, a cut on its
invariant mass is in order. These cuts are specified in the
input file (see discussion below). For calculations based
on a restricted set of Feynman graphs without photon exchange
(e.g. the \emph{CC03} diagrams) such cuts are of course not
necessary.
 
\newpage    
 
\item\hed{treatment of ISR:}
ISR is implemented in the form of two structure functions,
{\it i.e.\/} two energy fractions $x_1$ and $x_2$ are generated,
but no bremsstrahlung $p_T$. The four-fermion event is then
generated in the reduced-center-of-mass frame. The actual
photon structure functions used are the `type 2' ones
of the $W$-pair report.
 
\item\hed{treatment of FSR:}
No FSR is at the moment included.
 
\item\hed{treatment of final state decays:}
since the fermions are considered massless, they are stable
and no decay is provided: moreover, the fermions' density
matrix is strictly diagonal.
 
\item\hed{treatment of the Coulomb singularity:}
the Coulomb term can be easily implemented by multiplying
the appropriate $WW$ diagrams by the correct factor, but
is not yet included in the standard version.
 
\item\hed{treatment of anomalous couplings:}
a version of {\tt EXCALIBUR} is available which includes anomalous
triple-gauge-boson couplings. Six CP-conserving anomalous
contributions can be put to a nonzero value: these correspond to
the quantities $x_{\gamma}$, $y_{\gamma}$, $x_Z$, $\delta_Z$, $y_Z$,
and $z_Z$ defined in ref.~\cite{anomcoupl}.
For zero values of these numbers the minimal Standard Model
predictions are recovered.
 
\item\hed{treatment of fermion masses:}
as mentioned, these are zero, both in the matrix element
and in the phase space momenta.
 
\item\hed{treatment of hadronization:}
no interface with hadronization routines are provided in the
standard version; but since the momenta are completely specified
the necessary {\tt COMMON} can easily be constructed.
 
\item\hed{subsets of diagrams etc:}
since in {\tt EXCALIBUR} all diagrams and helicities are explicit,
it is simple, for a given final state, to select subsets
of diagrams or helicity combinations. There exists the
possibility to select, using the input file, only
those diagrams that correspond to the $WW$, $ZZ$,
$We\nu$, $Zee$ or $Z\nu_e\bar{\nu_e}$ final states, or
include all tree diagrams.
\end{enumerate}
 
\noindent{\bf Program layout}                       \\
The working of {\tt EXCALIBUR} can be divided into three parts:
initialization, generation, and evaluation. The two main
parts of the event generation stage are the choosing of
a random phase space point, and the computation of the
matrix element at that point.
 
The initialization is performed by the routine {\tt SETPRO}. It reads
the data from the input file, and determines from these which
are the Feynman tree graphs that will be considered. There are
two distinct diagram topologies: `abelian' graphs, with only
fermion-boson couplings, and `nonabelian' ones with also
triple-boson couplings. The program considers all possible
permutations of the external momenta over these diagrams,
and determines, by quantum numbers conservation, if they can contribute. Then,
 also the most significant phase space mappings
(so-called {\em channels\/}) are determined.
 
Upon the calling of an event, first the two energies
$x_{1,2}$ of the incoming $e^{\pm}$ are generated. Then,
in the center of mass frame after this ISR, one particular
channel is picked, by which uniform random numbers are
mapped into a phase space point.
The various channels are constructed from
a limited number of explicit mappings, each with its
own subroutine: this modular structure ensures
transparency of coding, easy debugging, and the possibility
of implemented additional channels when necessary. The probability
of picking a particular channel is given by its
{\em a-priori weight}: the final cross section is by construction
independent of the values of this weights.
After this, the event weight
is computed, as the ration of the matrix element squared
to the generated phase space density. For the computation
of the matrix element, we use the fact that every
contributing nonabelian
graph can, in the minimal standard model, be simply expressed
as a combination of two contributing abelian ones. These
are computed, for definite helicities, by spinor techniques.
The phase space density consists of a sum of the densities
appropriate to each contributing channel, weighted with
their a-priori weights.
At several points during a run of generating events, the
a-priori weights are optimized so as to approximate the
weight distribution with the minimum possible variance for
the available set of channels, as described in \cite{optimi}.
 
The evaluation stage consists of the estimate of the
average weight and its estimated error (and, in fact, the
estimated error on the error estimate). Also, the
distribution of all nonzero weights is plotted, together with
some information on the a-priori weight optimization.
More information can be found in \cite{excalit}.
 
\noindent{\bf Input parameters}                     \\
We have used the following sets of input parameters,
one for the tuned comparison with the other codes, and
one that reflects what (in our view) is the most accurate
prediction possible with {\tt EXCALIBUR}. They are given in the
table below.
\begin{center}\begin{tabular}{|c|c|c|} \hline\hline
parameter        &   `comparison'  &  `best' \\ \hline
Z mass (GeV)     & 91.1888         &  91.1546 \\
Z width (GeV)    &  2.4974         &   2.49646 \\
W mass (GeV)     & 80.23           &  80.02042   \\
W width (GeV)    &  2.0366         &   2.03302 \\
$\sin^2\theta_W$ & 0.231031        & 0.231031 \\
$1/\alpha$       & 128.07          & 128.07   \\
$\alpha_s$       & 0               & 0.103    \\ \hline \hline
\end{tabular}\end{center}
The following remarks are in order here. The `best values'
for the boson masses and widths are chosen so as to take into
account the running of the widths, using the transform
described in \cite{bardin}. The value of $\alpha$ is used for
the four-fermion system, but for the ISR the value 1/137
is of course used. The use of $\alpha_s$ is relevant
for four-quark and qq-two gluon final states, where the
QCD four-jet production diagrams are also included.
These values are set internally by the program.
In addition, there are a number of other input parameters, set
in the input file:\\
\begin{tabular}{ll}
{\tt NPROCESS} & the number of processes to be treated\\
{\tt N} &The number of events to be generated\\
{\tt ISTEPMAX}& the number of times the a-priori weights are
  to be optimized\\
{\tt OUTPUTNAME}& name of the output file\\
{\tt KREL}& the set of diagrams to be considered: 0 all diagrams,
 1 $WW$,\\ & 2: $ZZ$, 3: $We\nu$,
4: $Zee$, 5: $Z\nu_e\bar{\nu_e}$\\
{\tt LQED} &0: no ISR, 1: ISR included.\\
{\tt ROOTSMUL}& the total energy\\
{\tt SHCUT}& minimum invariant mass after ISR\\
{\tt ECUT}& minimum energy for the outgoing particles (4 values)\\
{\tt SCUT}& minimum invariant mass for outgoing particle
pairs (6 values)\\
{\tt CMAX}& maximum value of $\cos\theta$ between two
particles (14 values)\\
{\tt PAR}& labels of the produced fermions (4 character*3 values)\\
\end{tabular}
 
\noindent
All these values are reproduced in the output file.
 
\noindent{\bf Output}                               \\
The output prints the process considered, with the labeling
of the various particle momenta. Also a  complete list of
all abelian and nonabelian diagrams is given, and a list of
all generation channels that will be used.
Upon evaluation, information on the weight distribution is
given, and the results of the weight optimization procedure.
 
\noindent{\bf Availability}                         \\
The program is available from the authors upon
request, as well as from the CPC library.
 
 
\subsection{GENTLE/4fan}
%
\leftline{\bf Authors:}
 
\begin{tabular}{ll}
D. Bardin$^a$    & {\tt BARDINDY@CERNVM.CERN.CH} \\
M. Bilenky$^a$   & {\tt bilenky@ifh.de} \\
D. Lehner$^b$    & {\tt lehner@ifh.de} \\
A. Leike$^a$     & {\tt LEIKE@CERNVM.CERN.CH} \\
A. Olchevski$^a$ & {\tt OLSHEVSK@VXCERN.CERN.CH} \\
T. Riemann$^a$   & {\tt riemann@ifh.de}
\end{tabular}
\\ \\
\indent $ ^a$ {\sc Fortran} code {\tt gentle\_4fan.f} \\
\indent $ ^b$ {\sc Fortran} code {\tt gentle\_nc\_qed.f}
 
 
\noindent{\bf Description of the package}
 
\noindent
The {\tt GENTLE/4fan} package is designed to compute selected total
four-fermion production cross-sections and final-state fermion pair
invariant mass distributions for charged current (\emph{CC}) and
neutral current (\emph{NC}) mediated processes within the Standard
Model (SM).
For the \emph{CC03} subprocess, the W production angular distribution
is also accessible.
In the \emph{NC} case, SM Higgs Production is included.
The phase space integration is carried out by a semi-analytical
technique, which is described below.
The {\tt GENTLE/4fan} package is written in {\tt Fortran}.
It consists of two branches.
The basic branch {\tt gentle\_4fan.f} contains all features of the
package but complete initial-state radiation (ISR) to \emph{NC} processes.
The subroutine {\tt fourfan.f} called by {\tt gentle\_4fan.f} performs
the computation of \emph{NC} cross-sections and is described
in~\cite{gentle_4fan}.
The (as yet) independent branch {\tt gentle\_nc\_qed.f} includes complete
ISR to \emph{NC02} and \emph{NC08} and will soon be merged into
{\tt gentle\_4fan.f}.
 
\noindent{\bf Program features:}
\begin{enumerate}
\item\hed{Method of integration:}
The package is a {\em semi-analytical}~one. Without (with) ISR,
the phase space is parame\-trized by five (seven) angular variables
and the final state fermion pair invariant masses (plus the reduced
center of mass energy squared).
All angular variables are integrated analytically.
The resulting formulae are input to the package.
Invariant masses are subsequently integrated numerically with a
self-adaptive Simpson algorithm.
Optionally, for the \emph{CC03} subprocess, the W production angle may
also be numerically integrated.
The method is numerically stable and usually very fast.
 
\item\hed{Possible final states:}
The package may treat all four-fermion final states which do not
contain identical particles, electrons, or electron neutrinos. This
means that the package accesses all final states that are described by
{\em annihilation} and {\em conversion} type Feynman diagrams
(see~\cite{wwteup} for a classification):
\begin{enumerate}
  \item[(1)] \emph{CC03} (with complete ISR)~\cite{gentle_nunicc}
  \item[(2)] \emph{NC02}, \emph{NC08} (with complete ISR)~\cite{gentle_nuninc}
  \item[(3)] \emph{CC9}, \emph{CC10}, \emph{CC11}~\cite{gentle_unicc11}
  \item[(4)] \emph{NC06}, \emph{NC10}, \emph{NC24},
    \emph{NC32}~\cite{gentle_nc24}
  \item[(5)] \emph{NC} + Higgs~\cite{gentle_nc24h}
\end{enumerate}
Via flags, cross-sections for subsets of Feynman diagrams may be
extracted.
 
\item\hed{Cuts}
Cuts may be imposed on invariant masses of fermion pairs and on the
invariant mass of the final state four-fermion system.
Using the structure function approach in {\tt gentle\_4fan.f}, cuts on
the electron/positron momentum fraction can be imposed.
For the \emph{CC03} subprocess, cuts on the W production angle are
enabled.
 
\item\hed{Initial state radiation}
ISR is implemented into the package.
{\em Universal}~ISR is present for all processes~\cite{gentle_unicc11}.
In addition, the package includes complete, i.e. {\em universal}~and
{\em non-universal}~ISR for the \emph{CC03}, \emph{NC02}, and \emph{NC08}
processes~\cite{gentle_nunicc,gentle_nuninc}.
{\em Non-universal}~ISR does not contribute to {\em annihilation}
diagrams.
It may be argued that {\em non-universal}~ISR is very small,
${\cal O}(10^{-3})$, for {\em conversion}-{\em annihilation}
interferences.
The speed of the package is slowed down, if {\em non-universal}~ISR is
included, due to its complex analytical structure.
 
\item\hed{Final state radiation}
Final state radiation is not implemented.
 
\item\hed{Treatment of final state decays}
Final state decays are not accounted for.
 
\item\hed{Treatment of the Coulomb Singularity}
The Coulomb singularity is included according to reference~\cite{BBD}.
 
\item\hed{Treatment of the Anomalous Couplings}
Anomalous couplings are not included.
 
\item\hed{Treatment of masses}
In general, final-state masses are neglected in the matrix
elements. Where needed, however, masses are retained in the phase
space. In addition, masses of heavy particles coupling to the Higgs
boson are taken into account where appropriate.
 
\item\hed{Hadronization}
No interface to hadronization is foreseen.
 
\end{enumerate}
 
\noindent{\bf Input parameters}
 
\noindent
All input parameters are set inside the {\tt Fortran} code.
{\tt gentle\_4fan.f} uses the following flags, set in the subroutine
{\tt WWIN00}:
 
\begin{tabular}{ll}
 
 {\tt IBCKGR}: & \emph{CC03} case ({\tt IBCKGR}=0) or \emph{CC11} case
                 ({\tt IBCKGR}=1) \\
 {\tt IBORNF}: & Tree level ({\tt IBORNF}=0) or ISR corrected
                 ({\tt IBORNF}=1) quantities \\
 {\tt ICHNNL}: & \emph{CC03} ({\tt {ICHNNL}}=0), \emph{CC11} with
                 specific final state $\left[ l_1\nu_1 l_2\nu_2
                 ({\tt {ICHNNL}}=1), l\nu q{\bar q} \right. $ \\
               & $\left. ({\tt {ICHNNL}}=2,3), \;\;
                 q_1{\bar q}_1 q_2{\bar q}_2~
                 ({\tt {ICHNNL}}=4)\right]$, and
                 inclusive \emph{CC11} ({\tt ICHNNL}=5)  \\
 {\tt ICOLMB}: & Inclusion of Coulomb singularity ({\tt
                 ICOLMB}=1,...,5) or not ({\tt ICOLMB}=0) \\
               & Recommended value: {\tt ICOLMB}=2 \\
 {\tt ICONVL}: & Flux function ({\tt ICONVL}=0) or structure
                 function apporach ({\tt ICONVL}=1) \\
               & Recommended value: {\tt ICONVL}=0  \\
 {\tt IGAMZS}: & Constant $Z$ width ({\tt IGAMZS}=0) or $s$-dependent
                 $Z$ width ({\tt IGAMZS}=1) \\
 {\tt IINPT}:  & Input for tuned comparison ({\tt IINPT}=0) or
                 preferred Input ({\tt IINPT}=1) \\
 {\tt IIQCD}:  & Naive inclusive QCD corrections are included
                 ({\tt IIQCD}=1) or not ({\tt IIQCD}=0) \\
 {\tt IMMIM}:  & Minimal number of a moment requested by {\tt IREGIM}\\
 {\tt IMMAX}:  & Maximal number of a moment requested by {\tt IREGIM}\\
 {\tt IONSHL}: & On-shell ({\tt IONSHL}=0) or off-shell heavy bosons
                 ({\tt IONSHL}=1) \\
 {\tt IPROC}~: & \emph{CC} case ({\tt IPROC}=1) or \emph{NC} case
                 ({\tt IPROC}=2, call to {\tt fourfan.f} is initialized)
                  \\
 {\tt IQEDHS}: & Determination of the {\em universal} ISR radiator: \\
               & \indent ${\cal O}(\alpha)$ exponentiated
                 ({\tt IQEDHS}=--1,0); \\
               & \indent ${\cal O}(\alpha)$ exponentiated plus
                 different ${\cal O}(\alpha^2)$ contributions ({\tt
                 IQEDHS}=1,...,4) \\
               & Recommended value: {\tt IQEDHS}=3 
\end{tabular}
 
\newpage    
 
\begin{tabular}{ll}
 
 {\tt IREGIM}: & Calculation of the total cross-section ({\tt
                 IREGIM}=0), the moments of the radi- \\
               & ative loss of final state four-fermion invariant mass
                 ({\tt IREGIM}=1), the moments \\
               & of the radiative energy loss ({\tt IREGIM}=2), the
                 moments of the $W$ mass shift \\
               & $\left(\sqrt{s_+} \!+\! \sqrt{s_-} \!-\! 2
                 M_W\right)$ ({\tt IREGIM}=3), and the first moments of
                 $\cos\left( n\theta_W \right)$, \\
               & $n=1,...,4$ ({\tt IREGIM}=4) \\
 {\tt IRMAX}~: & Maximum value of {\tt IREGIM} \\
 {\tt IRSTP}~: & Step in a DO loop over {\tt IREGIM} \\
 {\tt ITVIRT}: & {\em Non-universal} virtual ISR included
                 ({\tt ITVIRT}=1) or not ({\tt ITVIRT}=0) \\
 {\tt ITBREM}: & {\em Non-universal} bremsstrahlung included
                 ({\tt ITBREM}=1) or not ({\tt ITBREM}=0) \\
 {\tt IZERO}~: & See equation (4.5) of~\cite{gentle_unicc11}.
                 Recommended value: {\tt IZERO}=1 \\
 {\tt IZETTA}: & See equation (4.21) of~\cite{gentle_unicc11}.
                 Recommended value: {\tt IZETTA}=1
\end{tabular}
 
\noindent
In the {\tt gentle\_nc\_qed.f} branch, only the flags {\tt IBORNF, IONSHL,
  ITVIRT, ITBREM} are used. The additional flag {\tt IBOSON} in
{\tt gentle\_nc\_qed.f} distinguishes between the \emph{NC02} and the {\tt
  NC8} process.
 
\noindent
The center of mass energy squared is chosen by setting the variable
{\tt IREG} and the parameters {\tt ISMAXA} or {\tt ISMAXB} in the main
program.
The following input may be changed by the user:
\begin{center}
\begin{tabular}{rcccl}
     {\tt GFER} & = &  $G_\mu$ & = &
        1.16639 $\times 10^{-5}$ GeV$^{-2}$, the Fermi coupling constant \\
     {\tt ALPW} & = &  $\alpha(2 M_W)$ & = &
        1/128.07, the running fine structure constant at $2\,M_W$ \\
     {\tt AME}  & = &  $m_e$ & = &
        0.51099906 $\times 10^{-3}$ GeV, the electron mass \\
     {\tt AMZ}  & = &  $M_Z$ & = & 91.1888 GeV, the $Z$ mass, \\
     {\tt AMW}  & = &  $M_W$ & = & 80.230 GeV, the $W$ mass \\
     {\tt GAMZ} & = &  $\Gamma_Z$ & = & 2.4974 GeV, the $Z$ width \\
     {\tt ALPHS}& = &  $\alpha_{_S}(2M_W)$ & = & $0.12$ 
\end{tabular}
\end{center}
 
\noindent{\bf Output}
 
\noindent
The following derived quantities are computed in {\tt gentle\_4fan.f}
and printed in the output:
\begin{eqnarray}
  {\tt GAMW} & = & \Gamma_W \; = \;
    \frac{9}{6\sqrt{2}\pi} \, G_{\mu}M_W^3
    \left( 1+\frac{2 \alpha_{_S}(2M_W) }{3\pi} \right)
    \nonumber \\
  {\tt SIN2W} & = & \sin^2 \theta_W \; = \; 1 - M_W^2/M_Z^2
    \nonumber \\
  {\tt GAE} & = & - \frac{e}{4s_Wc_W} \; = \;
    - \frac{\sqrt{4\pi\alpha(2M_W)}}{4s_Wc_W}
    \nonumber \\
  {\tt GVE} & = & {\tt GAE} \cdot (1-4s_W)
    \nonumber \\
  {\tt GWF} & = & \frac{g}{2\sqrt{2}} = - {\tt GAE} \cdot \sqrt{2} c_W
    \nonumber \\
  {\tt |GWWG|} & = & \sqrt{4\pi\alpha(2M_W)}
    \nonumber \\
  {\tt |GWWZ|} & = & {\tt |GWWZ|} \cdot \frac{c_W}{s_W}
    \nonumber
\end{eqnarray}
{\tt GVE} and {\tt GAE} are the electron vector and axial vector
couplings, {\tt GWF} is the fermion-$W$ coupling, and {\tt |GWWG|} and
{\tt |GWWZ|} are the trilinear gauge boson couplings for the photon
and the $Z$ respectively.
Further the output repeats the flag settings.
After the cross-section calculation, the following output is printed:
\begin{eqnarray}
  {\tt SQS}  & = & \sqrt{s}
    \nonumber \\
  {\tt XSEC0} & = & \sigma_{\rm tot}(s)  \hspace{.5cm} {\rm in~nanobarns}
\end{eqnarray}
In addition, the calculated {\tt MOMENTS} are printed. In the first
column {\tt IREGIM} is printed.
The second column is arranged in blocks of three lines each.
The first line contains the integer $n$.
The second line contains the $n^{th}$ moment of the physical quantity
indicated by {\tt IREGIM}.
The third line contains the dimensionless $n^{th}$ moment obtained
through division of the $n^{th}$ moment by the proper power of
$\sqrt{s}/2$.
 
\noindent
Although variable names are slightly different,
{\tt gentle\_nc\_qed.f} uses the same derived quantities as
{\tt gentle\_4fan.f}.
For one run, {\tt gentle\_nc\_qed.f} outputs the used flag values
together with the fermion code numbers {\tt IFERM1/IFERM2}, the color
factors {\tt RNCOU1/RNCOU2}, the masses {\tt AM1/AM2}, and the
invariant pair mass cuts {\tt CUTM12,CUTM34} for the final state
fermion pairs.
In addition, the lower cut {\tt CUTXPR} on the ratio of the
four-fermion invariant mass squared over the center of mass energy
squared, $s'/s$\ is output.
The main output, however, is an array of center of mass energies and
the corresponding total cross-sections.
 
\noindent{\bf Availability}                         \\
The codes are available from the authors upon E-Mail request or via WWW
 
\begin{tabular}{lcl}
  \hspace*{.5cm}
  {\tt gentle\_4fan.f} & from & {\tt http://www.ifh.de/}
        $\tilde{ }$ {\tt bardin/gentle\_4fan.uu} \\
  \hspace*{.5cm}
  {\tt gentle\_nc\_qed.f} & from & {\tt http://www.ifh.de/}
        $\tilde{ }$ {\tt lehner/gentle\_nc\_qed.uu}
\end{tabular}
%
 
 
\subsection{grc4f 1.0}
\leftline{\bf Authors:}
\begin{tabular}{ll}
J. Fujimoto       &  junpei@minami.kek.jp       \\
T. Ishikawa       &  tishika@gal.kek.jp         \\
T. Kaneko         &  kaneko@minami.kek.jp       \\
K. Kato           &  kato@sin.cc.kogakuin.ac.jp \\
S. Kawabata       &  kawabata@minami.kek.jp     \\
Y. Kurihara       &  kurihara@minami.kek.jp     \\
D. Perret-Gallix  &  perretg@cernvm.cern.ch     \\
Y.Shimizu         &  shimiz@minami.kek.jp       \\
H.Tanaka          &  tanakah@minami.kek.jp      \\
          e-mail: &  grc4f@minami.kek.jp
\end{tabular}
 
\noindent{\bf Program features}
 
\noindent
The program {\tt grc4f} is a Monte Carlo generator for
all final 4-fermion states generated  by  {\tt GRACE}\cite{grace}.
 
\noindent Several experimental cuts are implemented in default.\\
\noindent QED radiative corrections are implemented with structure
 functions for the ISR;
 in several processes QED parton shower (QEDPS) \cite{isr}
 is also an option, also for FSR.\\
\noindent Other final-state decays are implemented using
\texttt{JETSET}, \cite{jetset}.
 Color base information (related to the issue of color
 reconnection) is available.\\
\noindent The Coulomb term, and anomalous couplings, are both
 implemented.\\
\noindent Fermion masses can be kept nonzero everywhere.

 
 
\noindent{\bf Program layout}
 
\noindent{\bf Integration}
 
\noindent
The numerical integration of the differential cross section
over the phase space is carried out by the program {\tt BASES}
\cite{bases}.
The probability information is automatically produced and
saved in the file {\tt bases.data}, according to which the event
generation is done. An example is as follows:
 
\begin{quote}{\footnotesize \begin{verbatim}
      call bsinit                      initialization of BASES/SPRING.
      call userin                      initialization of parameters.
      call bases( func, estim, sigma, ctime, it1, it2 )  integration
      lun = 23
      open(lun,file='bases.data',status='unknown',form='unformatted')
      call bswrit( lun )               saving the information to a file.
      close ( lun )
\end{verbatim}}\end{quote}
 
In the arguments of subroutine {\tt bases}, {\tt func} is the
name of a function program, {\tt estim} is the cumulative estimate
of the integral, {\tt sigma} is the standard deviation of
the estimate of the integral, {\tt ctime} is the computing time in
seconds and {\tt it1} and {\tt it2} is the number of iterations
made in the grid optimization step and integration step.
 
\noindent{\bf Event generation}
 
\noindent
The event generation program {\tt SPRING}\cite{bases} samples a hypercube
according to {\tt bases.data},
and tests if this point is accepted by comparing the probability at
the point to the maximum probability in the hypercube.
When {\tt SPRING} accepts a point, the event corresponding to the
point is generated with weight one.
An example is as follows:
 
\begin{quote}{\footnotesize \begin{verbatim}
      implicit real*8 (a-h,o-z)
      parameter( nextrn = 6 )
      common /sp4vec/ vec(4,nextrn)
      ....
      real*4  p,v
      common /lujets/ n,k(4000,5),p(4000,5),v(4000,5)
      .....
      call bsinit                      initialization of BASES/SPRING.
      call userin                      initialization of parameters.
      lun = 23
      open(lun,file='bases.data',status='old',form='unformatted')
      call bsread( lun )               reading the probability information.
      close( lun )
 
      call gr2lnd                      setting parameters for JETSET from GRACE.
 
*===> Event generation loop
      mxtry  = 50                      number of maximum trials.
      mxevnt = 10000                   number of events.
      do 100 nevnt = 1, mxevnt
         call spring( func, mxtry )
        ( Four-momentum is stored in array vec.)
        ( The event information is converted into common block /lujets/.)
  100 continue
\end{verbatim}}\end{quote}
 
\noindent{\bf Input parameters}
 
\noindent
In the program {\tt grc4f} the menu modes are supported using the
command interpreter {\bf KUIP}\cite{kuip} developed at CERN and
the identical environment to {\bf PAW++}\cite{pawpp} is furnished to users,
who select the menu and type parameters in menu windows.
 
\begin{itemize}
\item Selection of 4 fermion process.
\item Center of mass energy:$\sqrt s$
\item Mass and width of all particles.
\item Experimental cuts
\begin{itemize}
\item Minimum and maximum angle cuts for each particles
(in the laboratory frame)
({\tt coscut}).
\item Minimum and maximum energy cuts for each particles({\tt engyct}).
\item Minimum and maximum invariant mass cuts({\tt amasct}).
($Q_{1} = (p_{3}+p_{4})^{2}$, $Q_{2} = (p_{5}+p_{6})^{2}$)
\item Resonance mass and width in case of $1/Q_{i}$-singularity.
\end{itemize}
\item Flag for Coulomb term.
\item Flag for anomalous couplings in some processes.
\item Selection of the calculation:
 no-radiation case, structure functions, or QEDPS.
\item Parameters for integration step: number of  iteration steps
and  number of sample points.
\item Parameters for event generation step: maximum number of
trials and number of events.
\end{itemize}
The general parameters in {\tt GRACE} can be found in the
{\tt GRACE} manual\cite{grace}
(spin polarization, graph selection and so on).
 
\noindent{\bf Output:}
 
\begin{itemize}
\item  Total cross section, the standard deviations
and the convergence behavior in the integration steps.
\item Histograms:
\begin{itemize}
\item $d\sigma /dE_{i}, i=3,4,5,6$:Energy distributions of each final particles
\item $d\sigma /d\cos\theta_{i}, i=3,4,5,6$
\item Invariant Masses $Q_{1}$ and $Q_{2}$.
\end{itemize}
\item Scatter plots:
\begin{itemize}
\item $\cos\theta_{i}$ -- $E_{i}$
\item $Q_{1}$ -- $Q_{2}$.
\end{itemize}
\end{itemize}
 
The contents of histograms and scatter plots are copied into the
{\tt HBOOK} format file\cite{hbook}.
 
\noindent{\bf Availability}                         \\
By anonymous ftp to
ftp location: {\tt /kek/minami/grc4f} at {\tt ftp.kek.jp}
 
 
\subsection{KORALW 1.03}
\leftline{\bf Authors:}
\begin{tabular}{ll}
M. Skrzypek   & skrzypek@hpjmiady.ifj.edu.pl  \\
S. Jadach     & jadach@cernvm.cern.ch         \\
W. P\l{}aczek & placzek@hephp02.phys.utk.edu  \\
Z. W\c{a}s    & wasm@cernvm.cern.ch
\end{tabular}
 
\noindent{\bf Description}                          \\
This program includes not only QED  effects in the initial state
but also in leptonic decays of $W$ and secondary decays, i.e. in the
 $\tau$ lepton decays. Hadronization of quarks is also performed.
The effects of spin are included in combined $W$-pair production
and decay. The $\tau$ polarization is also taken into account
in its decays.
Any experimental cut  and apparatus efficiency may be introduced
easily by rejecting some  of the generated events.
 
 
\noindent{\bf Program changes from version 1.02 to 1.03}
 
\noindent
Here we describe the main properties of the
generator {\tt KORALW}. We do not present the program, which
was published in \cite{koralw:1995a}-\cite{koralw:1995b}.
The present version 1.03 features all properties of the previous
version 1.02:
\begin{itemize}
\item
The matrix element for $W$-pair production and
$W$-pair decay into four fermions (the \emph{CC03} group)
with a proper $W$-spin treatment
and finite $W$ width,
\item
All $W$ decay channels into pairs of leptons or quarks,
\item
Initial-state multi-photon emission in the full photon phase space
(i.e. with finite transverse photon momenta),
\item
Simulation of the
decay of polarized $\tau$ leptons (from $W$ decay)
in all possible channels, taking into account spin polarization and
QED bremsstra\-hlung \cite{ref:Jadach}.
\item
Photon emission by
leptons in $W$ decay, up to double bremsstrahl\-ung
\cite{ref:Barberio}.
\item
Arrangement of
quarks from $W$ decay into colored strings and  fragmentation
into hadrons according to the LUND model using {\tt JETSET}~\cite{jetset}.
\item
Massive kinematics with exact four-momentum conservation for the
entire $W^-W^+$ production and decay process.
\end{itemize}
In version {\bf 1.03} the following four major
improvements have been introduced:
\begin{itemize}
\item
  Coulomb correction, in a form useful close to the $WW$
threshold.It is taken from ref. \cite{Khoze} and it can be activated
in straightforward way, as explained in the program documentation.
Starting from the present {\tt KORALW} version {\tt 1.03}, the
{\tt KeyCul} component of the program input parameter  {\tt NPAR(1)} is thus
{\em not} dummy anymore.
\item
{\tt KORALW} now includes an
interface to the external library calculating
the correction-weight due to
a more complete matrix element (so called background processes).
At present, an interface to the {\tt GRACE} library \cite{grace}
calculating multi-diagram matrix elements is available.
On occasion, one may wish to replace the matrix element by a different
one, for instance including special combinations of anomalous
couplings.
Due to the modular structure of {\tt KORALW} and, in particular,
due to the full factorizability of the
approximate QED matrix element into a Born matrix element and the QED part,
it is straightforward
to replace the existing Born-level matrix element  with any other one,
provided that the external library is able to calculate the corresponding
matrix elements out of the
externally generated four-momenta.
To this end an external program, calculating
the ratio of the matrix element squared of the particular choice to the basic
matrix
element squared of the program, has to be provided by the user.
 
A pre-defined interface, now included in {\tt KORALW}, will activate those
 routines
with the help of {\tt Key4f} component of {\tt KORALW} input parameter
{\tt NPAR(4)= 100*KeyACC +10*Key4f +KeyMix}. For {\tt Key4f=0} no
external matrix element
is
included and for {\tt Key4f=1 } it is active. The new position  of the weight
switch,
{\tt KeyWgt=NPAR(3)} is also introduced. For {\tt KeyWgt=2} the program
works as for the old and not modified {\tt KeyWgt=0} setting,
but the external weights are calculated and transmitted
to the common block {\tt wgtall}.

In our distribution directory (see section 4 of program documentation)
the additional fortran file  is introduced in the
directory {\tt interfaces}.
On the user side, his own directory has to replace the directory
{ \tt ampli4f}.
The following two routines have to be provided: {\tt AMPINI(XPAR,NPAR)}
which should initialize the
external matrix element library. Standard {\tt KORALW} input parameter matrices
{\tt XPAR} and {\tt NPAR} can be used there for the initialization purposes.
The
{\tt SUBROUTINE AMP4F(Q1,IFBM1,Q2,IFBM2,}
{\tt P1,IFL1,P2,IFL2,P3,IFL3,P4,} {\tt IFL4,}
{\tt WTMD4F,WT4F)}
should  calculate ratio {\tt WTMD4F}, of the new matrix element squared,
and the one of the standard {\tt KORALW}. The
{\tt Q1,IFBM1,Q2,IFBM2,P1,IFL1,P2,IFL2,P3,} {\tt IFL3, P4,IFL4} denote
respectively four momenta and identifiers (accordingly to the
PDG conventions \cite{PDG:1990}) of initial state effective
beams and the final state fermion states before final state bremsstrahlung
generation. The additional vector weight {\tt WT4F(I), I=1,9} may optionally
be
filled by routine
{\tt  AMP4F}. It is not used in the program but only
transmitted to the {\tt KORALW} optional weights common block
{\tt wgtall} as {\tt wtset(40+I)}. The {\tt WTMD4F} is set into
{\tt wtset(40)}.
 
An example of the interfaced external matrix-element, based on the
{\tt GRACE} code \cite{grace}, can be obtained upon request
from the authors of {\tt KORALW}. In the distribution version we include
a dummy {\tt ampli4f} library. It  sets the
external weight to $1$ and prints a warning message.
 
We found it useful to introduce
the {\tt KeyWu} switch which controls the  level of sophistication of the
$W$ width implementation.
Like for the $Z$ ({\tt KeyZet}) case {\tt KeyWu}=0,1,2 denotes respectively
$(s/M_W)\Gamma_W$, constant and zero $W$ width.
Note that
{\tt NPAR(2)=}\\{\tt 100000*KeyWu +10000*KeyRed}
{\tt +1000*KeySpn+100*KeyZet +10*KeyMas +KeyBra}.
\item
Anomalous couplings for the $WWV$, $V=Z,\gamma$ vertices in the
built-in matrix element are parameterized by $2\times7$ variables
$g_1^V, g_4^V,g_5^V, \lambda_V, \kappa_V, \tilde\lambda_V,
\tilde\kappa_V$ as defined in \cite{ref:Hagiwara}. They can be
reached by {\tt KeyACC} component of {\tt KORALW} input parameter
{\tt NPAR(4)=100*KeyACC+10*Key4f} {\tt +KeyMix}. {\tt KeyACC=1} activates
their values as set by the user via {\tt KORALW} input parameter vector
{\tt xpar} (see routine {\tt KORALW} for more details)
and prints them to the output. {\tt KeyACC=0}
enforces the Standard Model values.
\item The semianalytical part of the program
{\tt KORWAN} was enlarged with two functions {\tt s1wan(s1)} and
 {\tt s1s2wan(s1,s2)} for the one and two dimensional distribution of the
single
or double W invariant masses. These functions require
standard initialization of the {\tt KORWAN} routine
with the input parameters as explained in {\tt KORALW} manual.
Optionally, if the {\tt KORWAN} input parameter {\tt keymod} is increased by
10000
the
calculations in {\tt KORWAN} are not executed and the initialization is
performed
only.
 
\end{itemize}
 
\noindent
Still remaining limitations of the program are:
\begin{itemize}
\item
A simplified matrix element for the QED photon emission,
\item
Lack of electroweak non-QED corrections\footnote{Most
    probably these corrections
    are small in comparison with the experimental
    precision and it is not necessary to include them in the Monte Carlo
    program -- it is enough if they are in the auxiliary
    semi-analytical program.},
\item
A simplified ``color arrangement'' for four quark jets.
\end{itemize}
The above and other shortcomings of the program will be systematically
addressed
in the forthcoming versions of the program.
 
\noindent{\bf Availability}                         \\
The Version: 1.03 is available from\\
www: http://hpjmiady.ifj.edu.pl/programs/programs.html
 
\subsection{LEPWW}
\label{sec:LEPWW}
\newcommand{\noi}{\noindent}
\leftline{\bf Author:}
\begin{tabular}{ll}
    F.C. Ern\'e &
     z63@nikhef.nl
\end{tabular}
 
\noindent{\bf Description}
 
\noindent
The original {\tt LEPWW} event generator\cite {ref:Kleiss} contains \emph{CC03}
and \emph{NC02} tree-level
diagrams for the processes $e^-e^+\rightarrow u\bar u u\bar u$,
$e^-e^+\rightarrow u \bar u d\bar d$ and $e^-e^+\rightarrow u\bar u d\bar d$,
with massless fermions and $W$ and $Z$ poles. Its present name and version
is `egwwv208.car' in the L3 event generator library. A FORTRAN file is
available.
 
\noindent{\bf Features of the program}
 
\noi A complete set of final state fermions is available.
 
\noi Order $\alpha$ initial-state radiation, allowing transverse momentum,
is implemented following the procedure in the REMT routines\cite{ref:Berends}.
 
\noi Final state radiation from electrons, muons and $\tau$'s can be
switched on optionally, according to the {\tt PHOTOS}
package\cite{ref:Barberio}.
 
\noi For $\tau$ decay final lepton states of definite helicity are projected
out, which allows decay through an adapted version of the {\tt TAUOLA}
routines\cite{ref:Jadach}.
 
\noi Non-SM couplings have been implemented with the parameterization
of Hagiwara et al\cite{ref:Hagiwara}.
 
\noi Quark fragmentation proceeds through {\tt JETSET} routines\cite{jetset}.
 
\noi QCD effects on the boson widths and branching ratios can be taken into
 account.
 
\noi No Coulomb term is implemented.
 
\noi The program aims at a 1 to 2\% precision in the description of total
and differential processes. The program has been available throughout
the LEP2 workshop. The development has been completed.
 
\noindent{\bf Input parameters: data cards}
 
\noindent\begin{tabular}{ll}
FAW, FAZ &  fudge factors for W and Z width \\
PROC &  Generate WW or ZZ                  \\
DKW1,DKW2 & Decay of $W^+,W^-$ into $q\bar q$, $e\nu$, $\mu\nu$,
$\tau\nu$                                              \\
DKZ1,DKZ2 & Decay of $Z1,Z2$ into $q\bar q$, $\nu\bar\nu$,
$e^+e^-$, $\mu^+\mu^-$, $\tau^+\tau^-$                 \\
IRAD,FRAD & Flags for initial and final-state radiation \\
WMAX & Maximum weight                                  \\
F1G-F7Z & Fourteen variables for the Triple Boson Vertex \\
LEP2 & LEP2 workshop parameters; it overrules the other data cards
\end{tabular}
 
\noindent{\bf Availability}                         \\
\verb+http://www.fys.ruu.nl/~dieren/LEPWW.html+
 
\subsection{LPWW02}
\leftline{\bf Authors}
\begin{tabular}{ll}
Ramon Miquel & miquel@alws.cern.ch \\
Michael Schmitt & schmitt@vxaluw.cern.ch
\end{tabular}
 
\noindent{\bf General description}                  \\
{\tt LPWW02} is a
Monte Carlo program for the simulation of four-fermion final states at LEP2.
It contains the Feynman diagrams with two
resonating W's and Z's and features, among other things,
initial- and final-state radiation,
Coulomb singularity effects and effective
couplings. It is interfaced to the {\tt JETSET} package to handle gluon
radiation,
hadronization and decays.
 
The generator is based on a complete Monte Carlo calculation of the cross
section for the process $e^+e^-\to f_1\bar{f}_2f_3\bar{f}_4$ through a pair
of heavy bosons, WW and/or ZZ~\cite{kleiss}.
Initial-
and final- state radiation are incorporated with structure
functions. The Monte
Carlo algorithm
for event generation uses two subgenerators to generate the WW and
ZZ topologies. Suitable approximants are used in the generation step to increase
its efficiency using the importance sampling technique. At the end, a rejection
algorithm ensures that the unweighted events produced are distributed according
to the exact matrix element.
A complete description of the physics in the
program, with results
and comparisons with other calculations is available~\cite{lpww02}.
 
\noindent{\bf Features of the program}
 
\begin{itemize}
\item {\tt LPWW02} is a Monte Carlo event generator of unweighted events.
Any cut can be
applied to the generated events.
\item The accessible final states are those that can be produced in $e^+e^-$
collisions from intermediate states consisting on two W bosons or two Z bosons:
$u\bar{d}\mu^-\bar{\nu}_\mu$,
$u\bar{u}\mu^+\mu^-$,
$u\bar{u}d\bar{d}$,... In flavor configurations like the last one, the
interference between the WW and ZZ diagrams is properly taken into account.
In a given run, the user can either specify a fixed final state or get directly
the correct flavor mix for events produced through two W's and/or two Z's.
\item Initial state radiation is simulated using the structure-function
approach~\cite{ref:KF,ref:NITR}.
The Born-like cross section
at the reduced center-of-mass energy after initial-state radiation is
convoluted with the structure functions of the
electron and positron, which take into
account their probabilities to radiate.
The electron structure function, $D_e(z,s)$, taken from
ref.~\cite{ref:NITR}, includes soft-photon exponentiation and
leading-logarithmic corrections up to $\cal{O}$($\alpha^2$).
The structure function approach is used
in the collinear approximation and, hence, the
photon
direction is assumed to be that of the incoming beams. Consequently, no
real photon four-momenta are generated inside the experimentally accessible
regions of phase space.
Since the radiation not only changes the effective
center-of-mass energy of the event, but also the center-of-mass momentum
with respect to the laboratory system, a boost is applied to the
generated particles to take this into account.
\item
We employ the {\tt PHOTOS}
package~\cite{ref:Barberio} to simulate radiation from final
state electrons and muons.
Radiation from quarks is taken care of by the {\tt JETSET}
package~\cite{jetset}
Radiation from taus or their decay products is
neglected.
The algorithm in {\tt PHOTOS} provides full kinematic
information for the splitting $f\goto f^\prime\gamma$.
It is based on an implementation of $\cal{O}$ $(\alpha^2)$
bremsstrahlung calculation in the leading-log approximation.
This means that final-state radiation
does not influence the total cross section calculation in any way.
\item In the first stage, the program produces a final state consisting on
four-fermion plus a number of photons.
The interface with {\tt JETSET} takes care of hadronization
and subsequent decays of
hadrons. {\tt JETSET} also takes care of decaying the tau leptons.
\item We have implemented the Coulomb correction in the
production of two W's following ref.~\cite{ref:Fadin}. It is numerically
equivalent to the treatment of ref.~\cite{BBD}.
\item At this time, the possibility of anomalous couplings is not contemplated
in the program.
\item The fermions are generated with their appropriate masses. However the
matrix element is computed in the massless limit.
\item {\tt LPWW02} is interfaced with {\tt JETSET}.
\item It is straight-forward to get the information on the contributions from
different sets of diagrams in view of a possible simulation of the effect of
color recombination.
\end{itemize}
 
\noindent{\bf Program layout}
 
The structure of the program can be summarized as follows:
\begin{itemize}
\item Initialization. It includes the
computation of the maximum weight for the
rejection algorithm that will be used later
and the initialization of the {\tt PHOTOS} package
used for final-state radiation.
\item Event Loop. A fixed number of unweighted events are generated. There are
a number of steps:
\begin{itemize}
\item The electron and positron effective energies at collision point after
radiation are generated.
\item The final state flavor is chosen randomly according to some
approximate probabilities that take into account Cabibbo mixing. Alternatively,
the final state can be fixed to a particular combination of flavors.
\item One of two subgenerators is chosen randomly to generate the event
kinematics. One of them maps the peaks for the WW channel, the other for the ZZ
channel.
\item The exact matrix element squared is computed. A weight is assigned to each
event according to the ratio of the exact matrix element squared
to the approximate weights used in the generation stage,
including the ones for choice of flavor composition and
initial-state radiation.
\item A rejection algorithm is applied to the final weight to get
unweighted events.
\item The four momenta are given their corresponding masses, readjusting the
kinematics of the event. The event is boosted
to the lab frame according to the incoming electron and positron
effective energies.
\item {\tt PHOTOS}
is called to provide final-state radiation off electrons and muons
only.
\item {\tt JETSET} is invoked to take care of hadronization,
decays and final state
radiation off quarks or hadrons.
\item Four-vectors are stored in the standard Lund common block.
\end{itemize}
\item Final: The cross section is computed with statistical error. A summary of
the run is given.
\end{itemize}
 
\noindent{\bf Input Parameters and Flags}
 
The following is a description of the input parameters and flags
together with the values used for the tuned comparisons:
\begin{itemize}
\item {\tt XMZ=91.1888}, mass of the Z (GeV).
\item {\tt XMW=80.23}, mass of the W (GeV).
\item {\tt ALFA0=137.0359895}, $ 1/ \alpha_{_{QED}}(0)$. Used for the photon
radiation.
\item {\tt ALFA=128.07}, $ 1/ \alpha_{_{QED}}(s)$.
\item {\tt GF= 1.16639E-5}, Fermi constant.
\item {\tt ALFAS=0.}, $\alpha_s(M_W^2)$. Set to zero for the tuned comparisons.
\item {\tt WWUSER=2.03367}, user value for W width. Ignored if UWFLAG=0.
\item {\tt ZWUSER=2.4974}, user value for Z width. Ignored if UWFLAG=0.
\item {\tt IRFLAG=1}, generate initial-state radiation (1) or not (0).
\item {\tt CSFLAG=0}, include the Coulomb correction (1) or not (0)
\item {\tt BWFLAG=1}, Breit-Wigner with mass-dependent (1) or
constant (0) width.
\item {\tt ASFLAG=0}, apply $\alpha_s$ correction for widths (1) or not (0).
\item {\tt FRFLAG=0}, generate final-state radiation (1) or not (0)
({\tt PHOTOS}).
\item {\tt IZFLAG=0}, include contributions from ZZ diagrams (1) or not (0).
\item {\tt ILFLAG=0}, invoke {\tt JETSET} for showers,
fragmentation, and decay (1)
or not (0).
\item {\tt UWFLAG=1}, use total W and Z widths from the user (1) or the SM (0).
\end{itemize}
 
The preferred values would differ from the previous ones in the following:
\begin{itemize}
\item {\tt ALFAS=0.12}
\item {\tt CSFLAG=1}
\item {\tt ASFLAG=1}
\item {\tt FRFLAG=1}
\item {\tt IZFLAG=1}
\item {\tt ILFLAG=1}
\item {\tt UWFLAG=0}
\end{itemize}
 
\noindent{\bf Output}
 
The program's output consists on the result of the cross section for the
required final state. An estimate of the
statistical error is also provided.
The four-momenta of the generated particles are available
in the event loop through the standard Lund common block.
 
\noindent{\bf Availability of the program}          \\
{\tt LPWW02} is available from the authors.
 
 
\subsection{PYTHIA 5.719 / JETSET 7.4}
\leftline{\bf Author:}
\begin{tabular}{ll}
Torbj\"orn Sj\"ostrand & torbjorn@thep.lu.se
\end{tabular}
 
\noindent{\bf Description}
 
\noindent {\tt PYTHIA/JETSET} is a general-purpose
event generator for a multitude of processes in e$^+$e$^-$, ep and
pp physics \cite{TSpyCPC,TSpymanual}. The emphasis is on the detailed
modeling of hadronic final states, i.e. QCD parton showers, string
fragmentation and secondary decays. The electroweak description is
normally restricted to improved Born-level formulae, and so is not
competitive for high-precision studies.
 
\noindent{\bf Features of the program}
 
\begin{itemize}
 
\item Monte Carlo event generator.
 
\item By default any final state allowed for a process is included in
the generation, but it is possible to select a specific combination of
final states with large flexibility.
 
\item Several cuts are available, if desired. Examples include the mass
ranges for the hard scattering process and for resonances. It is not
possible to set cuts directly on the four final fermions, however.
 
\item ISR is implemented in a two-stage process. First structure functions
are used to select $x_1$ and $x_2$ values for the hard scattering.
Currently the structure function is the one recommended for LEP~1
\cite{TSKleiss}, but it would be easy to expand to more alternatives.
Thereafter a backwards evolution scheme is used to reconstruct explicit
sequences of e$\to$e$\gamma$ branchings, including $p_{\perp}$ recoils.
The algorithm used is essentially the same as originally developed for QCD
applications \cite{TSinitial}.
 
\item FSR is implemented inside each gauge boson system separately.
For a W this means as it would have been obtained in the formal limit
$\Gamma_{\mbox{\scriptsize W}} \to 0$. Again a parton-shower description
is used, with explicit matching to the first-order matrix elements,
as for final-state QCD radiation \cite{TSfinal}. Quarks can radiate both
photons and gluons.
 
\item For the hard process e$^+$e$^-\to$W$^+$W$^-$, only $x_1$, $x_2$,
the two W masses and one relative angle are selected
\cite{gentle_unicc11},~\cite{gentle_nunicc}.
FS decays are considered in a second step, using the formulae of
\cite{TSdecay} to calculate the conditional probability for a set of
four decay angles (two for each W). The philosophy is the same for
other processes.
 
\item Several optional Coulomb formulae are available \cite{TScoulomb};
the recommended one is the first-order expression in \cite{TScoulFKM}.
 
\item No anomalous couplings.
 
\item Finite fermion masses are included in the phase-space factors for
partial widths.
 
\item Hadronization comes built-in.
 
\item Since the program does not include interference e.g. between the
WW and ZZ processes, each individual event is uniquely assigned to a
specific process, and this information is available to the user.
 
\end{itemize}
 
\noindent{\bf Program layout}
 
   At initialization, coefficients are optimized in the analytical
expressions subsequently used to select kinematical variables
(i.e. phase-space points will be picked more often in those regions
where the matrix elements are peaked), and the
corresponding maxima of differential cross sections are found.
For each event, a process type and a phase-space point is selected by
hit-or-miss Monte Carlo. That is, events come with unit weight (but an
option with weighted events exists). The maximum found in the
initialization is increased if one encounters a larger differential
cross-section value. (Formally this introduces an error in
the method, but when the increase occurs early in the run and/or is small,
this error is negligible.) The cross-section information is improved with
increasing statistics. After its selection, the hard scattering is
gradually dressed up, by the addition of initial-state radiation,
resonance decays, final-state radiation and hadronization.
 
     Note that $\Gamma_{\mbox{\scriptsize Z}}$ is not set independently
in {\tt PYTHIA}; rather it is given by electroweak relations and is thus
too small when one asks for $\alpha_{\mbox{\scriptsize s}} =0$.
 
   Each event is listed in full in {\tt COMMON/LUJETS/}
(optionally also in {\tt COMMON/HEPEVT/}),
so any experimentally definable quantity can be extracted.
Also other pieces of event information is available in common blocks.
A table of cross sections can be obtained, but this does not include
error estimates.
 
\noindent{\bf Availability and documentation}       \\
   The master copies of the programs, documentation and sample main programs
are available at web address http://thep.lu.se/tf2/staff/torbjorn/.
 
   The main reference is \cite{TSpyCPC}. A full manual and physics
description (over 320 pages) is \cite{TSpymanual}. An overview,
with a table of the most interesting subprocesses, is given
in the QCD generators section of this report.
 
 
\subsection{WOPPER 1.4}
\leftline{\bf Authors:}
\begin{tabular}{ll}
Harald Anlauf & anlauf@crunch.ikp.physik.th-darmstadt.de \\
Thorsten Ohl  & Thorsten.Ohl@Physik.TH-Darmstadt.de
\end{tabular}
 
\noindent{\bf General description:}                 \\
    \texttt{WOPPER} is a fairly standard Monte Carlo event generator
    for \emph{un}weighted $e^+e^-\to4f$ events
    \cite{wopper1}-\cite{wopper3}.  Emphasis is put on
    leading logarithmic radiative corrections to $W^\pm$ pair production
    (i.e.~doubly resonant four-fermion production at LEP2).  An
    extension to singly resonant four-fermion production is being tested
    and will be released as \texttt{WOPPER} version 1.5.
    \texttt{WOPPER} is interfaced with fragmentation and hadronization
    Monte Carlos to allow full simulation of event samples at LEP2.
 
\noindent{\bf Features:}
\begin{itemize}
      \item \texttt{WOPPER} is a Monte Carlo event generator with
        \emph{un}weighted events, suitable for full simulation of
        event samples.
      \item All possible four-fermion final states are generated.
      \item All cuts can be applied to the final states.
      \item Initial state QED radiation is implemented in leading
        logarithmic approximation.  The leading logarithms $\propto
        (\alpha/\pi)(\ln (s/m_e^2) - 1)$ from collinear and soft
        emission are summed to all orders in a parton shower algorithm
        using the first order non-singlet splitting functions.  A
        finite $p_T$ for photons and the hard scattering center of
        mass system is generated according to the~$1/pk$ pole.
      \item Final-state QED radiation is not implemented.
      \item Decays of final states are left to external packages.
        Standard interfaces are implemented.
      \item Coulomb corrections are implemented with finite width
        according to ref.~\cite{BBD}.
      \item Anomalous couplings are not implemented.
      \item Finite fermion masses are implemented in the kinematics,
        but the matrix elements are calculated in the massless limit.
      \item Fragmentation and hadronization are left to dedicated QCD
        Monte Carlos.  The standard $W^+W^-$--QCD event generator
        interface is implemented.
      \item Currently, only charged current diagrams are implemented,
        therefore information on color reconnection is neither needed
        nor available.
\end{itemize}
 
 
\noindent{\bf Algorithm:}
    \begin{itemize}
      \item \texttt{WOPPER}'s initialization phase starts with
        calculating the coupling constants from the input
        parameters according to the value of \texttt{scheme}.
        The maximum of the total hard cross section
        $\sigma(s,k_+^2,k_-^2)$ for off-shell $W^\pm$ pair
        production is determined to allow the generation of
        \emph{un}weighted events.  NB: $k_\pm^2$ do not really
        correspond to off-shell $W^\pm$'s for singly resonant
        contributions.
      \item For event generation, an off-shell $W^\pm$ pair is
        produced with the invariant mass reduced and the center of
        mass system boosted from radiative corrections.  This pair is
        subsequently decayed, keeping all angular correlations among
        the four decay fermions.
      \item A Monte Carlo estimate of the total cross section
        based on the events generated so far can be requested at
        any time.  In particular, it is produced in the clean-up
        phase.
    \end{itemize}
 
\noindent{\bf Input parameters:}
\begin{enumerate}
    \item \textbf{Tuned comparison:}
        \begin{itemize}
          \item \texttt{scheme}: $1$, i.e.~use $G_F$, $M_W$ and
            $\alpha_{_{QED}}(2M_W)$ as input and calculate
            $\sin^2\theta_W = \pi\alpha_{_{QED}}(2M_W)/(\sqrt2 G_F M_W^2)$
            as well as $\Gamma_W = G_F M_W^3
            (3 + 2\alpha_{_{QCD}}(2M_W)/\pi)/(\sqrt8\pi)$.
          \item \texttt{mass1z}: $M_Z=91.1888$
          \item \texttt{gamm1z}: $\Gamma_Z=2.4974$
          \item \texttt{mass1w}: $M_W=80.23$
          \item \texttt{gfermi}:
            $G_F=1.16639\cdot10^{-5}\mathop{\textrm{GeV}}^{-2}$
          \item \texttt{ahpla}: $1/\alpha_{_{QED}}(2M_W)=128.07$
          \item \texttt{alphas}: $\alpha_{_{QCD}}=0$
          \item \texttt{ckmvus}: $V_{us}=0$
          \item \texttt{ckmvcb}: $V_{cb}=0$
          \item \texttt{ckmvub}: $V_{ub}=0$
          \item \texttt{coulom}: \texttt{false}, i.e.~no Coulomb
            correction
        \end{itemize}
 
    \item\textbf{Preferred input:} the input used in the
        \textit{``Best You Can Do''} event samples is identical to
        the one used in the tuned comparison, except for
        \begin{itemize}
          \item \texttt{alphas}: $\alpha_{_{QCD}}(M_Z)=0.123$
          \item \texttt{ckmvus}: $V_{us}=0.2196$
          \item \texttt{ckmvcb}: $V_{cb}=0.0400$
          \item \texttt{ckmvub}: $V_{ub}=0.0032$
          \item \texttt{coulom}: \texttt{true}, i.e.~apply Coulomb
            correction
        \end{itemize}
    \end{enumerate}
    In addition to the above $G_F$-scheme, the following schemes are
    available:
      \begin{itemize}
        \item $\mathop{\texttt{scheme}}=-1$:
          like $\mathop{\texttt{scheme}}=1$, but for $\Gamma_W$, which
          is taken from the input parameter \texttt{gamm1w}
        \item $\mathop{\texttt{scheme}}=2$: use \texttt{sin2w}
          ($\sin^2\theta_W$) as input and calculate $G_F =
          \pi\alpha_{_{QED}} / (\sqrt2 \sin^2\theta_W M_W^2)$
        \item $\mathop{\texttt{scheme}}=-2$:
          like $\mathop{\texttt{scheme}}=2$, but for $\Gamma_W$, which
          is taken from the input parameter \texttt{gamm1w}
        \item $\mathop{\texttt{scheme}}=3$: use \texttt{sin2w}
          ($\sin^2\theta_W$) and \texttt{gfermi} ($G_F$) as independent
          input parameters and force $\alpha_{_{QED}}(s)=\alpha_{QED}(0)$
        \item $\mathop{\texttt{scheme}}=-3$:
          like $\mathop{\texttt{scheme}}=3$, but for $\Gamma_W$, which
          is taken from the input parameter \texttt{gamm1w}
      \end{itemize}
 
\noindent{\bf Output:}                              \\
After startup and initialization,
    \texttt{WOPPER} prints a version number and a description of the
    selected input parameter scheme to standard output. Additional
    \texttt{print} commands can be used to print some or all internal
    flags and parameters.  Generated events are stored in the standard
    \texttt{/HEPEVT/} common block and a user routine (by default
    ``\texttt{call hepawk('scan')}'') is called.
    At the end of the run, the total cross section and an error
    estimate is available in the last \texttt{/HEPEVT/} record.
 
\noindent{\bf Availability:}                        \\
The \texttt{WOPPER} distribution can be
    obtained directly from the authors or from the internet
    \begin{itemize}
      \item WWW:
        \texttt{http://crunch.ikp.physik.th-darmstadt.de/}\goodbreak
        \texttt{monte-carlos.html\#wopper}
 
  \item Anonymous FTP from \texttt{crunch.ikp.physik.th-darmstadt.de},
        \goodbreak in the directory \texttt{pub/ohl/wopper}
    \end{itemize}
    Ready-to-run versions are available in the experimental LEP2
    collaborations.
 
 
\subsection{WPHACT}
 
\noindent
{\bf W} W and Higgs  Physics with {\bf PHACT}  \\ \\
\leftline{\bf Authors:}
 
\begin{tabular}{ll}
E.~Accomando & accomando@to.infn.it            \\
A.~Ballestrero & ballestrero@to.infn.it
\end{tabular}
 
\noindent{\bf General description}
 
\noindent
{\tt WPHACT} is a program created to study  four-fermion,
WW and Higgs  physics
at present and  future $e^+ e^-$ colliders. In its present form, it can
compute all SM processes with four fermions in the final state. For NC
processes involving b quarks, and no electrons in the final state,
finite b masses can be fully taken into account.
 
Full tree-level matrix elements for all CC and NC processes are computed
by means of subroutines which make use of the helicity formalism of ref.
\cite{method}. Their code has been written semi-automatically through
 the set of routines PHACT \cite{phact} ({\bf P}rogram for {\bf H}elicity
{\bf A}mplitudes {\bf C}alculations with {\bf T}au matrices)  which implements
 the  method in a fast and efficient way.
 
In the above  formalism, eigenstates of the fermion propagators are
used to simplify   matrix expressions. These eigenstates are chosen to be
generalizations of the spinors used in ref.\cite{ks}. Essentially,
the numerator of fermion propagators are diagonalized in the massless lines
and have very simple expressions in the massive ones.  The computation
of fermion lines reduces to evaluating  the matrices corresponding to
insertions of vector or scalar lines and combining them together.
This is performed most efficiently with the so-called
\emph{tau matrices} \cite{method}.
 The program PHACT writes automatically the optimized
fortran code necessary for every insertion and every combination,
given the names of the vectors, couplings, etc.  From various comparisons made,
we have been convinced that in fact the codes for the amplitudes written in
this way run very fast, and this is the case also for {\tt WPHACT}.
 
Different phase spaces, with different
random number mappings, are employed  in order to take into account
  the peak structure of the  resonating diagrams for the different processes.
  The adaptive routine  {\tt VEGAS}\cite{vegas}
is used for integrating over the
  phase space.
 
For additional information, see also the section on event generators
for Higgs physics.
 
\noindent{\bf Features of the program}
 
\noindent
    {\tt WPHACT} is a Monte Carlo program.
  For all phase spaces used, all momenta are explicitly
   computed in terms of
the  integration variables. This implies that any cut
   can be implemented, and it can be easily used also as an event generator.
   The events obtained in
this way are of course weighted. {\tt VEGAS} is an
   adaptive routine,
which normally runs a few iterations (good efficiency is
   normally obtained with about three iterations), seeking for a better
   grid of the integration space. If one doesn't want to generate too many
   events, it is better to  use the events
of the last iteration. Distributions
   for any variable can also be implemented.
   Even if various distributions have already been produced, and
examples are
   available, no automatic implementation of distributions has yet been
   introduced.
\par
   All SM final states with four fermions can be calculated.
    No W's or Z's or Higgs  are allowed in the final state.
    They are always appropriately considered as virtual particles.
\par
   Any cut can be performed.
   Initial state QED radiation is included through Structure Functions
  ${\cal O} (\alpha ^2)$.  FSR is not implemented.
    The Coulomb term is implemented with the approach of ref. \cite{BBD}.
    Anomalous couplings are available.  No interface to hadronization
    is available.
\par
    So far the only fermion masses which can be different from zero are
 those of quarks in NC  processes relevant for Higgs production, like e.g.
$e^+ e^- \rightarrow b \bar b b\bar b$,
$ e^+ e^- \rightarrow \nu_{e} \bar\nu_{e}  b\bar b$, etc.
The nonzero masses are fully taken into account both
in the matrix element and in the phase space. Just because of the helicity
formalism adopted,
the massive case does not cost much more than the massless
one in cpu time.
\par
    It is easy to obtain the contributions from different set of diagrams,
 as every diagram is
evaluated individually for all helicity configurations and then
summed to the others
before squaring and summing over helicity configurations.
Actually, in the
case of mixed CC and NC processes the two contributions are
evaluated and integrated separately.
  \par As far as speed is concerned, we give some indicative values about
 the running time on  ALPHA AXP 2100/4 OVMS:
  \par
   \ CPU time  per call for \emph{CC03} without ISR:
   \hskip 2cm  $5.6\times 10^{-5}$
    sec.
  \par
   \ CPU time  per call for \emph{CC11} with ISR:
   \hskip 2.3cm  $1.2\times 10^{-4}$
    sec.
   \par At Lep2 energies, 30 M  calls (about one hour)  are used
    to obtain \emph{CC11} with ISR
cross section with a typical estimated error of
    about  $1 \times 10^{-4}$.  The same process can be evaluated in about
    $2$ minutes with 1 M calls at permille level.
    For \emph{CC03} without ISR 20 M calls (20 minutes) give an estimated error
    of about $1 \times 10^{-4}$ and 1 M calls (1~minute) are necessary for
    permille precision. The same programs are about $5$ times slower on a
    VAXstation 4000/90.
 
\noindent{\bf Program layout}
 
\noindent
    The  variables by which the phase
spaces are described are  the W masses
    for CC contributions, the Z masses for NC contributions, together with
   the angle of the two virtual particles with respect to the beam,
   the decay angles in their rest frames, and $x_1$, $x_2$, the fractions
    of momenta carried by the electrons.  Appropriate change of variables
   to take care of peaks in $x_1$, $x_2$, $M_W$ or $M_Z$ lead to the real
integration variables. For every point chosen by the integration routine,
   the full set of four-momenta is reconstructed and passed to the
   subroutine which evaluates the differential cross section
with the helicity amplitude formalism. For every point in the integration
   variables, i.e. for every set of four momenta chosen,
{\tt VEGAS} gives a weight
   which must
be used together with the value of the cross section for producing
   distributions.
\par Four phase spaces are available and have been used  for the different
  matrix elements  contributions,  depending on the number of possible
  resonances.
Every single phase space  integrates   better that
  particular contribution it
has been constructed for. After various tests we
  however found that the phase space suitable for double resonant
  contributions
is quite precise also in evaluating all contributions together.
   It turns out to be faster than splitting the contributions and
  integrating them separately with automatic determination of the relative
  precision.
  At present
all contributions are normally evaluated together with one single
  kind of phase space. When mixed CC and NC are present, it is better to
  run the two
contributions separately (adding the interference to the biggest
  one), as the change
of variables necessary to take care of  the resonances
  depends on their masses.
 
\noindent{\bf Input parameters, flags, etc.}
 
\noindent
    Normal input parameters are $M_W$, $M_Z$, $\alpha$, $\alpha_S$.
    In the tuned comparisons  $sin^2 \theta_W$ has also been given as an
    input,  while it is usually derived from the relation
    $sin^2 \theta_W=1-M_W^2/M_Z^2$.
\par
    The main flag of the program is {\tt ich}, which
    chooses among different final states. Other flags
    allow to compute
    with (when their value $=1$) or without (when their value $=0$)
    ISR, Coulomb corrections and $\alpha_S$ corrections.
    They are respectively : {\tt isr, icoul, iqcd}.
    The last option refers at present only to CC10 processes.
    A flag ({\tt iterm}) allows using ({\tt iterm} $=1$) or not ({\tt iterm}
    $=0$) some iterations (normally one is enough) for thermalizing.
    The number of
    iterations ({\tt itmx}) and of  points for iteration ({\tt ncalls}) for
    the thermalizing phase as well as for the normal one and the accuracy
    required ({\tt acc}) are read from the input.
 
\noindent{\bf Output}
 
\noindent
The output is just the standard {\tt VEGAS} output,
from which one can read the
final result and estimated statistical error,  as well as the
result and error for every iteration. Results with big oscillations among
different iterations and corresponding big reported $\chi^2$
 simply mean that the number of evaluations per iteration was
not sufficient for the integrand, and have to be discarded.
 
\noindent{\bf Concluding remarks}
 
\noindent
As already stated, {\tt WPHACT} makes use of matrix elements which run fast.
Speed is in our opinion a relevant issue, not only because it allows to
perform complicated calculations, but also for rather short ones.
In Monte Carlos,  speed
corresponds to the possibility of generating in the same time many more events,
achieving a much better precision in integration.
\par The program , which does not make use of any library, has proved to be
  reliable over a vast range of statistical
   errors from the percent up to $10^{-5}$. Thus it can be used both to obtain
  very precise results   with high statistics runs  and to get fast answers.
 
\noindent{\bf Availability:}                        \\
The program is available from the authors or by anonymous ftp from \\
ftp.to.infn.it/pub/ballestrero.
 
 
\subsection{WTO}
\leftline{\bf Author:}
\begin{tabular}{ll}
Giampiero~Passarino & giampiero@to.infn.it
\end{tabular}
 
{\tt WTO} is a {\it quasi-analytical, deterministic} code for computing
observables related to the process $e^+e^- \to {\bar f_1}f_2{\bar f}_3f_4$. The
full matrix elements are used and in the present version the
following final states are accessible (see~\cite{wwteup} for a general
classification):
 
\begin{enumerate}
 
\item \emph{CC03}, \emph{CC11}, \emph{CC20}, \emph{NC21}, \emph{NC24},
\emph{NC32}, \emph{mix43}
 
\item \emph{NC23} (= \emph{NC21} + Higgs signal),
\emph{NC25} (= \emph{NC24} + Higgs signal)
 
\end{enumerate}
 
\noindent
Further extensions will be gradually implemented.
To fully specify {\tt WTO}'s
setup an option must be chosen for the renormalization
scheme (RS). One has the options commonly used for tuned comparisons or the
default, i.e.
 
\begin{eqnarray}
s_{_W}^2 &=& {{\pi\alpha(2\wm)}\over {{\sqrt 2}\gf\wm^2}},  \qquad
g^2 = {{4\pi\alpha(2\wm)}\over {s_{_W}^2}},
\label{wto_eq1} \\
s_{_W}^2 &=& 1 - {{\wm^2}\over {\zm^2}}, \qquad  \quad
g^2 = 4{\sqrt 2}\gf\wm^2
\label{wto_eq2}
\end{eqnarray}
 
\noindent
where $\alpha^{-1}(2\wm) = 128.07$ and $\gf$ is the Fermi coupling constant.
Final state QCD corrections are not taken into account in the present
version, except for the Higgs signal (NC21-NC25) where the pole quark
masses, $m_q(m_q^2)$, are in input. The code will compute the correct
running, up to terms ${\ord}(\alpha_s^2)$, i.e. $m_{b,c}(m_H^2)$ and
will include `effectively' a final state QCD correction.
 
The matrix elements are obtained with the helicity method described in
ref.\cite{cpb}.
The whole answer is written in terms of invariants, {\it i.e.}
\begin{eqnarray}
e^+(p_+)e^-(p_-) &\to& f(q_1){\bar f}(q_2)f'(q_3){\bar f}'(q_4),  \\
x_{ij}s &=& -\left(q_{i-2} + q_{j-2}\right)^2,  \quad
x_{1i}s = -\left(p_+ + q_{i-2}\right)^2,  \\
x_{2i}s &=& -\left(p_- + q_{i-2}\right)^2,  \quad
s_1s^2 = \epsilon\left(p_+,p_-,q_1,q_2\right), \dots
\end{eqnarray}
\noindent
and the integration variables are chosen to be $m_-^2 = x_{24}, \, m_+^2 =
x_{56}, \,M_0^2 = x_{45},  \, m_0^2 = x_{36}, \,m^2 = x_{35}, \, t_1 = x_{13},
\, t_{_W} = x_{13} + x_{14}$.
The convention for the final states in {\tt WTO} is: $e^+e^- \to 1+2+3+4$.
For CC processes $1=d, 2={\bar u}, 3=u', 4={\bar d}'$, with
$u = \nu,u,c$ and $d = l,d,s,b$. For NC processes the adopted
convention is $1=f, 2={\bar f}, 3=f'$ and $4={\bar f}'$.
Initial state QED radiation is included through the Structure Function approach
up to $O(\alpha^2)$. The code will return results according to three
(pre-selected) options, i.e $\beta^2\eta$ (default)~\cite{topv},
$\beta^3$~\cite{cpcww} and $\beta\eta^2$~\cite{excalit} where
$\beta = 2\,\frac{\alpha}{\pi}\,\left(\log\frac{s}{m_e^2} - 1\right),
\, \eta = 2\,\frac{\alpha}{\pi}\,\log\frac{s}{m_e^2}$.
QED corrections also include the Coulomb term correction~\cite{BBD}
for the \emph{CC03} part of the cross section.
When initial-state QED radiation is included, there are two additional
integrations over the fractions of the beam energies lost through radiation,
$x_{\pm}$.
This description of the phase space gives full cuts-availability through
an analytical control of the boundaries of the phase space. Upon
specification of the input flags it is therefore possible to cut on
all final state invariant masses, all (LAB) final state energies $E_i, i=1,4$,
all (LAB) scattering angles, $\theta_i, i=1,4$ all (LAB) final state angles,
$\psi_{ij}, i,j=1,4$.
 
Both the matrix elements and the phase space are given for massless fermions.
There is no interface with hadronization.
The integration is performed with the help of the NAG~\cite{nag} routine
D01GCF. This routine uses the Korobov-Conroy number theoretic approach with a
MC error estimate arising from converting the number theoretic formula for the
$n$-cube $[0,1]^n$ into a stochastic integration rule. This allows a `standard
error' to be estimated. Prior to a call to D01GCF the peak structure of the
integrand is treated with the appropriate mappings.
 
Whenever the program is called it will start the actual calculation of
one of the following observables: cross section or a pre-selected
sample of moments of distributions, for instance $<x^n_{\gamma}>$.
Since {\tt WTO} does not generate hard and
non-collinear photons, $E_{\gamma}$ is just the total radiated photon
energy. There is no adaptive strategy at work since the routine D01GCF,
being a deterministic one, will use a fixed grid. The evaluation of the
specified observable will be repeated NRAND times to give the final answer,
however there is no possibility to examine the partial results but only the
average and the resulting standard error will be printed. The error in
evaluating , say, a
cross section, satisfies $E < CK\,p^{-\alpha}\log^{\alpha\beta}p$,
where $p=$NPTS, $\alpha$ and $C$ are real numbers depending on the convergence
rate of the Fourier series, $\beta$ is a constant depending on the
dimensionality $n$ of the integral and $K$ is a constant depending on
$\alpha$ and $n$.
 
Numerical input parameters such as $\alpha(0), \gf, \zm, \wm, \dots$ are
stored in a BLOCK DATA.
There are various flags to be initialized to run {\tt WTO}.
Here follows a short
description of the most relevant ones:
 
\begin{description}
 
\item[NPTS] - INTEGER, NPTS=1,10 chooses the actual number of points for
applying the Ko\-ro\-bov-Con\-roy
number theoretic formulas. The built-in choices
correspond to to a number of actual points ranging from 2129 up to 5,931,551.
 
\item[NRAND] - INTEGER, NRAND specifies the number of random samples to be
generated in the error estimation (usually $5-6$).
 
\item[OXCM] - CHARACTER*1, the main decision branch for the process: [C(N)] for
CC, (NC).
 
\item[OTYPEM] - CHARACTER*4,Specifies the process,
i.e. \emph{CC03}, \emph{CC11}, \emph{CC20} for
CC processes and \emph{NC19, NC24, NC21, NC25, NC32} for NC processes.
 
\item[ITCM] - INTEGER, the type of observable requested ($0$ for cross section).
For \emph{CC11} ($e^+e^- \to \mu^-{\bar\nu}_{\mu}u{\bar d}$) a number of
distributions are available (for instance $<x_{\gamma}^n>$). If the n-th
moment of a distribution is requested then
 
\item[ITCNM] - INTEGER, must be set to $n$.
 
\item[OCOUL] - CHARACTER*1, controls the inclusion of the Coulomb correction
factor [Y/N].
 
\item[IOS] - INTEGER, two options [$1,2$] ($1=$default for
tuned comparisons)
for the renormalization scheme.
 
\item[IOSF] - INTEGER, three options [$1-3$] for the $\eta-\beta$ choice in the
structure functions.
 
\item[CHDM$\dots$] - REAL, Electric charges, third component of isospin for
the final states.
 
\end{description}
 
\noindent
{\tt WTO}
is a robust one call - one result code, thus in the output one gets a list
of all relevant input parameters plus the result of the requested
observable with an estimate of the numerical error. A very
rough estimate of the theoretical error (very subjective to say the least)
can be obtained by repeating runs with different IOS, IOSF options.
A rough estimate of the requested CPU time (on a VAXstation $4000\cdot 90$)
vs precision can be inferred from the following table which refers to
$\sigma(e^+e^- \to \mu^-{\bar\nu}_{\mu}u{\bar d})$ at $\sqrt{s} = 161\,$GeV
 
\begin{table}[hbtp]
\begin{center}
\begin{tabular}{|c|c|c|c|}
\hline
GENTLE & & & 0.1269543\\
$\sigma\,$(nb) & 0.1266300 $\pm$ 0.822D-03 & 0.1268430 $\pm$ 0.171D-03 &
0.1269526 $\pm$ 0.381D-05 \\
W/G($\%$)      & 0.26 & 0.09 & 1.$\times 10^{-3}$  \\
CPU    & 00:03:17.78 & 00:19:25.00 & 18:56:25.99 \\
\hline
\end{tabular}
\end{center}
\end{table}

\noindent
After initialization for the background process $e^+e^- \to {\bar\nu}_{\mu}
\nu_{\mu} {\bar b}b$ with $\zm - 25\,$GeV$ < M_{\nu\nu} < \zm + 25\,$ GeV,
$M_{{\bar b}b} > 30\,$ GeV and with the $b$ angle with respect to the
beams $> 20^o$, the typical output will look as follows:
 
\begin{quote}{\footnotesize \begin{verbatim}
This run is with:
 
NPTS         =  7
NRAND        =  6
 
E_cm (GeV) =          0.17500E+03
beta       =          0.11376E+00 sin^2     =          0.23103E+00
M_W  (GeV) =          0.80230E+02 M_Z (GeV) =          0.91189E+02
G_W  (GeV) =          0.20337E+01 G_Z (GeV) =          0.24974E+01
 
No QED Radiation
There are cuts on fs invariant masses, no cuts on fs energies,
cuts on scattering angles, no cut on fs angles
 
\emph{NC24}-diagrams : charges    -0.3333    0.0000
                isospin    -0.5000    0.5000
 
On exit IFAIL = 0 - Cross-Section
 
CPU time  41 min  28 sec, sec per call    =     0.415E-02
# of calls      =     599946
 
sigma       =       0.1489801E-02   +-          0.1930508E-05
 
Rel. error of      0.130 %
\end{verbatim}}\end{quote}
 
 
\subsection{WWF 2.2}
\leftline{\bf Author:}
\noindent\begin{tabular}{ll}
Geert Jan van Oldenborgh & gj@rulkol.LeidenUniv.nl
\end{tabular}
 
\noindent{\bf Description}
 
\noindent
This Monte Carlo is the beginning of a full one-loop Monte Carlo
\cite{wwf1}-\cite{wwf2}.
At the moment it includes
a tree level part ({\tt WWFT}, which participated in the tuned
comparisons),
hard and soft bremsstrahlung ({\tt WWFTSH}, exact matrix element,
resummed in the forward and backward region),
and the factorizable virtual graphs ({\tt WWFTSHV}, on request only).
We are working on the missing parts,
the non-factorizable loop graphs.
$t$-channel graphs for electrons in the final state,
and a shower algorithm for the forward/backward photons.
 
\noindent{\bf Features of the program}
 
\noindent
There are two forms of the program: an event generator ({\tt wwfax})
and `integrator' ({\tt wwfmc}), the latter has a parallel option ({\tt
wwfpvmmc}, {\tt wwfpvmslave}).  Interfaces to {\tt BASES/SPRING} are also
provided.
 
The program can generate all final states which are reachable through
two $W$ bosons.  The user can specify whether the final states should
be leptonic, semileptonic and/or hadronic, and which leptons should
be included in leptonic decays, for instance `all semi-leptonic and
leptonic channels with electrons and muons'.
All cuts can be implemented after the event is generated.  To
optimize event generation one can specify the minimum photon energy,
the minimum and maximum angle of photons to the beam, minimum angle
to charged particles, and the maximum virtuality of the $W$'s.
 
Two methods have been implemented to compute ISR:
structure functions (Leiden 2-loop and YFS 3-loop leading
logarithmic, with the possibility of giving the photon bunch a
one-photon spectrum $p_T$), and
the explicit 1-photon matrix element (for \emph{CC03} and \emph{CC11}
processes),
minus the leading
log  part of this matrix element, plus the resummed leading log structure
functions mentioned above.  In the latter case an estimate of the missing
virtual corrections is included, which makes it unsuitable for total cross
section predictions.
For FSR we use the exact one-photon matrix element; there is an
option to reduce the leading logarithmic part of this by an arbitrary
factor to compensate for the excess near jets (which are described by
on-shell quarks).
The default event generation routine calls {\tt JETSET} to do all the
hadronization and $\tau$ decays.  No polarization information is
passed as
yet, although all particles come from $W$ bosons and the helicities are
therefore fixed.
There is a {\tt JETSET} interface, which will soon
be adapted to the proposed
standard.  There is no possibility to get information about subsets
of diagrams yet, but this will be included in this interface.
 
We have the possibility to shift the Coulomb term from the virtual
corrections to the the tree level terms (and therefore include it in
the hard and soft radiation as well).  For this we take the one-loop
expression given in ref.
\cite{BBD}.
Anomalous couplings are implemented only at the tree level, we follow
the conventions of Jegerlehner \cite{Teupitzproceedings}.
In the hard radiation matrix element there is the option to include
the full effect of finite fermion masses; the default is to include
the leading effects only.  The tree level ME can also include some
mass effects.  The phase space is always taken massive.
 
\noindent{\bf Program layout}
 
\noindent
The `integrator' program {\tt wwfmc} is a stand alone program, which reads
its data from a file {\tt wwf.dat}, which defines the input parameters,
and {\tt vegas.dat}, which gives the parameters for the integration by
{\tt VEGAS} (adaptive weighted integration) or {\tt NVEGAS} (integrates many
quantities, like the tuned comparison data).
 
\begin{table}
\begin{center}
\small
    \begin{tabular}{|ccl|}
    \hline
    tuned & best & description\\
    80.23 & 80.26 & $W$ mass in GeV, LEP1 definition (running width) \\
    $-1$  & $-1$  & $W$ width, if $<0$ it is computed \\
    $-1$  & $-1$  & $Z$ mass, if $<0$ it is taken to be 91.188 GeV \\
    $-1$  & $-1$  & $Z$ width, if $<0$ it is taken to be 2.4974 GeV \\
    100   & 300   & Higgs mass (only used in virtual corrections) \\
    176   & 165   & top quark mass (only used in virtual corrections) \\
    2     & 0/2   & 0: constant width (use for hard \& virtual corrections)\\
          &       & 2: $s$-dependent width (preferred for tree level only) \\
    4     & 2     & renormalization scheme: 1: $\alpha$, 2: $G_\mu$
                    with $\alpha$ for soft radiation, 3: $G_\mu$ \\
          &       & 4: the tuned comparison scheme\\
    2     & 2     & 1: narrow-width approximation,
                    2: full off-shell calculation\\&& (not defined with
                       virtual), 3: pole scheme calculation\\
    1     & 1     & 1: fast massless matrix element,
                    2: slower massive matrix element\\
    0     & 0     & 0: include all diagrams\\
    0     & 0     & 0: include corrections both to production and decay\\
    0/1   & 0/1   & 0: only resonant tree level diagrams (\emph{CC03})\\
          &       & 1: same plus universal non-resonant diagrams
(\emph{CC11})\\
    0/1   & 0/1   & same for radiative graphs \\
    0     & .123  & $\alpha_s$\\
    2     & 0--7  & decay channel, sum of 1: leptonic, 2: semileptonic, 4:
hadronic\\
    0     & 0--7  & $W^+$ decay channels, sum of 1: $e^+\nu_e$, 2:
$\mu^+\nu_\mu$,
 
                    4: $\tau^+\nu_\tau$, 8: $u\bar{d}$\\
    2     & 0--7  & $W^-$ decay channels, sum of 1: $e^-\bar{\nu}_e$,
 
                    2: $\mu^-\bar{\nu}_\mu$, 4: $\tau^-\bar{\nu}_\tau$,
 
                    8: $\bar{u}d$\\
    0     & 0.01  & $E_{\gamma}^{\mathrm{min}}$ needed for hard/soft
cut-off\\
    0     & 0     & $\theta_{\gamma,f}^{\mathrm{min}}$ used to optimize event
 
                       generation\\
    0     & 0     & $\theta_{\gamma,e}^{\mathrm{min}}$ used to optimize event
 
                       generation\\
    180   & 180   & $\theta_{\gamma,e}^{\mathrm{max}}$ used to optimize event
 
                       generation\\
    0     & 0     & if $c>0$ generate $|\sqrt{s_\pm}-M_W| <c$ GeV\\
    0/1   & 0/1   & 0: no cuts, 1: canonical cuts, 2: require one observable
photon\\
    3     & 3     & 0: no extra initial-state radiation,\\
          &       & 1: use Leiden 2-loop structure functions,\\
          &       & 2: use YFS 3-loop structure functions.\\
    180   & 180/10& cone around beam pipe where radiation is exponentiated\\
          &       & (use 5--10 degrees when including explicit hard
radiation)\\
    1     & 1     & 1: use crude $p_T$ algorithms for ISR photons\\
    0     & 0     & 1: exclude leading logarithmic initial-state radiation\\
    0     & 0/20  & cone around final state particles where FSR is reduced\\
    0     & 0/0.4 & fraction of leading log final-state radiation off quarks
to leave out\\
    0     & 0/1   & 1: include explicit hard photon radiation matrix
element\\
    0     & 0/1   & 1: include explicit soft photon matrix element\\
    0     & 0     & 1: include loop graphs (not yet complete)\\
    1     & 1     & 1: include tree level matrix element\\
    0     & 1     & 1: include the Coulomb term in tree\\
    \hline
    \end{tabular}
\end{center}
\caption{Input file format of {\tt WWF 2.2}}
\label{tab:wwfinput}
\end{table}
 
\begin{flushleft}
The event generator is a set of three routines:\\
-- {\tt axinit}: preparation, this also establishes the maximum of the
function,\\
-- {\tt axeven}: generates one event\\
-- {\tt axexit}: finalization, prints statistics, gives cross section
   and weight per event.\\
The use of these routines is demonstrated in the program {\tt wwfax}.
The event generation does not use any adaptive strategies.
The event is presented in a subroutine {\tt wwfeve}, the default version
of which calls {\tt JETSET} and lists the event on standard output.
\end{flushleft}
 
\noindent{\bf Input}                                \\
The input parameters are expected to be in a file {\tt wwf.dat} with the
information described in table \ref{tab:wwfinput}

\noindent{\bf Output}
\begin{flushleft}
The program {\tt wwfax} (or the equivalent routines) will give call the
routine {\tt wwfeve} for each event generated; the default is to list the
event on standard output.  Some informative messages will also appear on
standard output:\\
-- while initializing: the current maximum, a measure of the progress
towards this maximum and the largest negative event found so far,\\
-- at the end of initialization: the maximum used and a summary of the
negative events,\\
-- while generating: error messages (mainly inaccuracies and negative
weights) and the numbers of events generated at powers of two,\\
-- at exit: the cross section, weight per event, efficiency, CPU time
used and a summary of the impact of the negative weight events.
The program {\tt wwfmc} integrates the cross section and the tuned
comparison quantities, and will dump these in this format.  One can make
plots by editing wwfill and the file {\tt h.dat}.
\end{flushleft}
 
\noindent{\bf Availability}                         \\
The programs can be obtained from\\
\texttt{ftp://rulgm4.LeidenUniv.nl/pub/gj},\\
\texttt{http://rulgm4.LeidenUniv.nl}\\
either as a compressed archive {\tt wwf.tar.gz} or separate files.  The
package includes a makefile and is known to compile without problems on
HP, DEC, Linux, NeXT and Sun workstations.
 
 
\subsection{WWGENPV/HIGGSPV}
\leftline{\bf Authors:}
\begin{tabular}{ll}
Guido Montagna   & montagna@pv.infn.it \\
\noindent
Oreste Nicrosini & nicrosini@vxcern.cern.ch, nicrosini@pv.infn.it \\
\noindent
Fulvio Piccinini & piccinini@pv.infn.it
\end{tabular}
 
\noindent{\bf Description:} \\
{\tt WWGENPV} and {\tt HIGGSPV}
are four-fermion Monte Carlo codes, originally
conceived for $W$-boson and Higgs-boson physics, respectively.
The present version of {\tt WWGENPV} is an upgrade of the published version.
A detailed description of the
formalism adopted and the physical ideas behind it can be found in the
original literature, namely ref.~\cite{cpcww} and references therein. A
detailed description of {\tt HIGGSPV} can be found in the report of the
``Event Generators for Discovery Physics'' Working Group, these
proceedings.
 
\noindent
The programs are based on the exact tree-level
calculation of several four-fermion final states. Any cut on the final
state configuration can be implemented. Initial- and final-state
QED corrections are taken into account at the leading logarithmic level
by proper structure functions, including $p_T / p_L$ effects. An
hadronization interface is at present available for \emph{CC03} processes, and
is under development~\cite{cpcwwup}. All the relevant presently known
non-QED corrections are also taken into account.
 
\noindent{\bf Features of the programs:}  \\
The codes consist of three Monte Carlo branches, in which the
importance-sampling
technique is employed to take care of the peaking behavior of
the integrand:
 
\begin{itemize}
 
\item {Unweighted event generation. The codes provide a
sample of unweighted events, defined as the components
of the four final-state fermions momenta,
plus the components of the initial- and final-state photons, plus
$\sqrt {s}$, stored into proper $n$-tuples.
The programs must be linked to CERNLIB for
graphical interfaces. }
 
\item {Weighted event integration. It is intended for
computation only. In particular, the codes return the values of several
observables together with a Monte Carlo estimate of the errors. The
programs must be linked to CERNLIB for the evaluation of few special
functions. }
 
\item {Adaptive integration. It is intended for
computation
only, but offering high precision performances. On top of importance
sampling, an adaptive Monte Carlo integration algorithm is used.
The program must
be linked to NAG library for the Monte Carlo adaptive routines. Full
consistency between non-adaptive and adaptive integrations has been
explicitly proven. Neither final-state radiation nor $p_T$ splitting
are taken into account in this branch. }
 
\end{itemize}
 
\noindent
The non-adaptive branches
rely upon the random number generator RANLUX.
 
\noindent
As far as the physical features are concerned, the most important items
are:
 
\begin{itemize}
 
\item Several Charged Current ({\tt WWGENPV})
and Neutral Current ({\tt HIGGSPV})
processes are available, namely \emph{CC11}, \emph{CC20},
\emph{NC21} (\emph{NC23} = \emph{NC21} + Higgs signals),
\emph{NC24} (\emph{NC25} = \emph{NC24} + Higgs signals), \emph{NC32},
\emph{NC48} (\emph{NC50} = \emph{NC48} + Higgs signals)
and all their subsets.
The extension to other classes is under development.
 
\item Any kind of cuts can be imposed.
 
\item Initial- and final-state photon radiation is implemented at the
leading logarithmic level in the structure function formalism.
The structure function used is explicitly written in~\cite{cpcww}.
Moreover, $p_T /p_L$ effects are taken into account.
 
\item The Coulomb correction is taken into account (see
\cite{cpcww} and references therein), together with flavor mixing and
the presently known  QCD corrections.
 
\item An interface to hadronization packages is available for \emph{CC03}
processes and the extension to other classes is under
development~\cite{cpcwwup}.
 
\item There is the possibility of getting
information on the contribution of subsets of the diagrams by setting
proper flags.
 
\end{itemize}
 
\noindent
At present, neither  final state decays nor anomalous couplings are
implemented. Moreover, finite fermion mass effects are partially taken
into  account only at the phase space boundary.
 
\noindent{\bf Program layout} \\
After the initialization of the Standard Model parameters and of
the electromagnetic quantities, the independent variables are
generated, according to proper importance samplings,
within the allowed range for an extrapolated set-up. The analytical
control of the phase-space boundaries allows to reach an efficiency
which, for an extrapolated set-up, is unitary, and remains very high
for a wide range of (reasonable) cuts.  By means of the
solution of the exact kinematics, the
four-momenta of the outgoing fermions
are reconstructed in the laboratory frame, together with the
four-momenta of all the generated photons. If the event satisfies the
cuts imposed by the user in SUBROUTINE CUTUSER,
the matrix element is called, otherwise it is set to zero.
 
\noindent
In the
generation branch, an additional random number is generated in order to
implement the hit-or-miss algorithm and if the event is accepted it is
recorded into an $n$-tuple.
 
\noindent
In the non-adaptive integration
branch, the integration of several (see below)
observables is performed
in a single run, by cumulating in parallel all the contributions to the
integrands.
 
\noindent
In the adaptive integration branch (ref.: NAG routine D01GBF), on top
of importance sampling the
integration routine automatically subdivides the integration region
into subregions  and iterates the procedure where the integrand is
found more variant. The program
stops when a required relative precision is satisfied.
 
\noindent{\bf INPUT parameters and flags ({\tt WWGENPV}):} \\
A sample of the input flags that can be used is  the following:
 
\noindent
{\tt OGEN = I} choice between integration [I] and generation [G]
branch
 
\noindent
{\tt RS = } c.m. energy (GeV)
 
\noindent
{\tt OFAST = N} choice between adaptive [Y] or non adaptive [N] branch
 
\noindent
{\tt NHITWMAX = } number of weighted events
 
\noindent
{\tt IQED = 1} choice for Born [0] or QED corrected [1] predictions
 
\noindent
{\tt ODIS = T} choice for a total cross section [T] or an invariant
mass distribution [W]
 
\noindent
{\tt OWIDTH = Y} $W$-boson width computed within the SM according to
LEP2 standard input [Y] or input the preferred value [N]
 
\noindent
{\tt NSCH = 2} Renormalization Scheme choice (three possible choices)
 
\noindent
{\tt ALPHM1 = 128.07D0} $1/\alpha$ value (LEP2 standard input)
 
\noindent
{\tt OCOUL = N} option for Coulombic correction [Y] or not [N]
 
\noindent
{\tt SRES = Y} option for \emph{CC11} [Y] or \emph{CC03} [N] diagrams
 
\noindent
A detailed account of the other relevant possibilities offered by the
code (namely, command files for generation and adaptive integration
branches) will be given elsewhere~\cite{cpcwwup}.
 
\noindent{\bf Description of the OUTPUT:}  \\
For all three branches the output contains
the values of the Standard Model
parameters and of the couplings appearing in the Feynman rules.
 
\noindent
In the generation branch, besides the output file containing the value
of the cross sections for unweighted events, together with a Monte
Carlo estimate of the error, also an $n$-tuple
containing the generated events is written.
 
\noindent
In the adaptive branch, the values of the cross section with
its numerical error plus (when ISR is included)
the energy and invariant
mass losses with their errors are then printed.
 
\noindent
In the non-adaptive branch,
together with the cross sections, the estimates of the moments
used in the tuned comparisons and of the histograms are also printed,
together with the Monte Carlo errors.
 
\noindent{\bf Availability: }                       \\
\noindent
The codes are available upon request to one of the authors.
%
\subsection{Summary}
 
We will now briefly summarize the features of the programs presented
in the previous subsections.  Table~\ref{tab:cast} gives an overview
over the features of the programs participating in the comparisons.
It is just intended as a brief digest and the short writeups in the
previous section should be consulted for reference.  Here is a
description of the columns of table~\ref{tab:cast}:
 
\begin{list}{}%
 {\setlength{\leftmargin}{2em}
  \setlength{\rightmargin}{2em}
  \setlength{\itemindent}{-1em}
  \setlength{\listparindent}{0pt}
  \setlength{\topsep}{3pt plus 2pt minus 2pt}       
  \setlength{\itemsep}{2pt plus 1pt}                
  \setlength{\partopsep}{2pt plus 1pt minus 1pt}    
  \renewcommand{\makelabel}{\hfil}}
\label{pg:table-desc}
  \item Type:\\
    one of the four types of programs:
    \emph{EG}: (unweighted) event generator,
    \emph{MC}: (weighted) Monte Carlo integration program,
    \emph{Int.}: deterministic integration program, and
    \emph{SA}: semi-analytical integration program.
  \item Diagrams:\\
    the subset(s) of Feynman diagrams implemented in the hard matrix
    element:
    \emph{CC03}: the three basic~$e^+e^-\to W^+W^-$ charged current
      diagrams from figure~\ref{fig:CC03},
    \emph{CC11}: the eleven charged current diagrams from
      figure~\ref{fig:CC11}, see table \ref{tab1};
    \emph{NC24} and \emph{NC21} subsets of neutral
      current diagrams, see table \ref{tab2};
    \emph{NNC}=\emph{NC32}/\emph{NC21}/\emph{NC48}/%
    \emph{NC4}$\times$\emph{16};  \\
    \emph{NCC}=\emph{CC11}/\emph{CC20}/\emph{NC32}/\emph{NC21}/%
       \emph{mix43}/\emph{NC48}/\emph{NC4}$\times$\emph{16};
    \emph{all}: all diagrams.
    We emphasize,
    that we have listed only those processes in this column for which
    participating codes have contributed at least one number,
    see also the tables in \cite{SMP-report}.  This
    entry may therefore differ from that presented in the program
    descriptions.
  \item ISR:\\
    the type of initial-state radiation implementation:
    \emph{SF}: structure functions;
    \emph{FF}: flux functions;
    \emph{REMT}: REMT routines, see subsection \ref{sec:LEPWW}.
    \emph{PS}: parton showers;
    \emph{YFS}: Yennie-Frautschi-Suura exponentiation; and
    \emph{ME}: matrix element (exact lowest order bremsstrahlung matrix
      element and infrared divergent virtual contributions);
    \emph{BME}: the one photon bremsstrahlung matrix
      element is available; no virtual contributions.
  \item FSR:\\
    the type of final-state radiation implemented, see also
    section~\ref{sec:FSR}; \emph{PH}: FSR is implemented by making use of
    {\tt PHOTOS} package; the other symbols are the same as in the ISR
    column.
  \item NQCD:\\
    naive, inclusive QCD correction to $W^\pm$ decays.  A~`$+$' does not
    imply that hard QCD radiation is implemented in the program (see
    page~\pageref{pg:QCD} for more details).
  \item Coul.:\\
    Coulomb correction (see page \pageref{pg:Coulomb} for more details).
  \item AC:\\
    availability of anomalous couplings in the three gauge boson
    vertices.  Since we have not compared predictions with anomalous
    couplings in this study, the entries in this column are identical
    to what is advertized in the program descriptions.
  \item $m_f$:\\
    treatment of fermion masses:
    \emph{$+$}: all fermion masses taken into account,
    \emph{$\pm$}: massless matrix elements with massive kinematics
   (mostly K\"{a}ll\'{e}n $\lambda$-functions),
      and finally
    \emph{$-$}: all fermions massless.
  It must be remarked here that `all' does not necessarily
  mean that nonzero masses have been included in all
  processes presented in the comparisons.
  \item Hadr.:\\
    availability of an interface to hadronization libraries.  With the
    exception of \PYTHIA/, no program includes hadronization code.  All
    rely on \texttt{HERWIG} or \texttt{JETSET} to perform this task.
The interface with hadronization packages and its interplay with
final-state QCD radiation deserves a longer comment. For some codes
a minus in this column is a direct consequence of the adopted strategy, e.g.
semianalytical codes were never meant for this interface.
\end{list}
\begin{table}[t]
  \def\arraystretch{1.2}
  \begin{center}
    \begin{tabular}{|c||c|c|c|c|c|c|c|c|c|}\hline
      Program    &Type&Diagrams&ISR    &FSR&NQCD&Coul.&AC &$m_f$&Hadr.
\\\hline
      \hline
      \ALPHA/    &MC  &all       &BME    &$-$&$-$&$-$  &$-$&$+$  &$-$
\\\hline
      \COMPHEP/  &EG  &all       &SF     &$-$&$-$&$-$  &$-$&$+$  &$-$
\\\hline
      \ERATO/    &MC  &CC11/CC20 &SF     &$-$&$+$&$-$  &$+$&$-$  &$+$
\\\hline
      \EXCALIBUR/&MC  &all       &SF     &$-$&$+$&$+$  &$+$&$-$  &$-$
\\\hline
      \GENTLE/   &SA  &CC11/NC32 &SF/FF  &$-$&$+$&$+$  &$-$&$\pm$&$-$
\\\hline
      \GRC4F/    &EG  &all       &SF/PS  &PS &$+$&$+$  &$+$&$+$  &$+$
\\\hline
      \HIGGSPV/  &EG  &NNC       &SF$(p_T)$     &$-$&$+$&$ $  &$-$&$\pm$&$-$
\\\hline
      \KORALW/   &EG  &CC11      &YFS    &PH &$+$&$+$  &$+$&$\pm$&$+$
\\\hline
      \LEPWW/    &EG  &CC03      &REMT
                                       &PH &$+$&$-$  &$+$&$-$  &$+$
\\\hline
      \LPWW02/   &EG  &CC03      &SF     &PH &$+$&$+$  &$-$&$\pm$&$+$
\\\hline
      \PYTHIA/   &EG  &CC03      &SF$+$PS&PS &$+$&$+$  &$-$&$\pm$&$+$
\\\hline
      \WOPPER/   &EG  &CC03      &PS     &$-$&$+$&$+$  &$-$&$\pm$&$+$
\\\hline
      \WPHACT/   &MC  &all       &SF     &$-$&$+$&$+$  &$+$&$+$  &$-$
\\\hline
      \WTO/      &Int.&NCC       &SF     &$-$&$+$&$+$  &$-$&$-$  &$-$
\\\hline
      \WWF/      &EG  &CC11      &SF$+$ME&ME &$+$&$+$  &$+$&$+$  &$+$
\\\hline
      \WWGENPV/  &EG  &CC11/CC20 &SF$(p_T)$ &SF$(p_T)$&$+$&$+$  &$-$&$\pm$&$+$
\\\hline
    \end{tabular}
  \end{center}
  \caption{\label{tab:cast}
    Overview of the participating programs.}
\end{table}
 
\section{Comparisons of CC Processes}
 
We now come to a detailed comparison of the Monte Carlo Event
Generators and semianalytical programs available for the study of
four-fermion processes at LEP2.
The next subsection contains our most comprehensive study of
\emph{CC10} processes.
Much shorter studies of \emph{CC11} and \emph{NC} processes are
presented in the following subsection and the next section.
Finally, the cross sections for \emph{all} four-fermion processes are
presented.
 
\subsection{\emph{CC10} processes}
 
 In a set of
 \emph{tuned comparisons} of \emph{CC} processes we have tested
the implementation of the \emph{CC10} family for a prescribed set of
approximations.
Because the \emph{CC03} set (cf.~fig.~\ref{fig:CC03}) is available in
\emph{all} programs, one of the \emph{tuned comparison} has been
restricted to this subset of all contributing diagrams.
 
It was then extended  to the process
$e^+e^- \to \mu^-\bar\nu_\mu u\bar d$,
where
from the \emph{CC11} set of diagrams only 10 contribute,
 because the photon does not
couple to the neutrino (cf.~fig.~\ref{fig:CC11}).
 
 
 In a second set of \emph{unleashed comparisons} all the contributors have
presented their preferred scenario for the  process
($e^+e^- \to \mu^-\bar\nu_\mu u\bar d$) or, in short, they have produced
the \emph{best prediction they can
give} at present. The latter comparison can
 show which
part of the spread in predictions is due to the different approximations used.
 
\subsubsection{Observables}
In comparing of predictions for exclusive observables, we have
concentrated on the prototypical ``semileptonic'' \emph{CC10} process
\begin{equation}
  e^+e^- \to \mu^-\bar\nu_\mu u\bar d\,,
\end{equation}
\noindent
which belongs to the \emph{CC11} family. This
 choice is also partially motivated
by the fact that the same process can be computed by restricting the
calculation to the \emph{CC03} class, thus allowing more codes to
participate. Moreover, it is known that at LEP~2 energies the ratio of
\emph{CC03/CC10} cross sections is very near to one, although the difference
is seen in some of the distributions.
 It should be mentioned that
 for the other semi-leptonic process
$e^+e^- \to e^-\bar\nu_e u\bar d$  even the total cross section can not be well
approximated by the \emph{CC03} limit.
 
The following simple observables have received particular attention,
because they are of prime importance for the measurement of the
properties of the charged intermediate $W^\pm$~bosons at LEP2.
\begin{itemize}
  \item The total cross section $\sigma$, with and without canonical
    cuts (see section~\ref{sec:canonical-cuts} for a precise definition).
  \item The moments of the production angle~$\theta_W$ of the~$W^+$ with
    respect to the~$e^+$-beam:
    \begin{equation}
      \langle \cos\theta_W \rangle_{1,2}
         = \frac{1}{\sigma} \int T_{1,2}(\cos\theta_W) d\sigma
    \end{equation}
    where the~$T_n(\cos\theta) = \cos(n\theta)$ are the Chebyshev
    polynomials $T_1 (x) = x$ and $T_2 (x) = 2x^2 - 1$.
    The distribution of the production angle will be used in some
    studies of the non-abelian $W^\pm$~couplings.  A precise
    description of the standard model prediction for this observable
    is therefore mandatory for this fundamental test of the
    non-abelian gauge structure of the standard model.
  \item From the invariant masses~$s_\pm$ of the hadronic~($W^+$) and
    leptonic~($W^-$) decay products we have constructed the following
    moments:
    \begin{equation}
      \langle x_m \rangle_{1,2}
          = \frac{1}{\sigma} \int
         \left(\frac{\sqrt{s_+}+\sqrt{s_-}-2M_W}{2E_B}\right)^{1,2} d\sigma
    \end{equation}
    These quantities will of course be of prime importance
    for the $W^\pm$-mass measurement.
  \item The moments of the sum~$E_\gamma$ of the energies of all
     radiated photons
    \begin{equation}
      \langle x_\gamma \rangle_1
          = \frac{1}{\sigma} \int
                \left(\frac{E_\gamma}{E_B}\right)^1 d\sigma
    \end{equation}
    For constraint fits of the $W^\pm$-mass, a precise knowledge of
    the energy lost by initial-state radiation is mandatory.  This
    quantity has to be described by all programs with high accuracy.
  \item{} Also, moments of the lost and visible photon
    energies~$E_\gamma^{\text{lost/vis.}}$. The latter are accessible only
    in programs which generate non-vanishing $p_{T}$ for ISR photons.
\end{itemize}
 
We have also looked at the following leptonic variables.
\begin{itemize}
  \item The moments of the production angle~$\theta_\mu$ of the~$\mu^-$ with
    respect to the~$e^-$-beam:
    \begin{equation}
      \langle \cos\theta_\mu \rangle_{1,2}
         = \frac{1}{\sigma} \int T_{1,2}(\cos\theta_\mu) d\sigma
    \end{equation}
 \item The moments of the decay angle~$\theta_\mu^*$ of the~$\mu^-$ with
   respect to the direction of the decaying $W^-$, measured in the
   latter's rest frame:
   \begin{equation}
     \langle \cos\theta_\mu^* \rangle_{1,2}
        = \frac{1}{\sigma} \int T_{1,2}(\cos\theta_\mu^*) d\sigma
   \end{equation}
   This is another quantity that can gainfully be used in the
   determination of the non-abelian $W^\pm$-couplings.
  \item The moments of the energy~$E_\mu$ of the~$\mu^-$:
    \begin{equation}
      \langle x_\mu \rangle_{1,2}
         = \frac{1}{\sigma} \int \left(\frac{E_\mu}{E_B}\right)^{1,2} d\sigma
    \end{equation}
\end{itemize}
 
\noindent However, the numerical
results will be given only for the first moments
of leptonic variables.
 
During early stages of the comparison effort, we have additionally
considered the third and fourth order moments of these observables.
It turned out, however, that these moments typically receive
very large statistical errors.
  They have therefore been dropped.  Together with the
moments, we have produced histograms for the observables.  Presenting
these histograms for all programs is next to impossible, however.  It
has turned out that the moments that have been just described are
much more powerful tools for the sake of comparison.  The histograms have
therefore been dropped, together with the higher order moments.
Towards the end
of the comparison effort, some codes have also performed a
study of various distributions, e.g. $d\sigma/dE_{\gamma},
d\sigma/ds_+(s_-)$ etc, where the relevant range of the variables has
been divided in a large number of bins (typically $\approx 50-100$).
Also for distributions we have registered a very good agreement,
showing among other things that moments can be reconstructed to
high precision from the distributions.
 
\subsubsection{Tuned Comparisons}
Our first task was to verify that all programs implement their
advertised features correctly within the given statistical and
numerical uncertainty, at least for \emph{CC03,CC10}.
  Obviously, this is only straightforward, if
all programs implement the same features.  This is not the case, of
course.  Therefore we have performed a set of so-called \emph{tuned
comparisons} in which only a common subset of features has been
enabled and identical inputs have been used, as far as possible.
 Actually a semi-tuned
comparison has also been attempted by several codes for \emph{all processes}
and the results will be described in subsection \ref{ayc_comp}.
 
Ideally, all programs would have options to emulate \emph{all} other
programs.  Then all programs should give the same results (up to Monte
Carlo errors), if running in the same mode and using the
same input.  This approach has been adopted in a
study~\cite{CERN-95-03} of electroweak radiative corrections at the
$Z$-resonance.
 
In the case at hand, this approach presents a more severe problem because
electromagnetic radiative corrections are implemented in a variety of
styles: some programs are using structure functions or flux functions,
while other programs employ
parton shower algorithms, see~\cite{sibling-report} for details.
  There are even hybrids of structure
functions and matrix elements available.  Since these algorithms are
central to the respective programs, it is not possible to exchange
them without destroying the identity of the programs. 
 In any case one should be aware that there are different
implementations of the QED corrections and that this issue is deeply
related to a quest for a fully gauge-invariant description of QED radiation
in 4f-processes; this goal has \emph{not} been achieved so far.
 
\subsubsection{Input parameters}
 
The choice of input parameters is related to the choice
of the electroweak renormalization scheme (RS).
Actually, we have at our disposal the usual set of precisely measured
parameters
 
\begin{equation}
\alpha(0),G_{_F}, M_{_Z},
\end{equation}
 
\noindent
and we want to include $M_{_W}$,~\cite{sibling-report}.
Given the fact that the $\cal{O}$($\alpha$)
electroweak corrections are not available for the \emph{off-shell}
case,we end up with an additional freedom in fixing the weak-mixing
angle and the $SU(2)_L$ coupling constant. There are at least two
\emph{natural} choices, one of which had been adopted for the tuned
comparisons, although it does not respect the proper Ward identities
(more a question of principle than  of numerical relevance).
In this scheme,
the effective weak mixing angle is determined
as
\begin{equation}
  \sin^2\theta_W = \frac{\pi\alpha(2M_W)}{\sqrt{2} G_F M_W^2}\,.
\end{equation}
 
\begin{table}[t]
  \def\arraystretch{1.2}
  \begin{center}
    \begin{tabular}{|c|c|}\hline
      Quantity       & Value                           \\\hline\hline
      $M_Z$          & $91.1888\GeV$                   \\\hline
      $\Gamma_Z$     & $2.4974\GeV$                    \\\hline
      $M_W$          & $80.23\GeV$                     \\\hline
      $\Gamma_W$     & $3G_F M_W^3/(\sqrt{8}\pi)$      \\\hline
      $\alpha(0)$    & $1/137.0359895$                 \\\hline
      $\alpha(2M_W)$ & $1/128.07$                      \\\hline
      $G_F$          & $1.16639\cdot10^{-5} \GeV^{-2}$ \\\hline
      $\alpha_{_{QCD}}$ & $0$                             \\\hline
      $V_{CKM}$      & $\mathbf{1}$                    \\\hline
    \end{tabular}
  \end{center}
  \caption{\label{tab:input}%
    Input parameters used in the \emph{tuned comparisons}}
\end{table}
 
In
order to achieve agreement in a tuned comparison, all programs have
to agree on the effective coupling constants entering
the hard matrix element; this has been controlled by printing out
these constants, for which all the codes have registered an agreement
up to computer precision: $g_V = -0.0141$, $g_A = -0.18579$, $g =
0.23041$, $g_{ZWW} = .057148$, $g_{\gamma WW} = 0.31324$.
 
The photonic corrections employed in the tuned comparisons
are only those corresponding to a leading-logarithmic
approximation of initial-state radiation, final-state
radiation being implemented in only a few programs so far
(for more details we refer to the section on FSR).
The non-logarithmic QED
radiative corrections have been fixed by demanding
that structure functions and parton showers should
use~$\beta = \ln(s/m^2) - 1$ instead of~$\eta = \ln(s/m^2)$.  Other
universal corrections should be left out, see
page~\pageref{pg:flux-function} for a brief discussion of flux functions.
Such pragmatic renormalization schemes are not easily reconciled with the
schemes used in~$\mathcal{O}(\alpha)$ calculations.  A complete
calculation of this kind is, however, not available and it is important
to resum the dominant contributions (cf.~\cite{sibling-report}),
therefore this pragmatic approach has been taken.
 
\subsubsection{Presentation}
 
\dofour{tuned_xsectn_n}{%
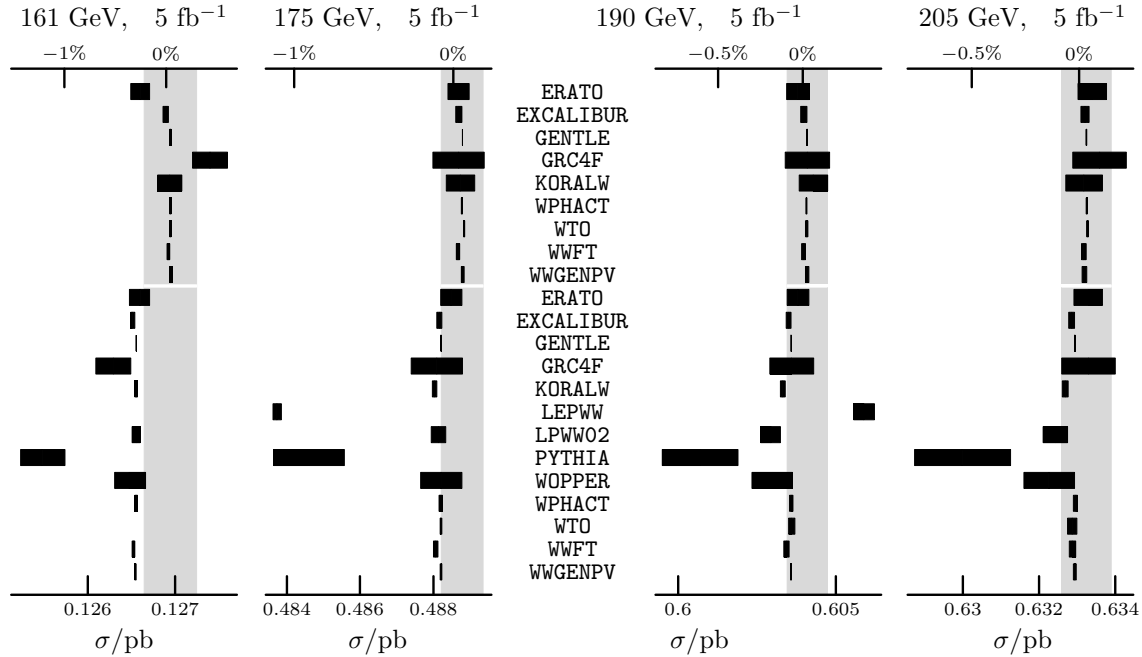
  Tuned predictions for the total cross section
  for~$e^+e^- \to \mu^-\bar\nu_\mu u\bar d$ without cuts.}
 
The comparisons are presented graphically in the style familiar from
the comparisons of experimental LEP1 results.  The predictions are
aligned vertically with horizontal error bars.  The scale at the
bottom of each plot gives the absolute value of the observables.
 
We provide also two tools to simplify the interpretation of the
results: at the top of each plot, a scale with the relative deviation
from some (insignificant) central value is drawn.  This can be used to
gauge the numerical accuracy of the results, which is of particular
importance for the tuned comparisons.  It should be noted, however,
that such a scale can be misleading for quantities that vanish in a
first approximation.  This `fine tuning' occurs
for~$\left\langle x_m\right\rangle$: further comments are given
below.
 
In addition there is a gray band drawn around the central value,
corresponding to a rough estimate of the experimental errors for a
suitable integrated luminosity.  This band is of particular importance
for the \emph{unleashed comparisons}, since it can be used by
experimentalists to gauge the theorists' predictive power in relation
to the experimental accuracy available at LEP2.
 
The results for both sets of Feynman diagrams are combined into one
plot for the \emph{tuned comparisons}.  The upper half corresponds to
the~\emph{CC10} set, while the~\emph{CC03} values are shown in the
lower half, separated by a thin white line.  This style of
presentation clearly shows the effect of the
\emph{incompleteness error} caused by leaving out a class of
diagrams.
For the interpretation of the \emph{incompleteness error} shown in the
plots, two competing effects must be taken into account: the~$e^+e^- \to
\mu^-\bar\nu_\mu u\bar d$ final state under consideration is known to be
less sensitive to ``background'' diagrams than final states with
electrons.  On the other hand, we have \emph{not} applied any
invariant mass cuts, which would reduce the contribution of
``background'' diagrams in an experimental analysis.
 
\subsubsection{Experimental Errors}
 
The statistical errors at an integrated luminosity of~$500\mathop{\rm
pb}^{-1}$ have been estimated by rescaling the errors from a high
statistics ($\mathcal{O}(10^7)$ events) simulation using
\WOPPER/\footnote{%
  A change of even a few percent in this error estimate would have no
  impact on our conclusions.  The choice of event generator is
  therefore completely irrelevant for our purposes and has been
  accidental.}.
For the error on the total cross section, we use the naive statistical
error
 
\begin{equation}
  \frac{\Delta\sigma}{\sigma} \approx \frac{1}{\sqrt N}
\end{equation}
 
\noindent
from the event count~$N = \sigma\cdot 500\mathop{\rm pb}^{-1}$ for
\emph{all} final states at~$500\mathop{\rm pb}^{-1}$.  This will
\emph{under}estimate the error on the cross section for
the~$\mu^-\bar\nu_\mu u\bar d$ final state by a factor of~$\approx 5$.
At the same time it is a more realistic number for a cross section
measurement in which events from a substantial fraction of all final
states will be counted.  The error on the moments is derived by
rescaling the statistical errors of the high statistics \WOPPER/ run
by
 
\begin{equation}
  \sqrt{\frac{N_{\text{generated}}}{N(500\mathop{\rm pb}^{-1})}}
    = \sqrt{\frac{\mathcal{L}_{\text{generated}}}{500\mathop{\rm pb}^{-1}}}\,.
\end{equation}
 
\noindent
Again, the event count for \emph{all} final states is used, but also
here the actual measurements will involve events of a variety of
final states.  The resulting relative errors are collected in
table~\ref{tab:errors}.  It must be kept in mind that these errors are
meant as order-of-magnitude estimates for gauging the accuracy of
the theoretical predictions only.  The actual measurement will be able to
reduce these errors by intelligent use of constraints. At the same
time, systematic errors will increase the experimental errors.
 
\begin{table}[ht]
  \def\arraystretch{1.2}
  \begin{center}
    \begin{tabular}{|c||r|r|r|r|}\hline
       $\sqrt s$ & 161 GeV& 175 GeV& 190 GeV& 205 GeV\\\hline
       \hline
       $\sigma$  &  2.4\% &  1.2\% &  1.1\% &  1.1\% \\\hline
       $\left\langle T_1(\cos\theta_W)\right\rangle$
                 &  6.8\% &  2.1\% &  1.4\% &  1.1\% \\\hline
       $\left\langle T_2(\cos\theta_W)\right\rangle$
                 &  5.3\% &  3.7\% &  5.9\% & 15.3\% \\\hline
       $\left\langle (x_m)^1\right\rangle$
                 &  3.2\% &  6.4\% & 38.1\% & 19.6\% \\\hline
       $\left\langle (x_m)^2\right\rangle$
                 &  7.4\% &  5.8\% &  4.5\% &  4.0\% \\\hline
       $\left\langle (x_\gamma)^1\right\rangle$
                 &  8.9\% &  2.9\% &  2.5\% &  2.4\% \\\hline
       $\left\langle (x_\gamma)^2\right\rangle$
                 & 26.4\% &  6.3\% &  4.1\% &  3.7\% \\\hline
       $\left\langle (x_\gamma^{\text{lost}})^1\right\rangle$
                 & 11.0\% &  3.7\% &  3.2\% &  3.0\% \\\hline
       $\left\langle (x_\gamma^{\text{lost}})^2\right\rangle$
                 & 32.7\% &  7.9\% &  5.2\% &  4.8\% \\\hline
       $\left\langle (x_\gamma^{\text{vis.}})^1\right\rangle$
                 & 14.9\% &  5.0\% &  4.2\% &  4.1\% \\\hline
       $\left\langle (x_\gamma^{\text{vis.}})^2\right\rangle$
                 & 45.2\% & 10.6\% &  7.0\% &  6.3\% \\\hline
       $\left\langle T_1(\cos\theta_\mu)\right\rangle$
                 &  4.1\% &  1.8\% &  1.3\% &  1.1\% \\\hline
       $\left\langle T_2(\cos\theta_\mu)\right\rangle$
                 &  3.5\% &  2.1\% &  2.4\% &  3.1\% \\\hline
       $\left\langle T_1(\cos\theta_\mu^*)\right\rangle$
                 & 16.6\% &  5.0\% &  3.2\% &  2.6\% \\\hline
       $\left\langle T_2(\cos\theta_\mu^*)\right\rangle$
                 &  4.4\% &  2.4\% &  2.2\% &  2.3\% \\\hline
       $\left\langle (x_\mu)^1\right\rangle$
                 &  0.4\% &  0.3\% &  0.3\% &  0.3\% \\\hline
       $\left\langle (x_\mu)^2\right\rangle$
                 &  0.8\% &  0.5\% &  0.5\% &  0.6\% \\\hline
     \end{tabular}
  \end{center}
  \caption{\label{tab:errors}%
    Estimated statistical errors
    at~$\mathcal{L}_0=500\mathop{\rm pb}^{-1}$.}
\end{table}
 
Some errors in table~\ref{tab:errors} appear suspiciously large, but
\label{pg:error-discussion}
their origin can be understood easily.  The
quantity~$\langle x_m\rangle =
  \langle\sqrt{s_+}+\sqrt{s_-}-2M_W\rangle/(2E_B)$ vanishes in the
narrow width approximation.  Therefore it is a \emph{fine tuned}
quantity for which the relative error can be of order one.  The
absolute error on~$\langle\sqrt{s_+}+\sqrt{s_-}\rangle$ is
about~$70\MeV$ ($200\MeV$ at $161\GeV$). Experimentalists expect that
the error on the~$W$ mass will be smaller by virtue of constraint fits.
The errors on the photonic observables at~$161\GeV$ are simply caused
by the small radiated energy and the small number of hard, observable photons
close to threshold.
 
In the plots below, the errors are presented for an integrated
luminosity of~$\mathcal{L}_0=500\mathop{\rm pb}^{-1}$.  If the
corresponding error is larger than the spread of the predictions,
$\mathcal{L}_0$ is multiplied by an appropriate power of ten.
According to the target set in~\cite{sibling-report}, our predictions
should have an error of less than one third of the expected
experimental error.  The spread of values in the plots below must
therefore be inside a gray band corresponding to~$5\mathop{\rm fb}^{-1}$.
 
At this point we should emphasize for the first time, that possible
discrepancies in the \emph{tuned comparisons} must \emph{not} be
mistaken for \emph{theoretical} errors.  They rather point to
\emph{incorrect} implementations and/or to still undiscovered bugs.
 
\subsubsection{Canonical Cuts}
\label{sec:canonical-cuts}
 
\dofour{tuned_xsectn_y}{%
  Tuned predictions for the total cross section
  for~$e^+e^- \to \mu^-\bar\nu_\mu u\bar d$ after canonical (\ADLO/)
  cuts.}
 
Canonical cuts (a.k.a.~\ADLO/) have been defined in collaboration with
ALEPH, DELPHI, L3 and OPAL.  The following \emph{acceptance cuts} define
an optimistic union of the phase spaces that the four collaborations
expect to cover:
\begin{itemize}
  \item the energy of light charged leptons ($e$, $\mu$) must be greater
    than $1\GeV$;
  \item light charged leptons ($e$, $\mu$) will be seen down to
    10~degrees from either beam;
  \item the energy of a jet must be greater than $3\GeV$.  For the
    purpose of our study, jets will be identified with quarks;
  \item jets can be detected in the entire $4\pi$ of solid angle;
  \item photons must have an energy of at least $100\MeV$ to be
    identified;
  \item photons will be seen down to 1~degree from either beam.
\end{itemize}
These cuts do not address the issue of $\tau$-identification.  For the
purpose of theoretical studies, $\tau$'s can be treated like the light
charged leptons~$e$ and~$\mu$.  It is understood that the programs
considered here will have to be interfaced to external $\tau$-decay
packages.
These \emph{acceptance cuts} are supplemented by the following set of
\emph{separation cuts}:
\begin{itemize}
  \item light charged leptons ($e$, $\mu$) must be separated by at least
    5~degrees from jets.  Jets will again be identified with quarks.
  \item the invariant mass of two jets that are resolved as two separate
    jets must be greater than $5\GeV$
  \item photons must be separated by at least 5~degrees from light
    charged leptons ($e$, $\mu$) and jets
\end{itemize}
$\tau$'s will again be treated like the light charged leptons~$e$
and~$\mu$. If any of the charged particles of our final state fails any
of these cuts, the event will be discarded.
 
Programs using the strict collinear limit for photons will count all
photons as lost and assign them to initial-state radiation.  If a
program generates photons with a finite~$p_T$, a more detailed treatment
is necessary.  Photons failing the separation cuts from charged final-state
particles will not simply be discarded.  Instead, their four
momentum is added to the closest charged particle.  Photons missing the
acceptance cut around the beam pipe will be counted as lost and will be
assigned to initial-state radiation.  The question if this procedure is
appropriate for dealing with final-state radiation will be discussed
below in section~\ref{sec:FSR}.  There the size of the separation cut
will be discussed in more detail.
 
These cuts serve two purposes.  Firstly they are important for
testing programs under more realistic conditions.  Secondly, they are
required to give well-defined predictions without the need for
internal technical cuts cutting out singular regions in phase space.
However, for final states involving photons and for programs using
massless fermions, some care must be taken in interpreting the results. Indeed,
the canonical cuts when applied to a final-state $l^+l^-$ allow for
a minimum invariant $l^+l^-$- mass of $87.2\,$MeV which is below $2\,m_{\mu}$.
 
Comparing figures~\ref{fig:tuned_xsectn_n}
and~\ref{fig:tuned_xsectn_y}, we observe that the effect of the
canonical cuts are rather small.  This shows that the effect of the
internal technical cuts are very similar for all programs under
consideration.
 
\subsubsection{``Unleashed'' Comparisons}
 
\dofour{bicd_xsectn_n}{%
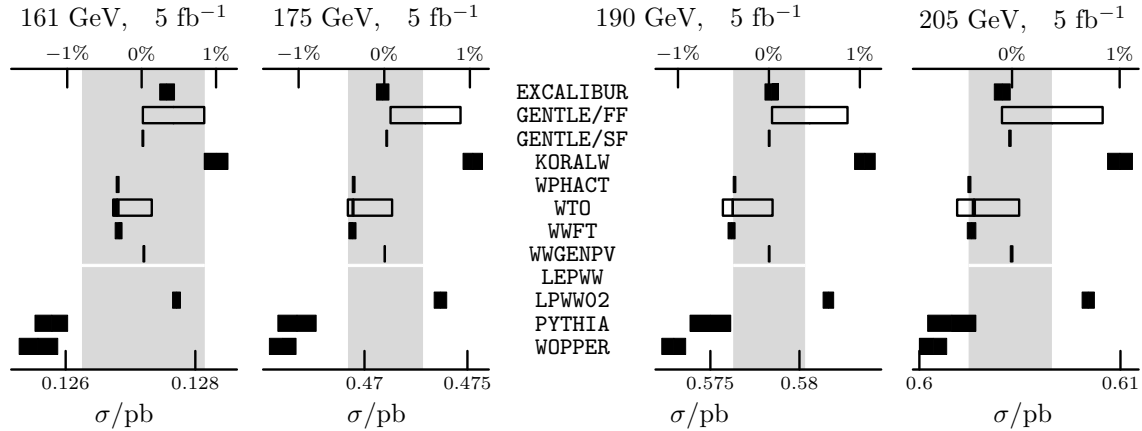
  Unleashed predictions for the total cross section
  for~$e^+e^- \to \mu^-\bar\nu_\mu u\bar d$ without cuts.
  The transparent, framed error bars are theoretical errors
  (cf.~page~\pageref{pg:th-error}).}
\dofour{bicd_xsectn_y}{%
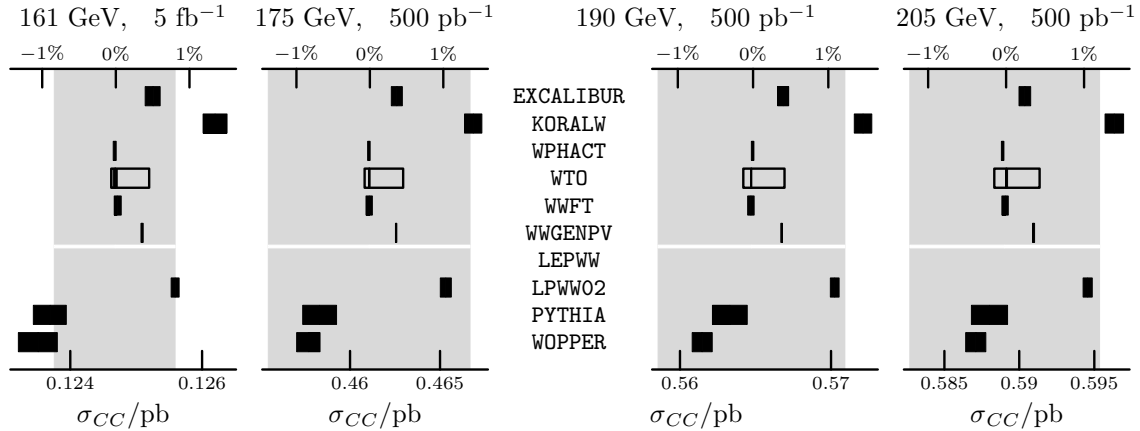
  Unleashed predictions for the total cross section
  for~$e^+e^- \to \mu^-\bar\nu_\mu u\bar d$ after canonical (\ADLO/)
  cuts.  The transparent, framed error bars are theoretical errors
  (cf.~page~\pageref{pg:th-error}).}
 
Some numerically important corrections to the total cross section have
been left out in the \emph{tuned comparisons}.  They have been studied
in separate set of comparisons.  In these \emph{unleashed
comparisons}, all program authors have been asked to provide the
\emph{``Best Prediction They Can Make''}.  It is of course clear that
this is a moving target and the data presented in this report must be
viewed as a snapshot of the situation at the end of 1995.  This is
different from the \emph{tuned comparisons}, which implement a fixed
set of approximations and input parameters. These predictions should
not change in time, unless bugs are found in some codes.
 
The \emph{Coulomb correction}
\label{pg:Coulomb}
(see~\cite{sibling-report} for a detailed formula) is well established
and can be implemented easily as a factor multiplying the part of the
cross section emanating from the \emph{CC03} subset of diagrams.
Using a narrow-width approximation exaggerates the effect of the
Coulomb correction.
 
The \emph{QCD corrections} to
the hadronic $W^\pm$ width, $\Gamma_W^{\text{hadr.}}$, must be properly
included in processes with
$q\bar{q}$ pair(s). We have adopted a \emph{naive QCD} factor (NQCD):
\label{pg:QCD}
\begin{equation}
\label{eq:naive-QCD}
  \Gamma_{W\to\text{hadr.}}^0
     \to \sum_{\bar qq} \left(\Gamma_{W\to\bar qq}^0 +
             \Gamma_{W\to\bar qq}^1 \right) +
         \sum_{\bar qqg} \Gamma_{W\to\bar qqg}^1
        = \sum_{\bar qq} \Gamma_{W\to\text{hadr.}}^0 \cdot
          \left(1 + \frac{\alpha_{_{QCD}}}{\pi}\right)
\end{equation}
It is certainly correct for inclusive quantities
like the total cross section without cuts if only the \emph{CC03}
diagrams are taken into account.
 
At the same time it is questionable for exclusive quantities and for
diagrams that can not be factorized in the production and decay of a
$W^+W^-$~pair.
Without a
complete~${\mathcal{O}}(\alpha_{_{QCD}})$ calculation including gluons in the
final state, we can not prove that the correction is really of this
magnitude in the presence of cuts. Similarly, we can not be sure about
the \emph{CC11} diagrams without a calculation of the QCD box diagram
corrections.
Here, we are faced with the very familiar
problem of whether we can shrink EW interactions to a point
in the presence of gluon emission.
 
On the other hand, for our set of canonical (\ADLO/) cuts with
complete~($4\pi$) coverage of jets, the \emph{``naive correction''}
could be very close to the truth for the \emph{CC03} diagrams.
Furthermore, even if the size of the correction to the \emph{CC11}
diagrams has not been calculated, we know that it is a
${\mathcal{O}}(\alpha_{_{QCD}})$~correction to a
${\mathcal{O}}(\Gamma_W/M_W)$~correction and it makes pragmatical sense to
include the overall NQCD correction anyway.  The
factor~(\ref{eq:naive-QCD}) has therefore been included by all programs
in the numbers below.
 
In connection with implementation of NQCD we emphasize that
the effect of NQCD on some moments, typically $\langle x_m\rangle_n$, is
quite large, i.e. of the order of few percent. For instance
both \WPHACT/ and \WTO/ have analyzed $\langle x_m\rangle_1$ with and without
the inclusion of NQCD. The latter has a net effect of changing
$\langle x_m\rangle_1$ of $1.5\%$ at $\sqrt{s} = 161\,$GeV and of $2.6\%$
at $\sqrt{s} = 175\,$GeV. This is a considerable correction
factor which, in general, calls for a better understanding
of the QCD corrections to have full reliability of the
order of magnitude of the effect.
 
Finally, the whole problem of the implementation of
NQCD must be seen in the light of describing the relationship between
the QCD matrix elements and the interface with hadronization.
Ideally, we would have at our disposal a chain of cross checking
programs starting from an exact semianalytical program, continuing
with less precise but more flexible integration programs and ending
with Monte Carlo event generators that can implement any cut and can
be be interfaced with hadronization.  In the last step
\emph{double-counting} should be carefully avoided.  It must be
kept in mind, however, that many hadronization codes will affect
differential distributions only, without correcting the total cross
section.  Therefore such corrections have to be put in by hand.
At the same time,
hadronization may suffer from its own problems, connected with
the identification of the proper \emph{color-singlet} structure which
is far from clear in the presence of complicated diagrams.
 
The \emph{QED corrections}:
Using the \emph{current-splitting trick}~\cite{gentle_nunicc}, it is
possible to identify a set of non-logarithmic universal QED radiative
corrections and to implement them in so-called \emph{flux functions}.
In order to assess the effect from these contributions, \GENTLE/ has
contributed two numbers to the \emph{unleashed comparisons}: one
(\texttt{GENTLE/SF}) using structure functions, like most other programs
and a second (\texttt{GENTLE/FF}) using flux functions.
This also allows us to understand the apparent deviation
of the \texttt{KORALW} number from the others: there, the
so-called YFS form factor has been included, which is
essentially  equivalent  with going from the SF to  the FF
description: indeed, the \texttt{GENTLE} result with FF
is in good agreement with the \texttt{KORALW} one.
\label{pg:flux-function}
 
The \emph{EW corrections} are
the theoretically most demanding problem. There is
a theoretical uncertainty from
having to choose a particular resummation scheme.  In the \emph{tuned
comparisons}, this uncertainty has artificially been removed by
demanding a particular choice of input parameters. In the
\emph{unleashed comparisons}, the spread of predictions \emph{can} point
to a theoretical uncertainty. This is, however, not due to EW uncertainties
because a sizeable fraction of the programs have used a
 scheme  very similar to the
\emph{tuned comparisons}.
 
The \emph{CKM quark mixing correction}
is a trivial correction arising from  non-trivial quark
mixing:
\begin{equation}
  \Gamma_W^{q\bar q} \propto |V_{q\bar q}|^2\,.
\end{equation}
Due to the unitarity of the CKM-matrix, the effects on the widths are
negligible.  If light quark flavors are summed over, as is required by
experimental procedures anyway, the effect on exclusive final states
will be small, except for the occasional $b$-quark.  Since the range
for $|V_{ud}|^2$ is larger than the uncertainties from other factors,
the plots in figures~\ref{fig:bicd_xsectn_n}
and~\ref{fig:bicd_xsectn_y} have been normalized
to~$|V_{ud}|^2 = 0.9518$.
 
   The \emph{fermionic masses} could, in principle,
    be included everywhere in the various calculations, but
    we point out that there are essentially three
    places where they become relevant. First of all,
    the electron mass
    in \textit{CC20},
    whenever the $e^-(e^+)$ scattering angle is considered
    without cuts (gauge invariance is also involved here).
Secondly,     whenever
    a charged
    fermion-antifermion pair occurs in the final state,  particular
    care should be devoted to study the threshold region in
    $\gamma^* \to f{\bar f}$.
    In the third place,
     the $b$-quark should be taken massive for a fully
    consistent study of Higgs boson production and of its background.
    For the last case,
    and for quarks in general,
    one should worry about
    which value to use, i.e. the pole mass or the running mass and, if
    the latter is chosen, at which scale. It is not at all an academic
    problem in view of the large difference between, say, $m_b(m_b)$
    and $m_b(M_W)$ or $m_b(m_H)$.

Programs that implement the complete \emph{CC10} set of diagrams have
contributed to the \emph{unleashed comparisons} as well as programs
restricted to the doubly resonant \emph{CC03} subset.  In the context
of a \emph{``Best Prediction They Can Make''} the comparison of
programs from both sets are justified.  In order to help the reader,
the \emph{CC10} programs have been collected at the top of each plot,
while the \emph{CC03} programs are shown at the bottom, separated by a
thin white line.
 
\subsubsection{Theoretical uncertainties}
 
\noindent
At the level of our present knowledge, it is impossible to expect
 a common treatment of the \emph{theoretical error},
something which is by definition highly subjective. However our preliminary
investigations
(mostly \GENTLE/ and \WTO/) have shown that even the most crude and
naive estimate of the theoretical error gives quite a wide spread of answers.
 
Ideally, a theoretical error should be inferred by estimating the differences
originating from different treatments of leading higher order effects as
well as from non-leading ones, whose size is notoriously much more difficult
to guess. Obviously, a theoretical error is bound to disappear whenever real
progress is achieved under the form of new and \emph{complete}
calculations. Most of the time, the potentialities claimed in the summary
table only refer to some \emph{naive} treatment of a particular effect. There
is no particular harm in that, as long as \emph{naive} estimates are kept
well separated from the \emph{precise} calculations. From this point of
view the extension from\emph{CC03} to \emph{CC10} (or even better
to \emph{CC20}) is a well-established piece of work while the inclusion of
final state QCD corrections is, at this stage, a \emph{naive} although
\emph{educated} guess.
 
By referring to a \emph{theoretical error} we can only admit a
very partial attempt to understand the missing components of our calculations.
Specifically, we can get a feeling of what is missing by allowing different
implementations of the SF approach ($\eta$-scheme versus $\beta$-scheme or even
the mixed one) and by judging in a very crude (and most probably
underestimated) way the effect of terms of order $\alpha \, \times$
\emph{constant}.
The same can be attempted by comparing the SF and the FF approaches.
In the end
the codes implementing SF have adopted the $\beta$-scheme for tuned
comparisons (although it violates gauge invariance), since there are
plausibility arguments showing that whenever the full answer is known
in other processes then the $\beta$-scheme gives the best numerical
approximation.
 
Very simple analyses of theoretical errors
have been performed by \GENTLE/ and \WTO/. They used different
sets of \emph{working options}.
 
\GENTLE/ ran over 6 options: 5 {\tt IZERO}$\times$\texttt{IQEDHS}
(see subsection
2.5) options using FF plus the standard SF treatment of ISR.
In this way, the error due to different treatment of ISR was simulated.
\GENTLE/ results for $\sigma\,$, $\langle E_{\gamma}\rangle$ and
$\langle 10x_{m}\rangle_1$ are presented in table \ref{gen_te}.
 
\begin{table}[hbtp]
\begin{center}
\begin{tabular}{|c|c|c|c|c|c|c|}
\hline
\hline
$E_{cm}\,$/{\tt IZERO-IQEQHS} & 0-0 & 0-1 & 0-2 & 0-3 & 1-3 & SF\\
\hline
\hline
\multicolumn{7}{|c|}{$\sigma$, $pb$}                            \\
\hline
161 & 0.13420 & 0.13366 & 0.13380 & 0.13379 & 0.13460 & 0.13364 \\
175 & 0.49598 & 0.49522 & 0.49562 & 0.49561 & 0.49862 & 0.49493 \\
190 & 0.60787 & 0.60801 & 0.60841 & 0.60838 & 0.61212 & 0.60758 \\
205 & 0.63483 & 0.63558 & 0.63592 & 0.63590 & 0.63984 & 0.63512 \\
\hline
\multicolumn{7}{|c|}{$\langle (m,E)_{\gamma}\rangle$, $GeV$}    \\
\hline
161 & 0.4671  & 0.4754  & 0.4746  & 0.4746  & 0.4749  & 0.4759  \\
175 & 1.1055  & 1.1267  & 1.1248  & 1.1249  & 1.1254  & 1.1271  \\
190 & 2.1052  & 2.1518  & 2.1473  & 2.1473  & 2.1488  & 2.1565  \\
205 & 3.1388  & 3.2084  & 3.2010  & 3.2010  & 3.2041  & 3.2223  \\
\hline
\multicolumn{7}{|c|}{$\langle 10x_{m}\rangle_1$}                \\
\hline
161 &-.38320  &-.38410  &-.38401  &-.38401  &-.38403  &-.38400  \\
175 &-.066431 &-.066714 &-.066684 &-.066684 &-.066695 &-.066701 \\
190 &-.012318 &-.012516 &-.012492 &-.012492 &-.012502 &-.012508 \\
205 & .015638 & .015478 & .015501 & .015501 & .015489 & .015450 \\
\hline
\hline
\end{tabular}
\end{center}
\caption{\GENTLE/ theoretical errors}
\label{gen_te}
\end{table}
 
Two comments are in order here. First, since for {\tt IQEDHS=0} only
${\cal O}(\alpha)$ exponentiated FF ISR corrections are used, while
for {\tt IQEDHS=1,2,3} different realizations of ${\cal O}(\alpha^2)$
are applied, one should consider the difference between
{\tt IQEDHS=0} and {\tt IQEDHS}$\geq${\tt 1} as an illustration
of the importance of ${\cal O}(\alpha^2)$ corrections rather than as an
estimate of theoretical errors. Second, in the FF method, one
may access only $\langle m_{\gamma}\rangle$, whose difference from
$\langle E_{\gamma}\rangle$ grows rapidly with energy, see \cite{wmass}.
So, in this case one should not consider the difference between FF and
SF calculations as a theoretical uncertainty. The
\GENTLE/ theoretical errors are exhibited
in figures~\ref{fig:bicd_xsectn_n}
and \ref{fig:bicd_mas_p1_y}
by a transparent, framed error bar.
 
\WTO/ ran over $6=2\times3$ {\tt IOS}$\times${\tt IOSF} options.
Two options, {\tt IOS}, for the renormalization of the weak sector,
see eqs. \ref{wto_eq1}-\ref{wto_eq2},
and three options, {\tt IOSF} for initial-state radiation structure functions
implementations, adopted respectively in \cite{excalit,topv,cpcww}.
\WTO/ results for $\sigma\,$ and $\langle E_{\gamma}\rangle$
are given in table \ref{wto_te}.
 
\begin{table}[hbtp]
\begin{center}
\begin{tabular}{|c|c|c|c|c|c|c|}
\hline
\hline
$E_{cm}\,$/{\tt IOS-IOSF} & 1-1 & 1-2 & 1-3 & 2-1 & 2-2 & 2-3 \\
\hline
\hline
\multicolumn{7}{|c|}{$\sigma$, $pb$}                            \\
\hline
161 &0.13206 &0.13204 &0.13250 &0.13201 &0.13198 &0.13244 \\
175 &0.49207 &0.49186 &0.49358 &0.49177 &0.49156 &0.49329 \\
190 &0.60240 &0.60192 &0.60404 &0.60188 &0.60139 &0.60352 \\
205 &0.62828 &0.62754 &0.62977 &0.62764 &0.62691 &0.62913 \\
\hline
\multicolumn{7}{|c|}{$\langle E_{\gamma}\rangle$, $GeV$}    \\
\hline
161 & 0.4688 & 0.4673  & 0.4674  & 0.4685  & 0.4669  & 0.4670 \\
175 & 1.1250 & 1.1219  & 1.1221  & 1.1251  & 1.1220  & 1.1222  \\
190 & 2.1579 & 2.1484  & 2.1489  & 2.1583  & 2.1488  & 2.1493  \\
205 & 3.2317 & 3.2119  & 3.2129  & 3.2324  & 3.2126  & 3.2135  \\
\hline
\hline
\end{tabular}
\end{center}
\caption{\WTO/ theoretical errors}
\label{wto_te}
\end{table}
 
 The largest uncertainty for
$\langle E_{\gamma}\rangle$ is
 of~$1.9, 3.2, 9.9, 20.5\,$MeV for $E_{cm} = 161, 175, 190,
205\,$GeV respectively.
 
 In figures~\ref{fig:bicd_xsectn_n},
\ref{fig:bicd_xsectn_y} and \ref{fig:bicd_ega_p1_y}
these uncertainties are exhibited by a transparent, framed error
bar drawn around the black statistical error bar.
\label{pg:th-error}
 
Inspecting figures~\ref{fig:bicd_xsectn_n} and \ref{fig:bicd_xsectn_y},
we see that the theoretical error derived this way nicely reproduces the
range in predictions defined by \WPHACT/ and \WWF/ at the low end and
\EXCALIBUR/, \GENTLE/ (structure function) and \WWGENPV/ at the high
end.  On the other hand we must not rush to the judgment that the
theoretical error will always be given by the spread in predictions from
different programs.  A detailed analysis like the one performed by \WTO/
is more reliable. In figure~\ref{fig:bicd_ega_p1_y} below, we will see
an example in which the theoretical error estimated from scanning the
options is slightly larger than the spread in predictions.
 
\subsubsection{Total Cross Sections}
 
\noindent
As can be seen in figures~\ref{fig:tuned_xsectn_n}
and~\ref{fig:tuned_xsectn_y}, the agreement among the programs is
generally good for the total cross sections.  As expected, the effect
of the \emph{CC11} diagrams is most notable at~$161\GeV$.  Even though
it will be hard to reach this level of experimental accuracy, the
programs that are still restricted to the \emph{CC03} subset should
aim at implementing a more complete subset.
 
For most energies, the predictions of \LEPWW/ have not been included
in the plots because they are too far off from the other programs.
This is caused by an insufficient implementation of initial state
radiation in this program, which is of mostly historical interest.
 
The agreement of the predictions of \PYTHIA/ with the rest
of the programs is unsatisfactory.
 
It should come as no surprise that the spread of predictions is larger
in the \emph{unleashed comparisons}.  It remains however at or below
the expected experimental accuracy of LEP2.
 
The qualitative pictures with and without cuts are very similar. For
this reason, we will show (with one exception) only results without
cuts for the tuned comparisons and only results with cuts for the
unleashed comparisons of exclusive observables below.
 
\subsubsection{$W$ Production Angle}
 
\dofour{tuned_ctw_t1_n}{%
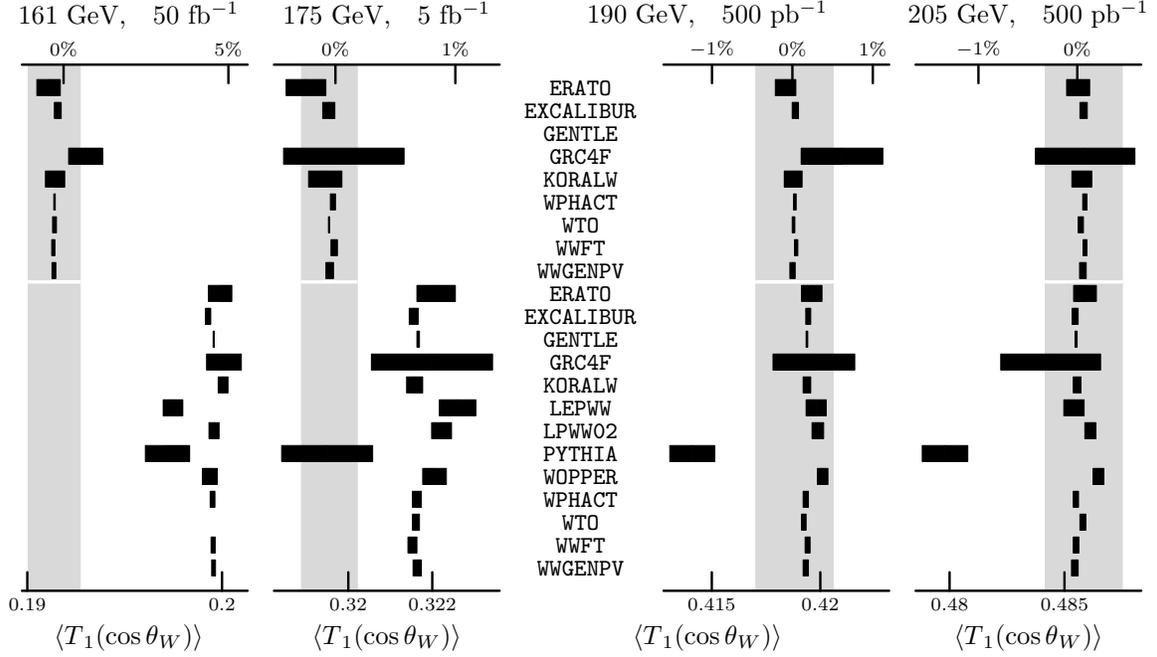
  Tuned predictions for the first Chebyshev polynomial of the
  $W$~production angle in~$e^+e^- \to \mu^-\bar\nu_\mu u\bar d$ without
  cuts.}
\dofour{bicd_ctw_t1_y}{%
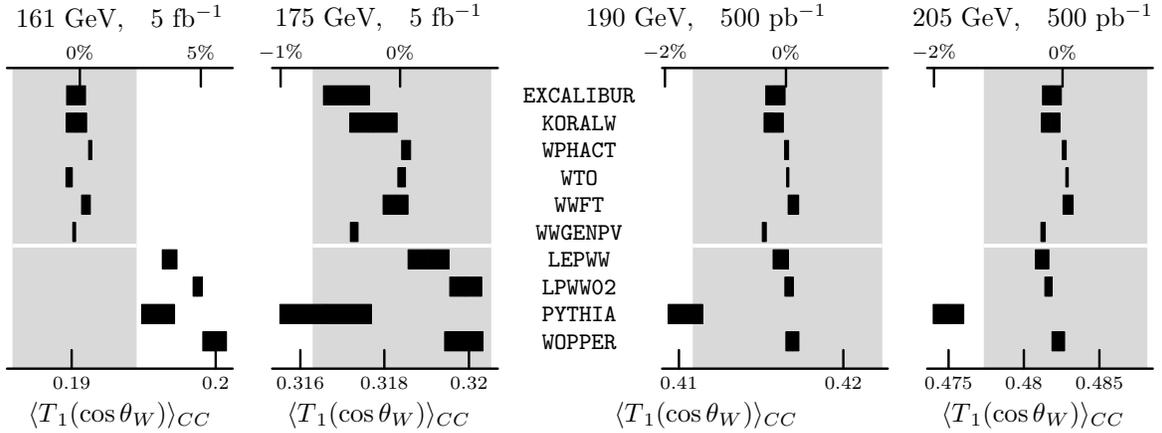
  Unleashed predictions for the first Chebyshev polynomial of the
  $W$~production angle in~$e^+e^- \to \mu^-\bar\nu_\mu u\bar d$ with
  canonical (\ADLO/) cuts.}
 
The trend observed in the total cross section continues in the moments
of the $W$~production angle.  The deviations of \PYTHIA/'s results are again
not acceptable for precision measurements.
 
\dofour{tuned_ctw_t2_n}{%
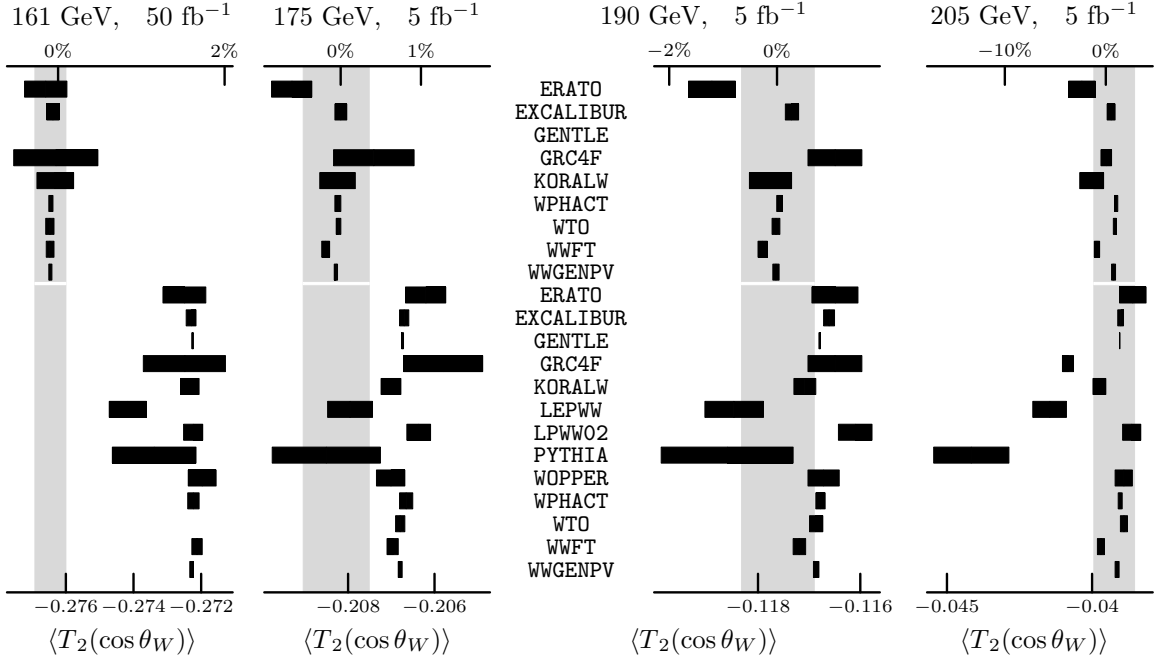
  Tuned predictions for the second Chebyshev polynomial of the
  $W$~production angle in~$e^+e^- \to \mu^-\bar\nu_\mu u\bar d$ without
  cuts.}
\dofour{bicd_ctw_t2_y}{%
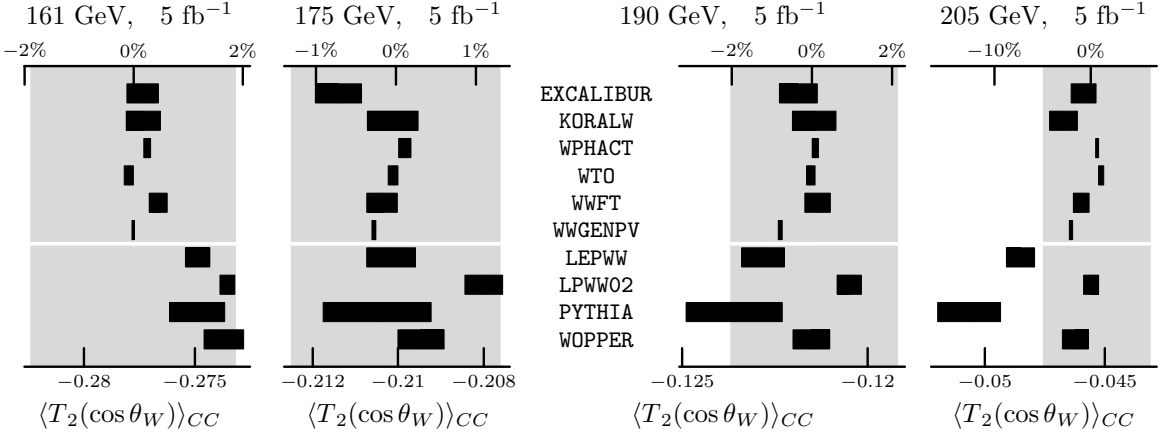
  Unleashed predictions for the second Chebyshev polynomial of the
  $W$~production angle in~$e^+e^- \to \mu^-\bar\nu_\mu u\bar d$ with
  canonical (\ADLO/) cuts.}
 
\subsubsection{Invariant Masses}
 
\dofour{tuned_mas_p1_n}{%
  Tuned predictions for the deviation of the sum of invariant
  $W$-masses from~$2M_W$ in~$e^+e^- \to \mu^-\bar\nu_\mu u\bar d$
  without cuts.}
\dofour{bicd_mas_p1_n}{%
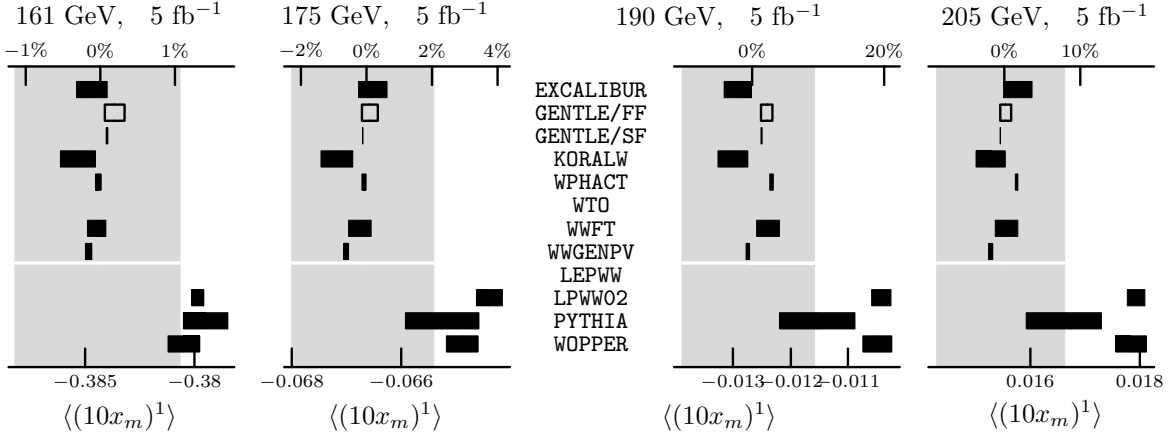
  Unleashed predictions for the deviation of the sum of invariant
  $W$-masses from~$2M_W$ in~$e^+e^- \to \mu^-\bar\nu_\mu u\bar d$
  without cuts.
  The transparent, framed error bars are theoretical errors
  (cf.~page~\pageref{pg:th-error}).}
\dofour{tuned_mas_p1_y}{%
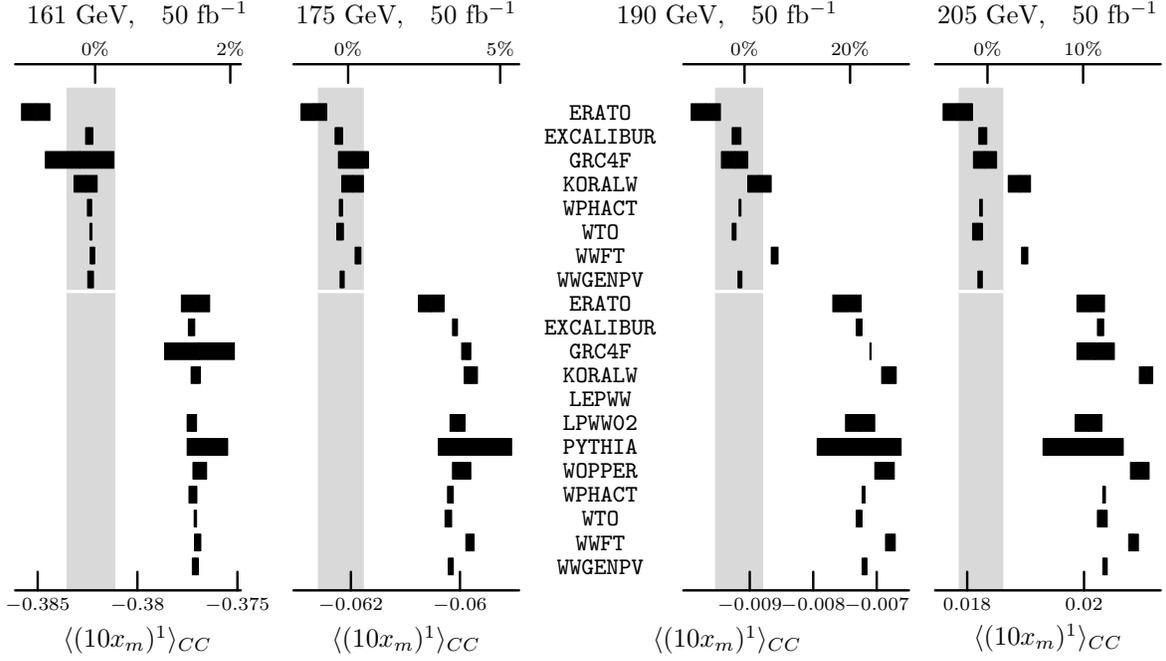
  Tuned predictions for the deviation of the sum of invariant
  $W$-masses from~$2M_W$ in~$e^+e^- \to \mu^-\bar\nu_\mu u\bar d$
  after canonical (\ADLO/) cuts.}
\dofour{bicd_mas_p1_y}{%
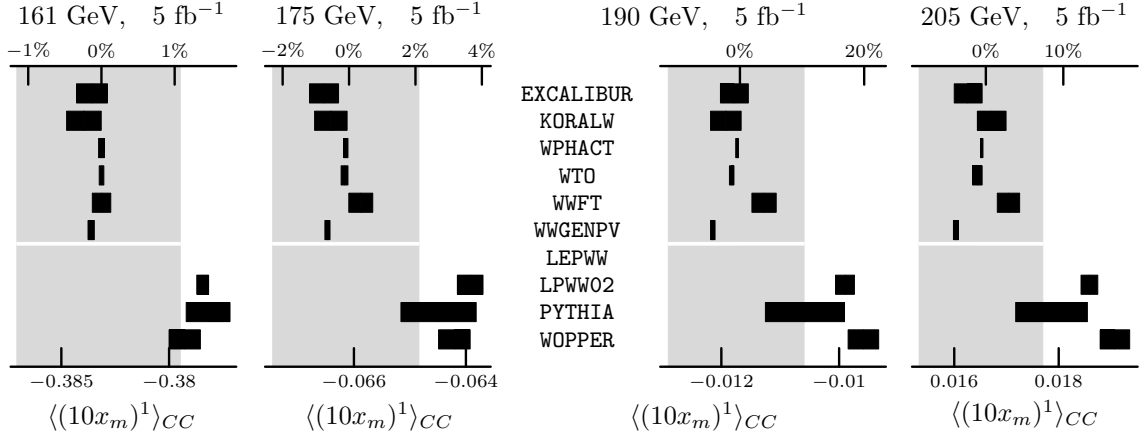
  Unleashed predictions for the the deviation of the sum of
  invariant $W$-masses from~$2M_W$
  in~$e^+e^- \to \mu^-\bar\nu_\mu u\bar d$
  after canonical (\ADLO/) cuts.}
 
The effect of the \emph{incompleteness error} of leaving out the
\emph{CC10} diagrams is of course most drastic in this observable.
While the effect will be reduced somewhat by the necessary invariant
mass cuts for reducing the non-$W^\pm$ background, all programs which are
still restricted to the \emph{CC03} set ought to attempt to lift this
restriction.
 
As has been discussed before, this observable vanishes in the zero
width approximation and we have to expect \emph{relative} errors which are
substantially larger than those for the other observables.
 
Comparing figures~\ref{fig:tuned_mas_p1_n} and~\ref{fig:tuned_mas_p1_y},
we observe a nontrivial effect of using a finite~$p_T$ for photons.  At
the higher energies,
 where a substantial number of hard photons is
radiated, the first moment of the invariant masses is slightly
higher for the programs with finite photonic~$p_T$ (\KORALW/, \WOPPER/
and \WWF/), when the \ADLO/ cuts are applied.  \WWGENPV/ gives also
finite~$p_T$ to the photons, but the numbers quoted in the
figures have been produced with an intermediate version of the code,
in which the $p_T$ is not transferred to the beam particles.
Hence, this small effect is absent in this particular case.
 
\dofour{tuned_mas_p2_n}{%
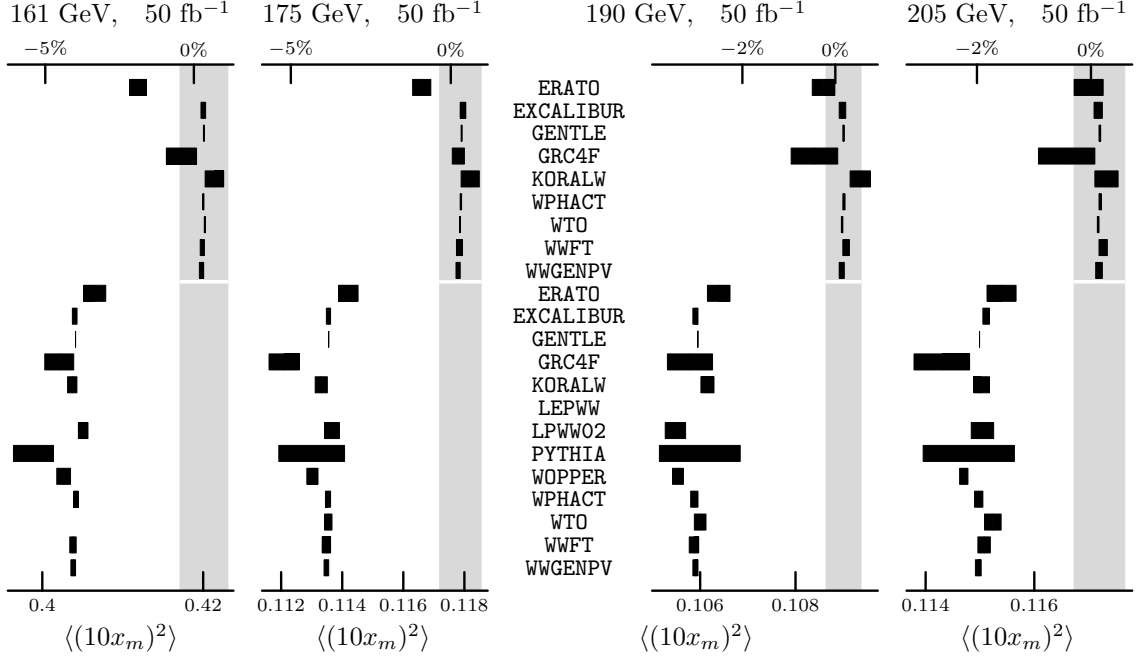
  Tuned predictions for the square of the deviation of the sum of
  invariant $W$-masses from~$2M_W$
  in~$e^+e^- \to \mu^-\bar\nu_\mu u\bar d$ without cuts.}
\dofour{bicd_mas_p2_y}{%
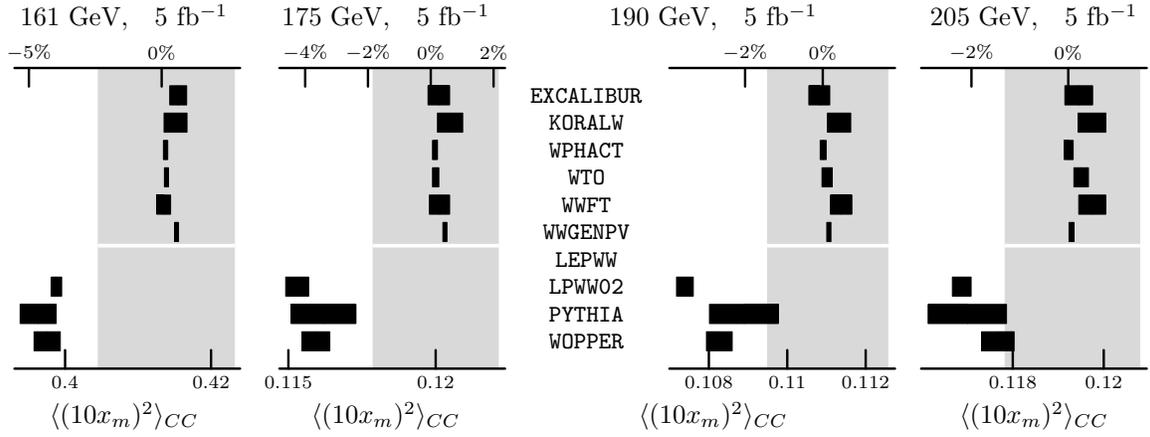
  Unleashed predictions for the square of the deviation of the sum of
  invariant $W$-masses from~$2M_W$
  in~$e^+e^- \to \mu^-\bar\nu_\mu u\bar d$ after canonical (\ADLO/)
  cuts.}
 
\subsubsection{$\gamma$ Energy}
\dofour{tuned_ega_p1_n}{%
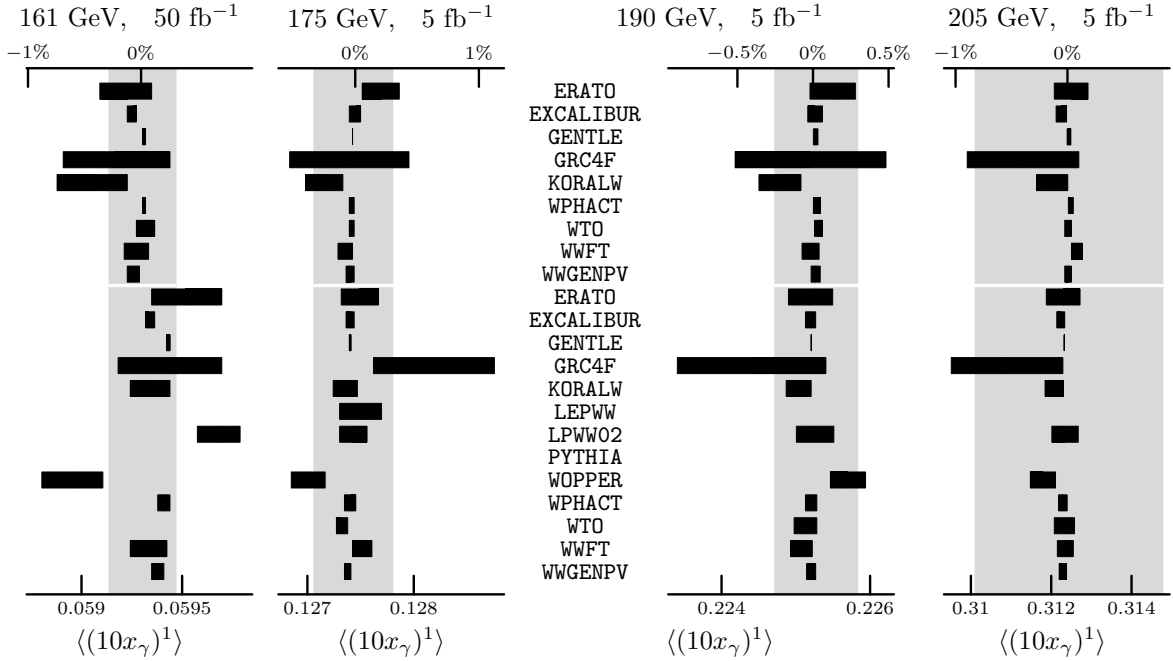
  Tuned predictions for the total radiated $\gamma$~energy
  in~$e^+e^- \to \mu^-\bar\nu_\mu u\bar d$ without cuts.}
\dofour{bicd_ega_p1_y}{%
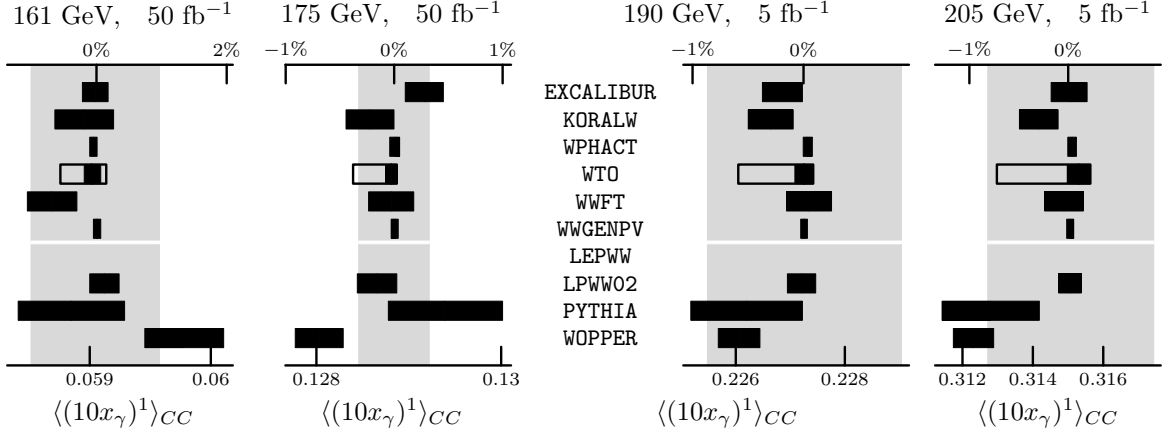
  Unleashed predictions for the total radiated $\gamma$~energy
  in~$e^+e^- \to \mu^-\bar\nu_\mu u\bar d$ after canonical (\ADLO/)
  cuts.  The transparent, framed error bars are theoretical errors
  (cf.~page~\pageref{pg:th-error}).}
 
The trend continues for the total energy radiated by photons.  Here, it
should be noted that the \emph{incompleteness error} caused by leaving
out the \emph{CC10} diagrams is most notable in the \emph{second}
moment, while it is hardly noticeable in the first moment.
 
We must keep in mind that this quantity is somewhat artificial and
has been used only for comparing the implementation of initial-state
radiation among programs which have finite $p_T$ and those who have
not.  Without the inclusion of final-state radiation, this quantity is
not measurable.
 
\dofour{tuned_ega_p2_n}{%
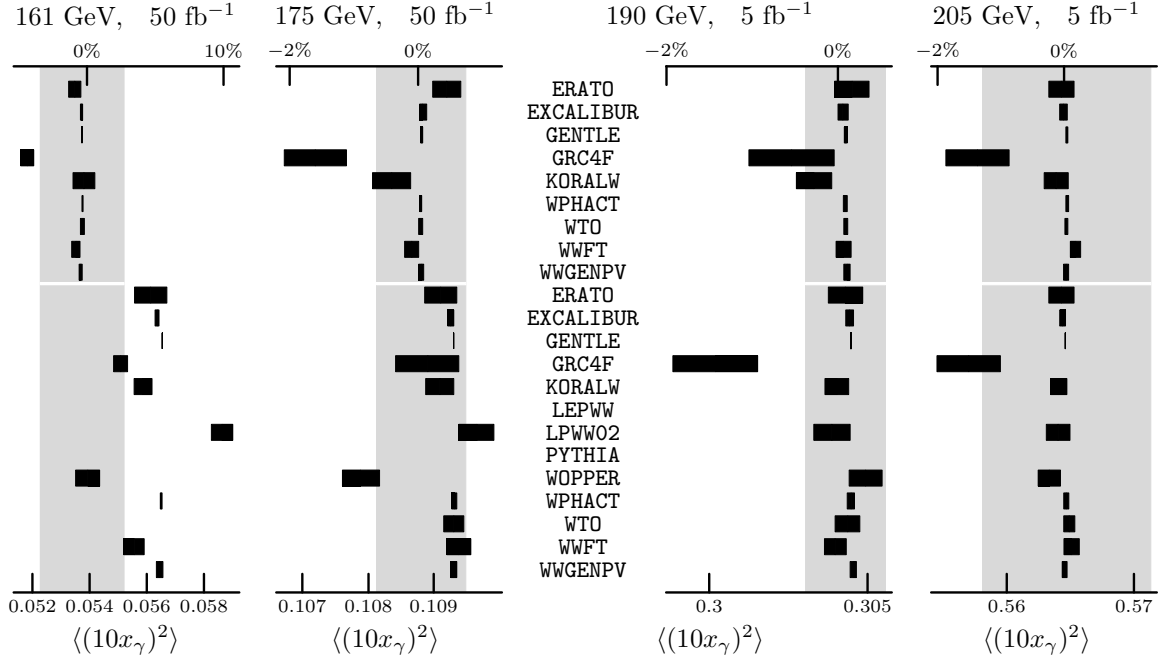
  Tuned predictions for the square of the total radiated $\gamma$~energy
  in~$e^+e^- \to \mu^-\bar\nu_\mu u\bar d$ without cuts.}
\dofour{tuned_ega_p2_y}{%
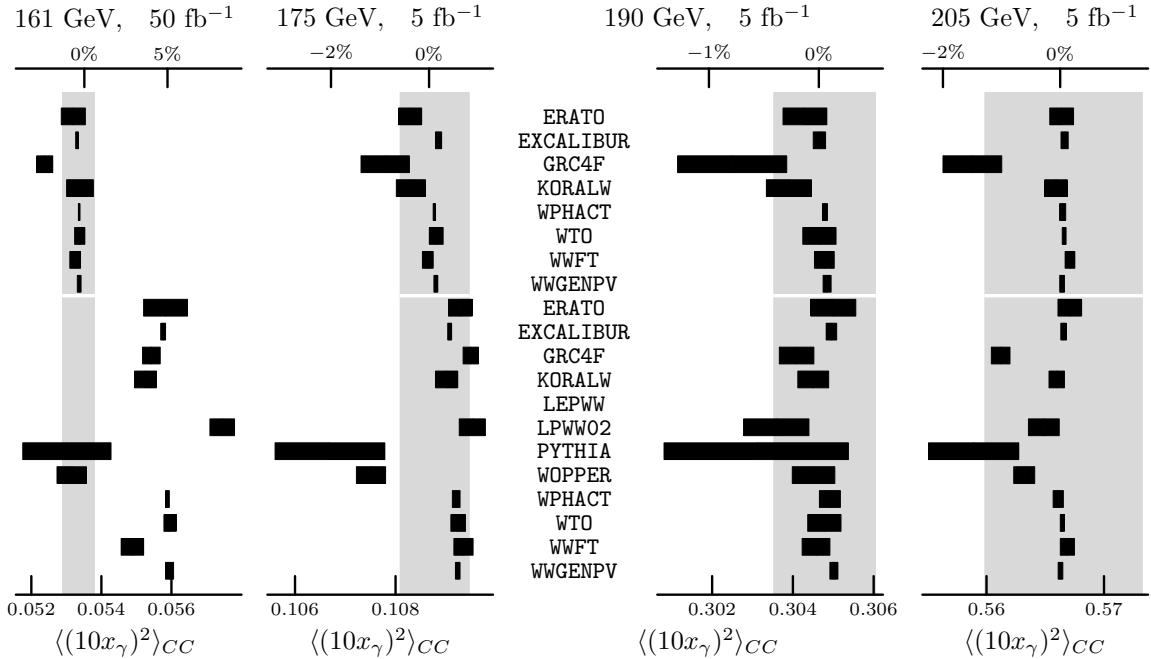
  Tuned predictions for the square of the total radiated $\gamma$~energy
  in~$e^+e^- \to \mu^-\bar\nu_\mu u\bar d$ after canonical (\ADLO/) cuts.}
 
\subsubsection{Leptonic Observables}
\dofour{tuned_ctm_t1_y}{%
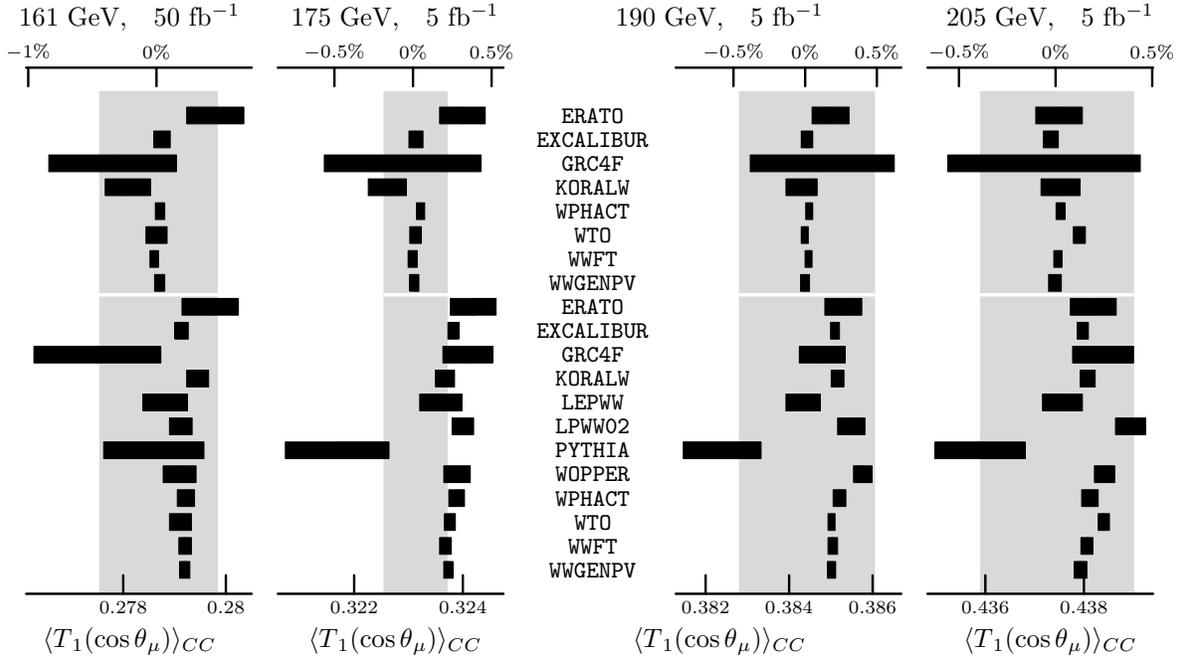
  Tuned predictions for the first Chebyshev polynomial of the
  $\mu$~production angle in the laboratory frame
  in~$e^+e^- \to \mu^-\bar\nu_\mu u\bar d$ after canonical (\ADLO/)
  cuts.}
\dofour{tuned_cts_t1_y}{%
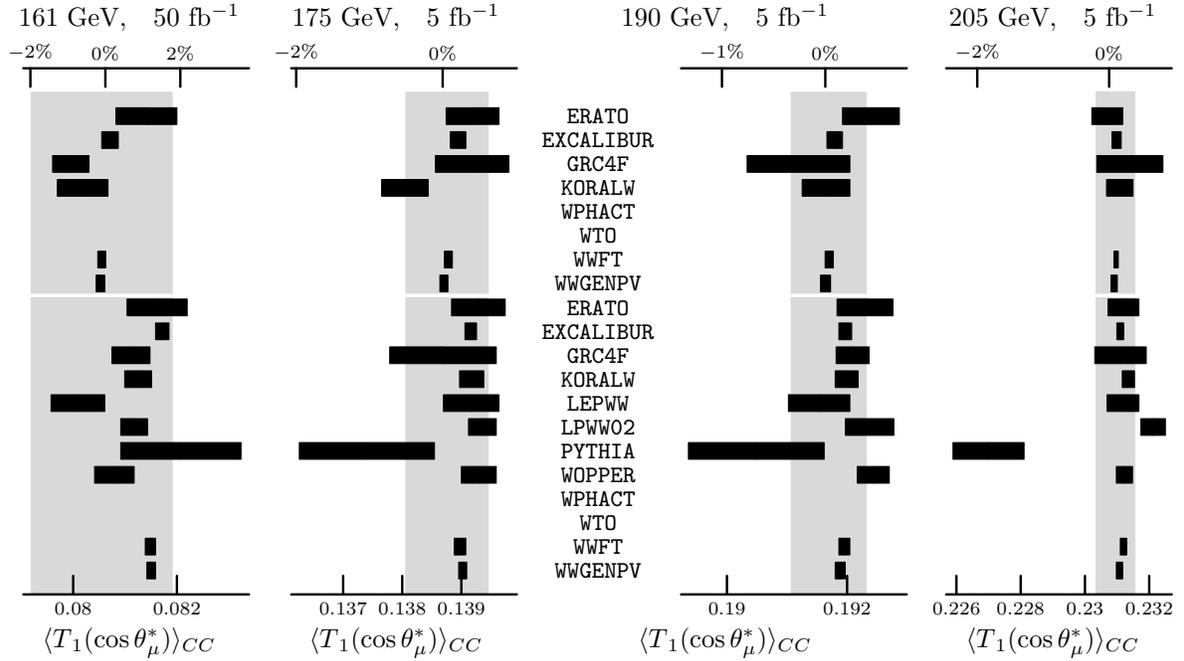
  Tuned predictions for the first Chebyshev polynomial of the
  $\mu$~decay angle in the rest frame of the~$W^-$
  in~$e^+e^- \to \mu^-\bar\nu_\mu u\bar d$ after canonical (\ADLO/)
  cuts.}
\dofour{tuned_emu_p1_y}{%
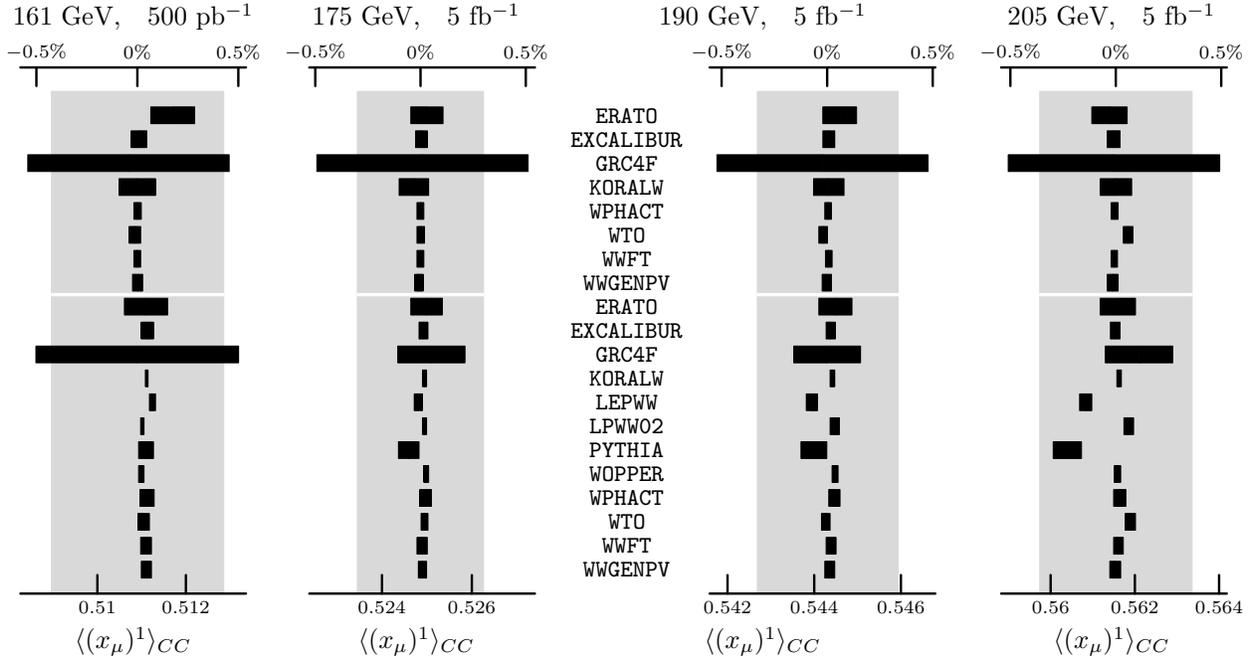
  Tuned predictions for the $\mu$~energy
  in~$e^+e^- \to \mu^-\bar\nu_\mu u\bar d$ after canonical (\ADLO/)
  cuts.}
 
The lepton angles and lepton energies are very well under control.
For the lepton energies, the effect of the \emph{incompleteness error}
from leaving out the \emph{CC11} diagrams is not even noticeable.
 
The \emph{incompleteness error} for the lepton angles is noticeable,
but hardly measurable.  \PYTHIA/'s predictions are significantly
different from the other programs.
 
\subsubsection{Visible $\gamma$ Energy}
 
\dofour{tuned_egv_p1_y}{%
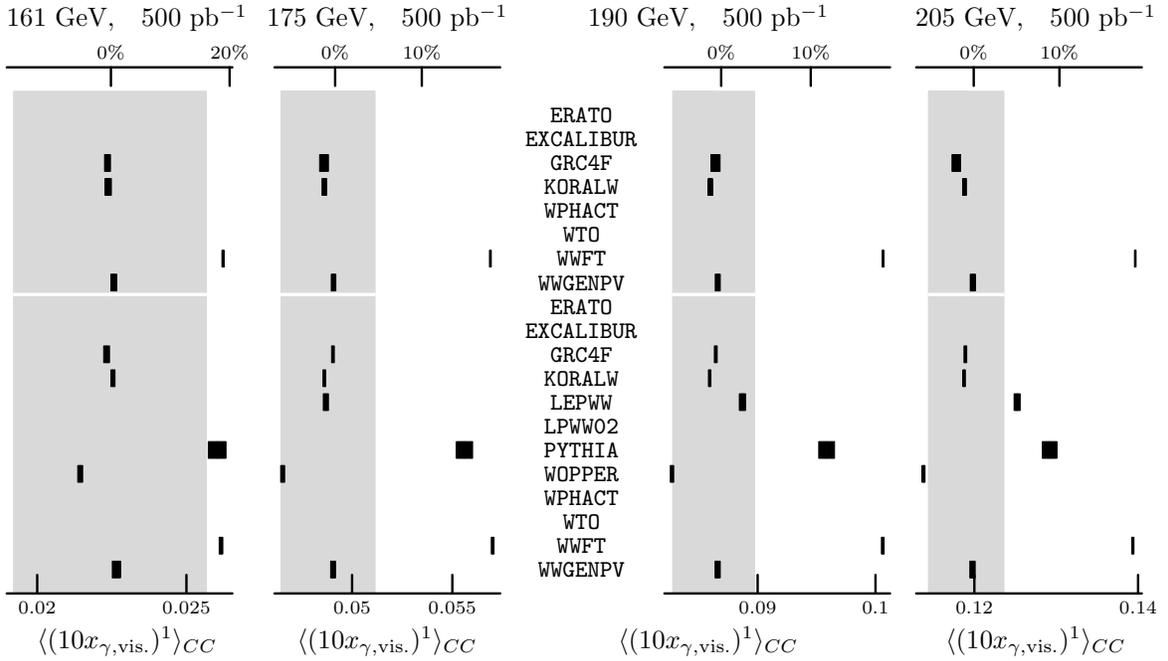
  Tuned predictions for the visible $\gamma$~energy
  in~$e^+e^- \to \mu^-\bar\nu_\mu u\bar d$ after canonical (\ADLO/) cuts.}
 
The situation for exclusive photonic observables is much less
satisfactory than the situation for the other observables studied.  This
should not be surprising, however.  The leading-logarithmic
approximation is theoretically justified using the renormalization group
and an operator product expansion for observables which are totally
inclusive in the photons.  A majority of programs implements this result
with structure functions and treats photons inclusively, treating
\emph{all} photons as emitted collinearly.
 
It is nevertheless possible to investigate the structure of the Feynman
diagrams contributing to the renormalization group evolution of the
structure functions.  This investigation shows that the leading
logarithms originate from a propagator pole
\begin{equation}
   \ln\left(\frac{s}{m_e^2}\right) = \int_{m_e^2}^{s} \frac{d(pk)}{pk}
\end{equation}
caused by the emission of almost collinear photons.  This observation
can be used to implement various parton shower algorithms for such
photons.  Another approach is to use $p_T$-dependent structure functions
that recover the $p_T$-dependence of the first-order matrix element.
 
In contrast to the structure function method which is unambiguously
defined by the renormalization group, these explicit resummations of
Feynman diagrams are not uniquely defined and can lead to differing
results.  These differences are reflected in our results.
 
\subsubsection{Final State Radiation}
\label{sec:FSR}
 
The canonical (\ADLO/) cuts are of calorimetric nature, i.e.~photons are
combined with nearby charged particles.  Therefore we should expect the
effect of final-state radiation to be very small and furthermore the
leading-logarithmic approximation to be sufficient.  Since some programs
have implemented final-state radiation, this assertion has to be checked.
 
We must, of course, again stress the fact that a theoretically meaningful
(i.e.~gauge invariant) separation of initial and final-state radiation
is \emph{not} possible in~$e^+e^- \to 4f+\gamma$.  The leading
logarithmic corrections, however, can be traced back to the mass
singularities in
initial-state radiation, and do form a gauge invariant subset.  From a
pragmatical point of view, it is also possible to calculate the
bremsstrahlung from the charged final-state particles.  The radiation
from off-shell intermediate states will likely contribute less than the
radiation from on-shell final states, because the latter contains
infrared and mass singularities.  Therefore one can argue that the
dominant radiative corrections will come from these diagrams.
 
This procedure has some pragmatical merit, but it should be kept in mind
that it could be justified only \emph{a posteriori}, after a full
calculation of the non-logarithmic terms is available.
 
At the time of the final meeting, a rather substantial effect for
exclusive observables was reported from a preliminary study using
the \ADLO/ cuts. The separation cut of 5~degrees for photons from
charged particles is rather tight, however.  For a realistic assessment
of the effect, a looser separation cut should be used.  A
study~\cite{wopper2} from 1994 (comparing version~1.1 of \WOPPER/ and
version~1.0 of \WWF/) had shown that about 20~degrees are required for
cutting the effect of final-state radiation at LEP2 energies.
 
Therefore, another study with modified canonical cuts has been
performed.  These cuts are identical to \ADLO/, except for the
photonic separation cuts.  In the results shown below, a photon is
counted as initial-state radiation if it is closer to a beam than to
any charged particle.  All other photons are counted as final-state
radiation and are combined with the closest charged particle.
 
In order to finish the study before the deadline, it was agreed to
perform only tuned comparisons, for the \emph{CC03} subset of diagrams.
 
The plots feature eight data sets:
\begin{itemize}
  \item \texttt{KORALW/FSR} and \texttt{KORALW}: results from
    \KORALW/, with and without final-state radiation, using the
    \emph{CC03} diagrams.  The final-state radiation is generated
    using the \texttt{PHOTOS} package~\cite{ref:Barberio}.
    \texttt{PHOTOS} has been modified to generate final-state radiation
    for quarks as well.
  \item \texttt{LPWW02/FSR} and \texttt{LPWW02}: results from
    \LPWW02/, with and without final-state radiation, using the
    \emph{CC03} diagrams.  The final-state radiation is generated
    using again the modified \texttt{PHOTOS} version.  \LPWW02/ does
    not include a finite~$p_T$ for the initial-state radiation.  This
    will reduce the effect from final-state radiation considerably.
  \item \texttt{WWF/FSR} and \texttt{WWF}: results from \WWF/, with
    and without final-state radiation, using the \emph{CC03} diagrams.
    \texttt{WWF/FSR} is the only data set in this study which uses a
    complete~$\mathcal{O}(\alpha)$ matrix element for hard radiation.
    The virtual corrections are not complete but the most important
    contributions have been included consistently by demanding
    the cancellation of infrared and mass divergences, leaving a
    theoretical uncertainty of~$\mathcal{O}(\alpha)$.
  \item \texttt{WWGENPV/FSR} and \texttt{WWGENPV}: results from
    \WWGENPV/, with and without final-state radiation, using the
    \emph{CC03} diagrams.  The final-state radiation is generated
    in leading-logarithmic approximation, using fragmentation
    functions (the final state equivalent of structure functions).
\end{itemize}
For some programs, another set of cuts has also been studied: \ADLO/
with a separation cut of 20~degrees.  These results will not be shown,
because they do not reveal anything unexpected.  They are inbetween
the results from fully inclusive and those from the \ADLO/
cuts, but closer to the former.
 
\dofour{fsr_xsectn_y}{%
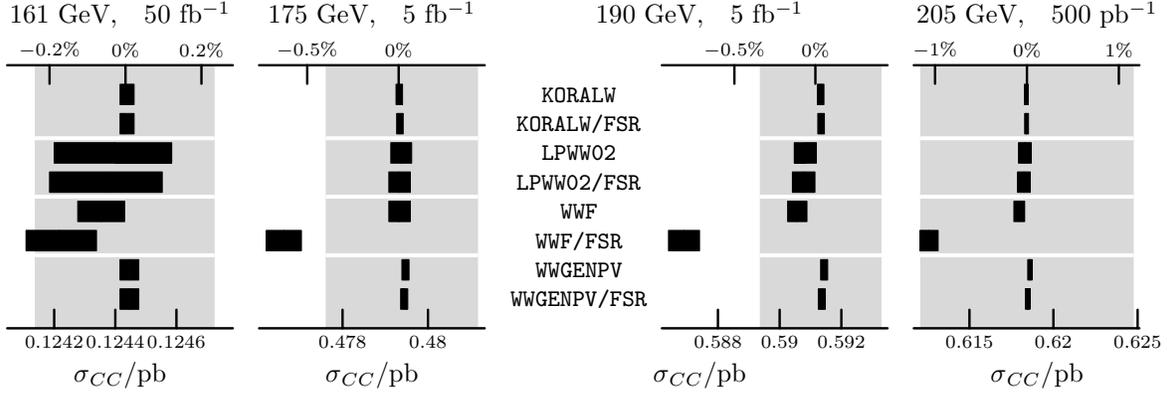
  The total cross sections with cuts are not affected by the inclusion
  of leading logarithmic final-state radiation.  See
  page~\pageref{pg:xsectn} for comments.}
For completely inclusive observables like the total cross section, we
should not expect any effect from final state parton showers, as
implemented in \texttt{PHOTOS} or in \texttt{WWGENPV}.
The sum of the probabilities for
radiating zero or~$N$ photons has to add up to one.  This expectation
is confirmed in figure~\ref{fig:fsr_xsectn_y}.  Since we are applying
acceptance cuts, a small residual effect will remain from charged
particles, that are ``kicked'' out of, or into, the acceptance cuts.
 
\label{pg:xsectn}
This is different for calculations including the
complete~$\mathcal{O}(\alpha)$ matrix element for hard radiation,
where non-trivial effects are possible.  The result from \WWF/ in
figure~\ref{fig:fsr_xsectn_y} shows that there is an uncertainty,
because the non -(infrared or mass)-divergent virtual contributions are
not taken
into account and the total cross section is expected to have a
theoretical error almost as big as the apparent deviation.
 
The phenomenologically most important issue is certainly the effect of
final-state radiation on the measured $W^\pm$~masses.  If a final-state
 particle radiates a sufficiently hard photon that is not
included in the corresponding ``jet'', a smaller invariant mass will
be measured.  We have to answer the question of whether this shift is
numerically important, and whether it is under control.
 
\dofour{fsr_mas_p1_y}{%
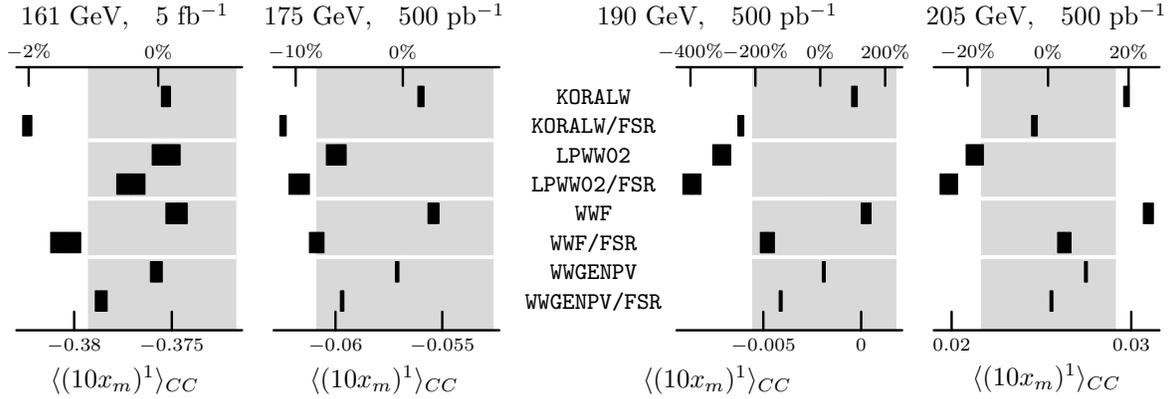
  The seemingly large shifts in~$\left\langle x_m \right\rangle$
  correspond to rather moderate shifts in the absolute values of the
  sum of invariant masses.  For the case of \WWF/ we have shifts
  of~$\approx90\MeV$.  See page~\pageref{pg:mass-shift} for comments.}
\label{pg:mass-shift}
{}From figure~\ref{fig:fsr_mas_p1_y}, we see that
both \KORALW/ and \WWF/ predict a shift in the sum of
invariant masses in the $80$--$90\MeV$ range.  Toggling options in
\WWF/, it can be verified that this shift is dominated by the leading
logarithms and that non-factorizable contributions are negligible.
 
On the other hand, \WWGENPV/ and \LPWW02/ predict smaller shifts of
$40\MeV$ and $30\MeV$, respectively.  For \texttt{LPWW02},
the difference can, presumably, be traced
back to the missing~$p_T$ in the initial-state radiation.
As for \texttt{WWGENPV}, the difference is probably due
to differences in the formulations.
 
As already observed in
figures~\ref{fig:tuned_mas_p1_n} and~\ref{fig:tuned_mas_p1_y}, a
finite~$p_T$ of the hard scattering system has a noticeable effect on
the invariant masses if \ADLO/ cuts are applied.
It must be noted, however, that these results are still very fresh,
and the work on this issue must be considered as still in progress.
Still, it can be said that all the $p_T$ codes give (apart from
small differences in particularly sensitive observables) consistent
results on the FSR issue.
 
Extrapolating the shift predicted by \KORALW/ and \WWF/ naively to a
single~$W^\pm$, we have an effect of about~$40\MeV$.  Measuring
exclusive photons and making use of constraints, the experiments
should be able to control this shift if event generators include final-state
 radiation in leading logarithmic approximation and initial-state
radiation with finite~$p_T$.  At the end of the day, the uncertainty
from final-state radiation will drop well below the anticipated
experimental resolution.
 
\dofour{fsr_ctw_t1_y}{%
  The programs based on leading logarithms show no measurable effect
  in the $W^\pm$~production angle.}
\dofour{fsr_ctm_t1_y}{%
  The programs based on leading logarithms show no measurable effect
  in the $\mu$~production angle.}
There is a hardly measurable effect of the hard-radiation matrix element
in \WWF/ on the $W^\pm$~production angle, as shown in
figure~\ref{fig:fsr_ctw_t1_y}.  This effect is of the order of~$1\%
\approx 4\alpha/\pi$ and corresponds to non-logarithmic contributions,
which can not be reproduced in the structure function and parton shower
calculations.
 
There is a similar effect of the hard-radiation
matrix element on the $\mu$~production angle, as shown in
figure~\ref{fig:fsr_ctm_t1_y}, where the~$\mu$'s are pulled towards the
forward direction.
 
For the decay angle of the~$\mu$ in the $W^\pm$'s~decay frame as well
as for its energy in the laboratory frame, there is
a tiny effect from final-state radiation, which is neither measurable
nor different for the LL programs from \WWF/.  It is completely
absent in \LPWW02/.
 
About one of the important quantities, the `lost'
photon energy, we want to remark the following.
All four programs that enter this comparison have studied
the total energy lost to `initial-state' radiation. This,
however, being not an unambiguously defined quantity, we
have settled on a definition as described above, where a
photon is deemed to be ISR if its angle with respect to
one of the beams is smaller than that with respect to any other
charged particle. We have studied the average value of both
the \emph{total} energy of emitted bremsstrahlung and that
of the \emph{lost} amount of energy. The total energy results
from the four programs are in a rather good agreement, with about twice
as much energy lost under ISR + FSR than under ISR alone.
If, however, we impose the cuts intended to define the more
meaningful `lost' bremsstrahlung energy, the agreement is
not so good at this moment. We ascribe this to yet remaining
differences in the cuts' implementation, and we refrain
from presenting a plot here, since we feel that it does not
adequately reflect the situation, which has to be
clarified in the near future.
 
Summing up, we see that the effects of final-state radiation are at
the level of the experimental resolution or below.  They have to be
studied in particular for a reliable determination of the
$W^\pm$~mass.  Therefore an inclusion of final-state radiation in the
event generators is desirable from a pragmatical point of view, even
before a theoretically satisfactory $\mathcal{O}(\alpha)$ matrix element
calculation is available.
 
It has, however, to be noted that the effect of final-state radiation
beyond the collinear approximation is crucially dependent on the
details of the cuts, and that the quantitative determination of
it has to rely on the use of those codes which implement such an
effect.
 
The differences between the leading logarithms and the
$\mathcal{O}(\alpha)$ matrix element for hard radiation in the total
cross section and some angular distributions will have to be
reevaluated when the virtual contributions in the latter
calculation will be complete.
 
\subsubsection{Conclusions}
Most Monte Carlo event generators, integration programs and
semi-analytic programs are ready for physics at LEP2, at least for the early,
low-luminosity stages.  However, once enough integrated luminosity has
been collected, only the high precision programs should be used:
\begin{itemize}
  \item Programs with \emph{incompleteness errors}, i.e.~omission of
    Feynman diagrams will have to be upgraded or retired.  This effort
    is known to be under way in some cases and users are encouraged to
    ask the authors for updated versions once in a while.
  \item We have concentrated on a typical \emph{CC10} process, which is
    dominated by the \emph{CC03} diagrams.  For processes with
    electrons in the final state, and also for processes like $u\bar u
    d\bar d$, the \emph{incompleteness errors} could be much larger.
    For these processes, the high-precision complete programs are
    relevant, unless fairly stringent invariant mass cuts are applied.
  Of course, to \emph{prove}  that  such cuts indeed
   allow for the use of an incomplete program, one has again
   to rely on a complete program  after all.
  \item
   For several observables, the effect of finite $p_T$ on both initial
   and final-state radiation is important. For these observables the
   programs implementing the effect of finite $p_T$ on photonic
   radiation are relevant, unless particular experimental cuts are
   applied.
  \item Authors of
  programs with bugs are encouraged to fix them.  \emph{At the very
    least, the results of this comparative study should be mentioned in
    the respective user manuals.}  Let us again repeat that deviations
    in the \emph{tuned comparisons} are \emph{not} theoretical errors
    but symptoms of bugs.
\item From the considerations of the effect of changes
 in the  theoretical approach (SF versus FF, or the use of
$\eta$ versus that of $\beta$  in the ISR), it is clear that
the theoretical error is \emph{not} much smaller than
the expected experimental one, at least for several important
quantities. Therefore we conclude  that the calculation
of the complete one-loop electroweak radiative
correction is of much more than purely academic interest.
\end{itemize}
 
In any case, it is safe to say that the
perfect, all-round Ultimate Monte Carlo
event generator for $W^\pm$-physics at LEP2 does \emph{not} exist.  In
all likelihood it will \emph{never} exist because different
implementation strategies lead to different strengths and weaknesses.
Usually this reflects more of the preferences and interests of the
respective authors than their ability to provide complete and bug-free
codes.
 
One important issue that has not been studied in detail by our group is
the implementation of \emph{anomalous couplings}~\cite{TGV-report}.
While a precise experimental determination of such couplings will in all
likelihood not be possible at LEP2, a similarly detailed analysis would
be valuable and might be performed in the future.
 
\clearpage
 
\subsection{\emph{CC11} processes \label{cc11_subs}}
%
%
\begin{table}[ht]\centering
\begin{tabular}{|c|c|c|c|c|}
\hline\hline
$E_{cm}$
&{\tt GE/4fan}
&{\tt WPHACT }
&{\tt WTO    }
&{\tt WWGENPV} \\
\hline\hline
\multicolumn{5}{|c|}{Born}                         \\
\hline\hline
  95&.52886(0) &.52890(10) &    ---    &.52895(8)  \\
 100&.63217(0) &.63220(10) &    ---    &.63218(6)  \\
 130&9.0560(0) &9.0559(5)  &    ---    &9.0560(7)  \\
 &{\it 9.0517(1)} &{\it 9.0522(4)}  &{\it 9.0530(25)} &{\it 9.0515(4)}\\
\hline
 160&.38447(0) &.38447(1)  &    ---    &.38446(1)  \\
 161&.53580(0) &.53581(2)  &    ---    &.53580(2)  \\
 175&1.77062(0)&1.77061(6) &    ---    &1.77061(6) \\
 176&1.80481(0)&1.80483(7) &    ---    &1.80483(7) \\
 &{\it 1.80445(2)}&{\it 1.80450(5)} &{\it 1.80446(4)} &{\it 1.80447(7)}\\
 190&2.04049(0)&2.04053(8) & 2.0403(1) &2.04048(10)\\
 205&2.05733(0)&2.05738(8) &    ---    &2.05743(10)\\
 &{\it 2.05631(2)}&{\it 2.05640(6)} &{\it 2.05637(8)} &{\it 2.05641(10)}\\
 300&1.49733(0)&1.49742(8) &    ---    &1.49735(7) \\
 500& .81482(0)& .81483(7) &    ---    & .81480(6) \\
1000& .32607(0)& .32607(5) &    ---    & .32602(6) \\
2000& .16684(0)& .16683(5) &    ---    & .16682(7) \\
 &{\it .10734(0) }&{\it .10737(7) }&{\it .10782(6) } &{\it .10727(5) }\\
\hline\hline
\multicolumn{5}{|c|}{With ISR}                     \\
\hline\hline
  95&.55170(1) &.55170(10) &.55190(70) &.55140(55) \\
 100&.57908(1) &.57910(10) &.57930(50) &.57937(34) \\
 130&7.5225(1) & 7.5221(7) &7.5219(13) &7.5214(15) \\
 &{\it 7.5187(1) }&{\it 7.5195(5) }&{\it 7.5215(15)} &{\it 7.5186(17)}\\
\hline
 160&.27563(1) &.27563(2)  &   ---     &.27563(3)  \\
 161&.38090(2) &.38090(2)  &.38092(4)  &.38092(4)  \\
 175&1.46646(1)&1.46649(6) &   ---     &1.46643(6) \\
 176&1.50459(2)&1.50457(9) &1.50464(10)&1.50453(7) \\
 &{\it 1.50430(2)}&{\it 1.50433(6)}&{\it 1.50423(12)}&{\it 1.50426(6)}\\
 190&1.81236(2)&1.81235(7) &1.81229(11)&1.81235(7) \\
 205&1.89984(2)&1.89986(12)&1.89995(8) &1.89996(10)\\
 &{\it 1.89897(2)}&{\it 1.89900(7)}&{\it 1.89896(34)}&{\it 1.89899(10)}\\
 300&1.51351(2)&1.51353(10)&1.51353(20)&1.51349(11)\\
 500& .86950(1)& .86956(9) & .86960(25)& .86956(14)\\
1000& .36514(1)& .36515(5) & .36554(49)& .36530(35)\\
2000& .18247(1)& .18250(4) &   ---     & .18247(13)\\
 &{\it .12800(0) }&{\it .12797(12)}&{\it .12858(48) }&{\it .12806(13) }\\
\hline\hline
\end{tabular}
\caption[.]{\emph{CC11} process. Cross sections are in
\emph{fb} for $E_{cm}=95, 100, 130$ GeV, in \emph{pb} for higher
energies. Numbers in {\it italics} correspond to constant $Z$ width.
\label{table_cc11} }
\vspace*{-7cm}
\end{table}
\normalsize
\clearpage
 
\begin{table}[t]\centering
\begin{tabular}{|l|c|c|c|}
\hline\hline
$E_{cm}$&       175           &        190          &        205     \\
\hline
\multicolumn{4}{|c|}{Born}                                           \\
\hline
{\tt ALPHA}    & 0.8152 $\pm$ 0.0004 & 9.505$\pm$0.005 & 12.505$\pm$0.006 \\
{\tt CompHEP}  & 0.8160 $\pm$ 0.0013 & 9.514$\pm$0.011 & 12.506$\pm$0.014 \\
{\tt EXCALIBUR}& 0.8162 $\pm$ 0.0011 & 9.514$\pm$0.008 & 12.499$\pm$0.010 \\
{\tt GENTLE/4fan}
           & 0.8157 $\pm$ .00001 & 9.511$\pm$.0001 & 12.500$\pm$.0001 \\
{\tt HIGGSPV}  & 0.8159 $\pm$ 0.0004 & 9.506$\pm$0.005 & 12.505$\pm$0.008 \\
{\tt WPHACT}   & 0.8150 $\pm$ 0.0008 & 9.509$\pm$0.006 & 12.501$\pm$0.007 \\
{\tt WTO}      & 0.8168 $\pm$ 0.0003 & 9.517$\pm$0.002 & 12.509$\pm$0.013 \\
\hline
\multicolumn{4}{|c|}{with ISR}                                       \\
\hline
{\tt EXCALIBUR}& 0.6478 $\pm$ 0.0004 & 7.371$\pm$0.003 & 10.789$\pm$0.004 \\
{\tt GENTLE/4fan}
           & 0.6481 $\pm$ 0.0001 & 7.370$\pm$0.001 & 10.791$\pm$0.001 \\
{\tt HIGGSPV}  & 0.6481 $\pm$ 0.0003 & 7.371$\pm$0.003 & 10.789$\pm$0.006 \\
{\tt WPHACT}   & 0.6482 $\pm$ 0.0006 & 7.367$\pm$0.007 & 10.784$\pm$0.008 \\
{\tt WTO}      & 0.6477 $\pm$ 0.0010 & 7.373$\pm$0.003 & 10.792$\pm$0.005 \\
\hline\hline
\multicolumn{4}{|c|}{Born}                                                \\
\hline
{\tt ALPHA}    & 0.7724 $\pm$ 0.0004 & 9.036$\pm$0.005 & 11.804$\pm$0.006 \\
{\tt CompHEP}  & 0.7732 $\pm$ 0.0014 & 9.058$\pm$0.012 & 11.834$\pm$0.016 \\
{\tt EXCALIBUR}& 0.7728 $\pm$ 0.0004 & 9.036$\pm$0.003 & 11.809$\pm$0.003 \\
{\tt HIGGSPV}  & 0.7728 $\pm$ 0.0003 & 9.034$\pm$0.006 & 11.814$\pm$0.006 \\
{\tt WPHACT}   & 0.7723 $\pm$ 0.0006 & 9.034$\pm$0.006 & 11.810$\pm$0.007 \\
{\tt WTO}      & 0.7739 $\pm$ 0.0002 & 9.042$\pm$0.002 & 11.818$\pm$0.001 \\
\hline
\multicolumn{4}{|c|}{with ISR}                                            \\
\hline
{\tt EXCALIBUR}& 0.6119 $\pm$ 0.0004 & 7.004$\pm$0.003 & 10.199$\pm$0.004 \\
{\tt HIGGSPV}  & 0.6128 $\pm$ 0.0003 & 7.002$\pm$0.004 & 10.199$\pm$0.005 \\
{\tt WPHACT}   & 0.6129 $\pm$ 0.0006 & 7.000$\pm$0.007 & 10.193$\pm$0.008 \\
{\tt WTO}      & 0.6128 $\pm$ 0.0010 & 7.007$\pm$0.002 & 10.203$\pm$0.006 \\
\hline\hline
\end{tabular}
\caption[]{Cross sections for the process
$e^+e^- \to \mu^+ \mu^- b {\bar b}$,
with invariant mass cuts:
$M_Z-15<m_{\mu\mu}<M_Z+15\;{\mbox{GeV}},\;\;m_{bb}>30\;{\mbox{GeV}},
\;\;m_b=0$. The two lower parts have additional cuts:
 lepton momenta $>10$ GeV, lepton polar angles with beams $>15^0$.
\label{table_1} }
\end{table}
\normalsize
 
\begin{table}[t]\centering
\begin{tabular}{|l|c|c|c|}
\hline\hline
$E_{cm}$&       175           &        190          &        205     \\
\hline
\multicolumn{4}{|c|}{Born}                                            \\
\hline
{\tt ALPHA}    & 1.5863 $\pm$ 0.0009 & 18.375$\pm$0.009& 24.138$\pm$0.012 \\
{\tt CompHEP}  & 1.5785 $\pm$ 0.0030 & 18.352$\pm$0.030& 24.180$\pm$0.039 \\
{\tt EXCALIBUR}& 1.5916 $\pm$ 0.0020 & 18.398$\pm$0.020& 24.141$\pm$0.015 \\
{\tt GENTLE/4fan}
           & 1.5878 $\pm$0.00002 & 18.381$\pm$.0002& 24.150$\pm$.0002 \\
{\tt HIGGSPV}  & 1.5876 $\pm$ 0.0011 & 18.376$\pm$0.014& 24.150$\pm$0.021 \\
{\tt WPHACT}   & 1.5868 $\pm$ 0.0013 & 18.383$\pm$0.011& 24.151$\pm$0.013 \\
{\tt WTO}      & 1.5864 $\pm$ 0.0024 & 18.378$\pm$0.002& 24.159$\pm$0.008 \\
\hline
\multicolumn{4}{|c|}{with ISR}                                       \\
\hline
{\tt EXCALIBUR}& 1.2770 $\pm$ 0.0008 & 14.243$\pm$0.008& 20.840$\pm$0.010 \\
{\tt GENTLE/4fan}
           & 1.2782 $\pm$ 0.0001 & 14.243$\pm$0.001& 20.838$\pm$0.002 \\
{\tt HIGGSPV}  & 1.2781 $\pm$ 0.0008 & 14.248$\pm$0.009& 20.846$\pm$0.014 \\
{\tt WPHACT}   & 1.2773 $\pm$ 0.0010 & 14.235$\pm$0.014& 20.827$\pm$0.017 \\
{\tt WTO}      & 1.2799 $\pm$ 0.0027 & 14.246$\pm$0.004& 20.833$\pm$0.005 \\
\hline\hline
\multicolumn{4}{|c|}{Born}                                            \\
\hline
{\tt ALPHA}    & 1.4204 $\pm$ 0.0008 & 16.767$\pm$0.008 & 21.784$\pm$0.010 \\
{\tt CompHEP}  & 1.4141 $\pm$ 0.0032 & 16.748$\pm$0.032 & 21.851$\pm$0.044 \\
{\tt EXCALIBUR}& 1.4197 $\pm$ 0.0009 & 16.750$\pm$0.008 & 21.782$\pm$0.010 \\
{\tt HIGGSPV}  & 1.4199 $\pm$ 0.0009 & 16.771$\pm$0.012 & 21.782$\pm$0.016 \\
{\tt WPHACT}   & 1.4197 $\pm$ 0.0014 & 16.775$\pm$0.013 & 21.785$\pm$0.015 \\
{\tt WTO}      & 1.4169 $\pm$ 0.0021 & 16.766$\pm$0.002 & 21.776$\pm$0.004 \\
\hline
\multicolumn{4}{|c|}{with ISR}                                        \\
\hline
{\tt EXCALIBUR}& 1.1423 $\pm$ 0.0008 & 12.995$\pm$0.008 & 18.812$\pm$0.010 \\
{\tt HIGGSPV}  & 1.1437 $\pm$ 0.0007 & 13.001$\pm$0.011 & 18.799$\pm$0.017 \\
{\tt WPHACT}   & 1.1430 $\pm$ 0.0010 & 13.001$\pm$0.009 & 18.813$\pm$0.018 \\
{\tt WTO}      & 1.1449 $\pm$ 0.0021 & 13.003$\pm$0.003 & 18.814$\pm$0.007 \\
\hline\hline
\end{tabular}
\caption[]{Cross sections for the process
$e^+e^- \to \nu_{\mu} {\bar \nu}_{\mu} b {\bar b}$
with invariant mass cuts:
$M_Z-25<m_{\mu\mu}<M_Z+25\;{\mbox{GeV}},\;\;m_{bb}>30\;{\mbox{GeV}},
\;\;m_b=0$. The lower parts have an addition cut of 20 degrees on the
angle of the $b$'s with respect to both beams.
\label{table_2} }
\end{table}
\normalsize
 
\begin{table}[t]\centering
\begin{tabular}{|l|c|c|c|}
\hline\hline
$E_{cm}$&       175           &        190          &        205           \\
\hline
\multicolumn{4}{|c|}{Born}                                                 \\
\hline
{\tt ALPHA}    & 1.3940 $\pm$ 0.0007 & 18.299$\pm$0.009 & 26.361$\pm$0.013 \\
{\tt CompHEP}  & 1.3909 $\pm$ 0.0029 & 18.309$\pm$0.031 & 26.470$\pm$0.051 \\
{\tt HIGGSPV}  & 1.3946 $\pm$ 0.0005 & 18.294$\pm$0.011 & 26.348$\pm$0.011 \\
{\tt WPHACT}   & 1.3955 $\pm$ 0.0010 & 18.314$\pm$0.012 & 26.384$\pm$0.017 \\
{\tt WTO}      & 1.3937 $\pm$ 0.0029 & 18.304$\pm$0.004 & 26.386$\pm$0.008 \\
\hline
\multicolumn{4}{|c|}{with ISR}                                             \\
\hline
{\tt HIGGSPV}  & 1.1444 $\pm$ 0.0004 & 14.053$\pm$0.009 & 22.490$\pm$0.012 \\
{\tt WPHACT}   & 1.1440 $\pm$ 0.0010 & 14.064$\pm$0.010 & 22.505$\pm$0.020 \\
{\tt WTO}      & 1.1483 $\pm$ 0.0028 & 14.068$\pm$0.003 & 22.508$\pm$0.009 \\
\hline\hline
\multicolumn{4}{|c|}{Born}                                                 \\
\hline
{\tt ALPHA}    & 1.2466 $\pm$ 0.0007 & 16.732$\pm$0.008 & 23.843$\pm$0.012 \\
{\tt CompHEP}  & 1.2430 $\pm$ 0.0031 & 16.761$\pm$0.034 & 23.965$\pm$0.054 \\
{\tt EXCALIBUR}& 1.2458 $\pm$ 0.0008 & 16.727$\pm$0.008 & 23.862$\pm$0.015 \\
{\tt HIGGSPV}  & 1.2463 $\pm$ 0.0005 & 16.715$\pm$0.009 & 23.822$\pm$0.013 \\
{\tt WPHACT}   & 1.2473 $\pm$ 0.0010 & 16.749$\pm$0.013 & 23.855$\pm$0.018 \\
{\tt WTO}      & 1.2457 $\pm$ 0.0023 & 16.735$\pm$0.004 & 23.855$\pm$0.006 \\
\hline
\multicolumn{4}{|c|}{with ISR}                                             \\
\hline
{\tt EXCALIBUR}& 1.0227 $\pm$ 0.0007 & 12.865$\pm$0.008 & 20.381$\pm$0.015 \\
{\tt HIGGSPV}  & 1.0239 $\pm$ 0.0004 & 12.853$\pm$0.008 & 20.306$\pm$0.042 \\
{\tt WPHACT}   & 1.0229 $\pm$ 0.0010 & 12.865$\pm$0.010 & 20.378$\pm$0.015 \\
{\tt WTO}      & 1.0263 $\pm$ 0.0022 & 12.864$\pm$0.003 & 20.377$\pm$0.008 \\
\hline\hline
\end{tabular}
\caption[]{Cross sections for the process
$e^+e^- \to \nu_e {\bar \nu}_e b {\bar b}$ under the same cuts as
table~\ref{table_2}.
\label{table_3} }
\end{table}
\normalsize
 
\noindent
  A few codes have performed a very precise ($\simeq 10^{-4}$)
  \emph{tuned comparison} of the total cross section of a \emph{CC11}
  process, $e^+e^-\to u \bar d s \bar c$, in a broad CM energy
  range, $130\div2000$ GeV, using the input parameters of tuned
  comparison, as in
  table~\ref{tab:input} both with \emph{running} and
  \emph{constant} $Z$ widths.
The results are given in table \ref{table_cc11}.
 
 An interesting conclusion can be drawn from comparing these
two cases. There is practically no difference between running at
constant $Z$ widths result at LEP2 energies, whereas at $E_{cm}=2000$
GeV the running $Z$ width results starts to blow up. This is an
illustration of gauge-invariance violation, see \cite{sibling-report}.
 
 This comparison was attempted at an early phase of our
work. The extreme accuracy served as a very efficient tool
for hunting down many tiny bugs. Furthermore, it demonstrates that
a level of precision of the order $10^{-4}$ is now within the reach
of not only semi-analytical but also adaptive Monte Carlo integrators.
 
\section{Comparisons of NC processes}
 
Here we present the results of the \emph{tuned} comparison
for three \emph{NC} processes
\emph{NC24, NC10, NC21}. We computed only cross sections at three c.m.s
energies: $175, 190$ and $205$ GeV with simple cuts.
Seven codes participated in this comparison.
 
We have concentrated on processes where a $b\bar b$~pair is produced
together with two leptons, since these can form an important
background for the production and decay of a light Higgs boson.
All cross sections are given in fb: since they are quite small, we
have not pursued detailed comparisons of other quantities as we have
done for the \emph{CC} processes.
 
{}From the tables it is apparent that the agreement among the various
codes is very good, both at the Born level and after inclusion of
ISR.  The cuts have been chosen so as to be more or less realistic in
an experimental Higgs search.
 
\clearpage
\section{All four-fermion processes \label{ayc_comp}}
 
\noindent
In the following two subsections we present the cross sections for many four
fermion processes at only one center-of-mass energy,
$\sqrt{s}=190$ GeV, in the massless approximation $m_f=0$,
with the Standard LEP2
Input, see table \ref{tab:input}. In the first subsection,
{\it all} 32 four-fermion processes are presented. They are calculated with
the standard Canonical Cuts.
The four-fermion processes are ordered in accordance with the classification
of tables \ref{tab1}-\ref{tab2}.
For historical reasons, the Born cross sections are presented in the
Report of the Working Group on Standard Model Processes,
\cite{SMP-report}. The
tables of the next subsection contain numbers computed
\emph{with} the ISR radiation (SF)
and \emph{with} gluon exchange diagrams for
non-leptonic processes.
 
Since this is a tuned comparison
all codes have used a fixed strong coupling constant,
$\alpha_{_S} =  0.12$.
Obviously, any further study of the non-leptonic processes
must include some educated guess on the scale of $\alpha_{_S}$,
e.g. $\alpha_{_S}(s_{\pm})$ (running) or
$\alpha_{_S}(2M_W)$ (fixed).
 
The precision of the computation is quite high, normally better than $.1\%$.
These numbers are supposed to provide benchmarks for future calculations
of four-fermion processes.
 
\subsection{AYC, Canonical Cuts}
 
%
%
\begin{table}[ht]\centering
\begin{tabular}{|c|c|c|c|c|c|c|}
\hline\hline
final state
&{\tt CompHEP}&{\tt EXCALIBUR}&{\tt grc4f}
&{\tt WPHACT}&{\tt WTO}&{\tt WWGENPV} \\
\hline\hline
   $\mmn\numb\nut\tpl$
&.1947(5)  &.1941(1) &.1941(2) &.1942(2) &.1941(0) &.1941(1) \\
\hline\hline
\ru{$\mmn\numb u\dbar$}
&.5917(11) &.5916(3) &.5919(5) &.5921(5) &.5919(0) &.5920(6) \\
\hline \hline
\ru{$u\dbar s\cbar$}
&1.791(5)  &1.788(1) &1.791(2) &1.789(1) &1.788(0) &1.789(1) \\
\hline \hline
\end{tabular}
\caption[.]{\emph{CC11, CC10, CC09} family.  Cross sections in pb.
\label{tableone} }
\vspace{-0.25cm}
\end{table}
\normalsize
%
%
\begin{table}[h]\centering
\begin{tabular}{|c|c|c|c|c|c|c|c|}
\hline\hline
final state
&{\tt CompHEP}&{\tt ERATO}
&{\tt EXCALIB}&{\tt grc4f}&{\tt WPHACT}&{\tt WTO}&{\tt WWGENPV}\\
\hline\hline
  $\emn\nueb\num\mpl$
&.2012(6) &   ---   &.2014(1)
&.2014(3) &.2015(1) &.2014(2) &.2013(4) \\
\hline\hline
\ru{$\emn\nueb u\dbar$}
&.6131(12)&.6139(6) &.6140(4)
&.6135(4) &.6135(6) &.6137(6) &.6134(12)\\
\hline \hline
\end{tabular}
\caption[.]{\emph{CC20, CC18} family. Cross sections in pb.
\label{tabletwo} }
\vspace*{-.25cm}
\end{table}
\normalsize
%
%
\begin{table}[h]\centering
\begin{tabular}{|c|c|c|c|c|c|}
\hline\hline
final state
&{\tt CompHEP}
&{\tt EXCALIBUR}&{\tt grc4f}&{\tt WPHACT}&{\tt WTO}\\
\hline\hline
  $\mpl\mmn\num\numb$
&.2018(8) &.2049(1)   &.2029(4)  &.2050(0) &.2032(3) \\
\hline\hline
 \ru{$u\ubar d\dbar$}
&1.967(8) &1.992(2)   &1.985(4)  &1.992(0) &1.980(6) \\
\hline \hline
\end{tabular}
\caption[.]{{\tt mix43} family. Cross sections in pb.
\label{tabletri} }
\vspace*{-.25cm}
\end{table}
\normalsize
%
%
\begin{table}[h]\centering
\begin{tabular}{|c|c|c|c|c|}
\hline\hline
final state
&{\tt CompHEP}&{\tt EXCALIBUR}&{\tt grc4f}&{\tt WPHACT} \\
\hline\hline
  $\emn\epl\nue\nueb$
&.2244(12) &.2294(2) &.2289(7)  &.2292(2) \\
\hline \hline
\end{tabular}
\caption[.]{{\tt mix56} process. Cross sections in pb.
\label{tablefor} }
\vspace*{-7cm}
\end{table}
\normalsize
\clearpage
%
%
\begin{table}[ht]\centering
\begin{tabular}{|c|c|c|c|c|c|c|}
\hline\hline
final state
&{\tt CompHEP}&{\tt EXCALIB}&{\tt grc4f}
&{\tt HIGGSPV}&{\tt WPHACT}&{\tt WTO} \\
\hline\hline
  $\mpl\mmn\tpl\tmn$
&13.19(9)   &13.38(3)  &13.28(4)   &13.32(1)  &13.33(2)  &13.26(14)\\
\hline
  $\nut\nutb\mpl\mmn$
&10.75(4)   &10.71(2)  &10.71(1)   &10.720(4) &10.72(1)  &10.76(13)\\
\hline
  $\num\numb\nut\nutb$
&6.366(8)   &6.377(3)  &6.373(4)   &6.377(5)  &6.376(1)  &6.375(0) \\
\hline\hline
   $\mpl\mmn u\ubar$
&27.09(9)   &27.29(5)  &27.20(2)   &27.22(2)  &27.24(3)  &27.16(24)\\
\hline
\ru{$\mpl\mmn d\dbar$}
&25.39(17)  &25.49(5)  &25.44(2)   &25.48(1)  &25.49(2)  &25.37(13)\\
\hline
   $\num\numb u\ubar$
&18.17(6)   &18.22(1)  &18.20(3)   &18.22(1)  &18.21(1)  &18.22(5) \\
\hline
\ru{$\num\numb d\dbar$}
&15.80(5)   &15.84(1)  &15.85(2)   &15.83(1)  &15.83(1)  &15.83(1) \\
\hline \hline
   $u\ubar c\cbar$
&210.7(15)  &206.8(7)  &208.3(4)   &207.8(2)  &208.0(2)  &208.9(5) \\
\hline
   $u\ubar s\sbar$
&203.6(13)  &203.5(8)  &203.7(6)   &203.0(2)  &203.2(2)  &204.4(5) \\
\hline
 \ru{$d\dbar s\sbar$}
&183.8(19)  &182.2(10) &181.0(4)   &181.2(2)  &181.3(2)  &182.6(5) \\
\hline \hline
\end{tabular}
\caption[.]{\emph{NC32, NC24, NC10, NC06} family. Cross sections in fb.
\label{tablefiv} }
\vspace*{-.25cm}
\end{table}
\normalsize
%
%
\begin{table}[h]\centering
\begin{tabular}{|c|c|c|c|c|c|c|}
\hline\hline
final state
&{\tt CompHEP}&{\tt EXCALIB}&{\tt grc4f}
&{\tt HIGGSPV}&{\tt WPHACT}&{\tt WTO}\\
\hline\hline
   $\nue\nueb\mpl\mmn$
&18.07(8)   &18.03(5)  &17.98(5)  &18.07(1)  &18.05(2)  &17.83(13)\\
\hline
   $\nue\nueb\num\numb$
&6.408(9)   &6.417(3)  & 6.408(5) &6.364(91) &6.416(1)  &6.439(5) \\
\hline\hline
  $\nue\nueb u\ubar$
&20.78(5)   &20.74(1)  &20.74(4)  &20.78(16) &20.72(3)  &20.95(9) \\
\hline
\ru{$\nue\nueb d\dbar$}
&16.12(4)   &16.48(1)  &16.48(2)  &16.37(17) &16.46(2)  &16.67(15)\\
\hline \hline
\end{tabular}
\caption[.]{\emph{NC21, NC12} family. Cross sections in fb.
\label{tablesix} }
\end{table}
\normalsize
%
%
\begin{table}[h]\centering
\begin{tabular}{|c|c|c|c|c|c|}
\hline\hline
final state &{\tt CompHEP}&{\tt EXCALIBUR}
&{\tt grc4f}&{\tt HIGGSPV}&{\tt WPHACT} \\
\hline\hline
  $\epl\emn\mpl\mmn$
&.1231(15)  &.1251(2)   &.1247(5)  &.1192(21) &.1253(2)  \\
\hline
  $\epl\emn\num\numb$
&.01421(8)  &.01426(2)  &.01421(2) &.01445(18)&.01429(2) \\
\hline\hline
  $\epl\emn u\ubar $
&.09070(76) &.09234(11) &.09226(12)&.09003(89)&.09244(14)\\
\hline
  $\epl\emn d\dbar $
&.04259(45) &.04427(6)  &.04425(4) &.04491(46)&.04429(8) \\
\hline \hline
\end{tabular}
\caption[.]{\emph{NC48} family. Cross sections in pb.
\label{tablesev} }
\vspace*{-.25cm}
\end{table}
\normalsize
%
%
\begin{table}[h]\centering
\begin{tabular}{|c|c|c|c|c|c|}
\hline\hline
final state
&{\tt CompHEP}&{\tt EXCALIBUR}&{\tt grc4f}&{\tt HIGGSPV}
&{\tt WPHACT}\\
\hline\hline
  $\mpl\mmn\mpl\mmn$
&    ---    &.006650(17)&.006643(30)&.006671(85)&.006622(13)\\
\hline
  $\num\numb\num\numb$
&.003176(7) &.003142(1) &.003141(4) &.003142(7) &.003142(1) \\
\hline\hline
  $u\ubar u\ubar$
&    ---    &.1017(3)   &.1020(5)   &    ---    &.1014(1)   \\
\hline
 \ru{$d\dbar d\dbar$}
&    ---    &.08765(38) &.08767(17) &    ---    &.08788(22) \\
\hline\hline
\end{tabular}
\caption[.]{\emph{NC4x16, NC4x12} family. Cross sections in pb.
\label{tableegh} }
\vspace{-.25cm}
\end{table}
\normalsize
%
%
\begin{table}[h]\centering
\begin{tabular}{|c|c|c|c|c|}
\hline\hline
final state
&{\tt CompHEP}&{\tt EXCALIBUR}&{\tt grc4f}&{\tt WPHACT} \\
\hline\hline
  $\epl\emn\epl\emn$
&    ---    &.1169(2)   &.1156(11)  &.1169(2)   \\
\hline
  $\nue\nueb\nue\nueb$
&.003194(18)&.003123(1) &.003128(3) &.003125(1) \\
\hline \hline
\end{tabular}
\caption[.]{\emph{NC4x36} and \emph{NC4x9} processes.
Cross sections in pb.
\label{tablenin} }
\vspace{-7cm}
\end{table}
\normalsize
\clearpage
\subsection{AYC, Simple Cuts}
 
%
%
\begin{table}[ht]\centering
\vspace*{-.25cm}
\begin{tabular}{|c|c|c|c|c|c|c|c|}
\hline\hline
final state
&{\tt  ALPHA }
&{\tt EXCALIB}
&{\tt GE/4fan}
&{\tt grc4f}
&{\tt WPHACT }
&{\tt WTO    }
&{\tt WWGENPV} \\
\hline\hline
\multicolumn{8}{|c|}{Born}           \\
\hline\hline
   $\mmn\numb\nut\tpl$
&.2264(2) &.2267(1) &.2267(0) &.2267(1)
&.2267(0) &.2267(0) &.2267(0)  \\
\hline\hline
\ru{$\mmn\numb u\dbar$}
&.6804(4) &.6801(4) &.6801(0) &.6799(2)
&.6801(1) &.6801(0) &.6801(0)  \\
\hline \hline
 \ru{$u\dbar s\cbar$}
&2.040(1) &2.040(1) &2.040(0) &2.040(1)
&2.041(0) &2.040(0) &2.040(0)  \\
\hline\hline
\multicolumn{8}{|c|}{With ISR}                                        \\
\hline\hline
   $\mmn\numb\nut\tpl$
&   ---   &.2013(1) &.2014(0) &.2014(1)
&.2014(0) &   ---   &.2014(0) \\
\hline\hline
\ru{$\mmn\numb u\dbar$}
&   ---   &.6036(4) &.6041(0) &.6041(3)
&.6041(0) &.6041(0) &.6041(1) \\
\hline \hline
 \ru{$u\dbar s\cbar$}
&   ---   &1.811(1) &1.812(0) &1.812(1)
&1.812(0) &1.812(0) &1.812(0) \\
\hline \hline
\end{tabular}
\caption[.]{\emph{CC11, CC10, CC09} family. Cross sections in pb.
\label{tabletwopr}}
\vspace*{-0.25cm}
\end{table}
\normalsize
 
%
%
\begin{table}[h]\centering
\begin{tabular}{|c|c|c|c|c|c|}
\hline\hline
final state
&{\tt ALPHA  }
&{\tt EXCALIB}
&{\tt grc4f}
&{\tt HIGGSPV}
&{\tt WPHACT } \\
\hline\hline
\multicolumn{6}{|c|}{Born}      \\
\hline\hline
  $\nue\nueb\mpl\mmn$
&12.40(1) &12.38(1) &12.37(1) &12.37(1) &12.38(1) \\
\hline
 $\nue\nueb\num\numb$
&8.335(4) &8.336(3) &8.335(6) &8.342(5) &8.339(1) \\
\hline\hline
  $\nue\nueb u\ubar$
&24.95(2) &24.92(1) &24.92(2) &25.01(3) &24.91(1) \\
\hline
\ru{$\nue\nueb d\dbar$}
&20.91(2) &20.92(1) &20.91(1) &20.90(3) &20.92(1) \\
\hline\hline
\multicolumn{6}{|c|}{With ISR}   \\
\hline\hline
  $\nue\nueb\mpl\mmn$
&   ---   &11.59(1) &11.59(1) &11.59(1) &11.60(0) \\
\hline
 $\nue\nueb\num\numb$
&   ---   &6.412(3) &6.408(5) &6.411(7) &6.416(1) \\
\hline\hline
  $\nue\nueb u\ubar$
&   ---   &21.87(1) &21.88(2) &21.94(2) &21.86(1) \\
\hline
\ru{$\nue\nueb d\dbar$}
&   ---   &16.75(1) &16.76(1) &16.74(2) &16.75(1) \\
\hline \hline
\end{tabular}
\caption[.]{\emph{NC21, NC12} family.  Cross sections in pb.
\label{tabletripr}}
\end{table}
\normalsize
 
\noindent
In this subsection, only those processes are
given that were treated within the semi-analytic approach with Simple
Cuts on the invariant mass of any charged fermion-antifermion pair.
The latter cut value is chosen to be 5 GeV.
Every table contains two sets of numbers
which are computed: \\
1. in the Born approximation and without gluon exchange diagrams for
non-leptonic processes; \\
2. with the ISR radiation (SF) and with gluon exchange diagrams for
non-leptonic processes. \\
 
\subsection{Conclusions}
 
We want to stress that many of the codes contributing to the ``all you
can'' comparison have been developed during this workshop.
The level of agreement documented in these tables demonstrates a
substantial progress in our understanding of the general $e^+e^-\to4f$
cross section.
 
 
%
%
\begin{table}[t]\centering
\begin{tabular}{|c|c|c|c|c|c|c|c|}
\hline\hline
final state
&{\tt  ALPHA }
&{\tt EXCALIB}
&{\tt GE/4fan}
&{\tt grc4f}
&{\tt HIGGSPV}
&{\tt WPHACT }
&{\tt WTO    }\\
\hline\hline
\multicolumn{8}{|c|}{Born, without gluon exchange diagrams} \\
\hline\hline
    $\mpl\mmn\tpl\tmn$
&10.06(9) &10.08(0) &10.07(0) &10.07(0) &10.07(0) &10.07(0) &10.14(7) \\
\hline
   $\nut\nutb\mpl\mmn$
&9.894(10)&9.872(3) &9.871(0) &9.875(4) &9.872(3) &9.873(3) &9.884(10)\\
\hline
   $\num\numb\nut\nutb$
&8.245(4) &8.242(3) &8.241(0) &8.240(4) &8.237(6) &8.241(1) &8.241(1) \\
\hline\hline
    $\mpl\mmn u\ubar$
&23.99(2) &24.04(1) &24.03(0) &24.04(2) &24.03(1) &24.04(1) & ---  \\
\hline
  \ru{$\mpl\mmn d\dbar$}
&23.46(2) &23.45(1) &23.45(0) &23.46(2) &23.45(1) &23.46(1) & ---  \\
\hline
   $\num\numb u\ubar$
&21.59(2) &21.59(1) &21.59(0) &21.58(1)&21.58(1) &21.59(1) &21.63(3) \\
\hline
 \ru{$\num\numb d\dbar$}
&20.00(2) &19.99(1) &19.99(0) &20.00(1)&20.00(1) &19.99(1) &20.00(1) \\
\hline \hline
     $u\ubar c\cbar$
&54.80(5) &54.75(2) &54.74(0) &54.73(4) &54.69(4) &54.74(2) & ---  \\
\hline
     $u\ubar s\sbar$
&51.83(5) &51.86(1) &51.86(0) &51.85(2) &51.85(5) &51.87(2) & ---  \\
\hline
   \ru{$d\dbar s\sbar$}
&48.30(5) &48.33(2) &48.33(0) &48.34(1) &48.27(6) &48.34(1) & ---  \\
\hline\hline
\multicolumn{8}{|c|}{With ISR, with gluon exchange diagrams}    \\
\hline\hline
    $\mpl\mmn\tpl\tmn$
&   ---   &10.29(0) &10.30(0) &10.29(1) &10.30(0) &10.30(0) & ---  \\
\hline
   $\nut\nutb\mpl\mmn$
&   ---   &9.279(3) &9.284(1) &9.278(7) &9.283(3) &9.284(4) & ---  \\
\hline
   $\num\numb\nut\nutb$
&   ---   &6.379(3) &6.376(1) &6.373(4) &6.377(5) &6.377(1) &6.379(2) \\
\hline\hline
    $\mpl\mmn u\ubar$
&   ---   &23.74(1) &23.76(0) &23.77(2) &23.75(1) &23.75(1) & ---  \\
\hline
  \ru{$\mpl\mmn d\dbar$}
&   ---   &22.31(1) &22.34(0) &22.33(1) &22.33(1) &22.34(1) & ---  \\
\hline
   $\num\numb u\ubar$
&   ---   &18.83(1) &18.84(0) &18.84(1) &18.85(1) &18.84(1) & ---  \\
\hline
 \ru{$\num\numb d\dbar$}
&   ---   &16.00(1) &15.99(0) &15.99(1) &16.00(1) &15.99(0) & ---  \\
\hline \hline
     $u\ubar c\cbar$
&   ---   &272.6(9) &272.3(0) &271.4(9) &272.1(1) &272.2(1) & ---  \\
\hline
     $u\ubar s\sbar$
&   ---   &267.0(10)&266.8(0) &266.5(6) &266.8(1) &266.8(1) & ---  \\
\hline
   \ru{$d\dbar s\sbar$}
&   ---   &240.7(11)&240.8(0) &240.5(6) &240.6(4) &240.8(1) & ---  \\
\hline \hline
\end{tabular}
\caption[.]{\emph{NC32, NC24, NC10, NC06} family. Cross sections in fb.
\label{tableforpr}}
\end{table}
\normalsize
 
However, this comparison revealed also some problems, e.g.:
some numbers still disagree within declared errors;
during the collection of these tables, some codes exhibited fluctuations
much larger than the statistical errors; we didn't attempt a comparison
of CPU times, needed by different codes to reach a given accuracy.
All these items deserve a more thorough study in the future.
 
%
\section*{Acknowledgments}
We have to thank Francesca Cavallari,  Jules Gascon,  Martin
Gr\"unewald, Niels Kjaer, and Jerome Schwindling for helping us
to define realistic \texttt{ADLO/TH} cuts, which have been used
extensively in the comparisons of our programs.
%

\end{document}